\documentclass[fleqn,usenatbib]{mnras}

\usepackage{newtxtext,newtxmath}

\usepackage[T1]{fontenc}

\DeclareRobustCommand{\VAN}[3]{#2}
\let\VANthebibliography\thebibliography
\def\thebibliography{\DeclareRobustCommand{\VAN}[3]{##3}\VANthebibliography}

\usepackage{graphicx}	
\usepackage{amsmath}	

\defcitealias{reguitti2022}{R22}

\title[The transitioning SN 2021foa]{SN 2021foa: the bridge between
SN IIn and Ibn}

\author[Anjasha Gangopadhyay et al.]{
Anjasha Gangopadhyay,$^{1,2}$\thanks{E-mail: anjashagangopadhyay@gmail.com}
Naveen Dukiya$^{3,4}$, Takashi J Moriya$^{5,6,7}$, Masaomi Tanaka$^{8}$, Keiichi Maeda$^{9}$, 
\newauthor
D. Andrew Howell$^{10,11}$, Mridweeka Singh$^{12}$, Avinash Singh$^{1,2}$, Jesper Sollerman$^{1}$, Koji S Kawabata$^{2}$, 
\newauthor 
Se{\'a}n J Brennan$^{1}$, Craig Pellegrino$^{15}$, Raya Dastidar$^{13,14}$,Tatsuya Nakaoka$^{2}$, Miho Kawabata$^{16}$, Kuntal Misra$^{3}$,
\newauthor 
Steve Schulze$^{17}$, Poonam Chandra$^{18,26}$, Kenta Taguchi$^{16}$, Devendra K Sahu$^{12}$, Curtis McCully$^{10,11}$, 
\newauthor
K. Azalee Bostroem$^{19}$, Estefania Padilla Gonzalez$^{10,11}$, Megan Newsome$^{10,11}$, Daichi Hiramatsu$^{20,28}$, 
\newauthor
Yuki Takei$^{21,22,23}$, Masayuki Yamanaka$^{24}$, Akito Tajitsu$^{25}$, Keisuke Isogai$^{25,27}$
\\
$^{1}$Oskar Klein Centre, Department of Astronomy, Stockholm University, AlbaNova, SE-106 91 Stockholm, Sweden \\
$^{2}$Hiroshima Astrophysical Science Centre, Hiroshima University, 1-3-1 Kagamiyama, Higashi-Hiroshima, Hiroshima 739-8526, Japan \\
$^{3}$Aryabhatta Research Institute of Observational Sciences, Manora Peak 263001, India\\
$^{4}$Department of Applied Physics, Mahatma Jyotiba Phule Rohilkhand University, Bareilly, 243006, India\\
$^{5}$National Astronomical Observatory of Japan, National Institutes of Natural Sciences, 2-21-1 Osawa, Mitaka, Tokyo 181-8588, Japan \\
$^{6}$Graduate Institute for Advanced Studies, SOKENDAI, 2-21-1 Osawa, Mitaka, Tokyo 181-8588, Japan \\
$^{7}$School of Physics and Astronomy, Monash University, Clayton, Victoria 3800, Australia \\
$^{8}$Astronomical Institute, Tohoku University, Aoba, Sendai 980-8578, Japan \\
$^{9}$Department of Astronomy, Kyoto University, Kitashirakawa-Oiwake-cho, Sakyo-ku, Kyoto 606-8502, Japan \\
$^{10}$Las Cumbres Observatory, 6740 Cortona Drive Suite 102, Goleta, CA, 93117-5575 USA \\
$^{11}$Department of Physics, University of California, Santa Barbara, CA 93106-9530, USA \\
$^{12}$Indian Institute of Astrophysics, Koramangala 2nd Block, Bangalore 560034, India \\
$^{13}$Instituto de Astrofísica, Universidad Andres Bello, Fernandez Concha 700, Las Condes, Santiago RM, Chile \\
$^{14}$Millennium Institute of Astrophysics, Nuncio Monsenor Sótero Sanz 100, Providencia, Santiago, 8320000 Chile \\
$^{15}$Department of Astronomy, University of Virginia \\
$^{16}$Department of Astronomy, Kyoto University, Kitashirakawa-Oiwake-cho, Sakyo-ku, Kyoto 606-8502, Japan \\
$^{17}$Center for Interdisciplinary Exploration and Research in Astrophysics (CIERA), Northwestern University, 1800 Sherman Ave, Evanston, IL 60201, USA \\
$^{18}$National Radio Astronomy Observatory, 520 Edgemont Road, Charlottesville, VA, 22903, USA \\
$^{19}$Steward Observatory, University of Arizona, 933 North Cherry Avenue, Tucson, AZ 85721, USA \\
$^{20}$Center for Astrophysics | Harvard \& Smithsonian, 60 Garden Street, Cambridge, MA 02138-1516, USA \\
$^{21}$Yukawa Institute for Theoretical Physics, Kyoto University, Kitashirakawa-Oiwake-cho, Sakyo-ku, Kyoto, Kyoto 606-8502, Japan \\
$^{22}$Research Center for the Early Universe, School of Science, The University of Tokyo, 7-3-1 Hongo, Bunkyo-ku, Tokyo, 113-0033 Japan \\
$^{23}$ Astrophysical Big Bang Laboratory, RIKEN, 2-1 Hirosawa, Wako, Saitama 351-0198, Japan \\
$^{24}$Amanogawa Galaxy Astronomy Research Center (AGARC), Graduate School of Science and Engineering, Kagoshima University, 1-21-35 Korimoto, \\
Kagoshima 890-0065, Japan \\
$^{25}$Okayama Branch Office, Subaru Telescope, National Astronomical Observatory of Japan, Kamogata, Asakuchi, Okayama, 719-0232, Japan \\
$^{26}$ National Centre for Radio Astrophysics, TIFR, Ganeshkhind, Pune 411007, India\\
$^{27}$Department of Multi-Disciplinary Sciences, Graduate School of Arts and Sciences, The University of Tokyo, 3-8-1 Komaba, Meguro, Tokyo 153-8902, Japan \\
$^{28}$The NSF AI Institute for Artificial Intelligence and Fundamental Interactions, USA \\
}
\date{Accepted XXX. Received YYY; in original form ZZZ}

\pubyear{2015}

\begin{document}
\label{firstpage}
\pagerange{\pageref{firstpage}--\pageref{lastpage}}
\maketitle

\begin{abstract}
We present the long-term photometric and spectroscopic analysis of a transitioning SN~IIn/Ibn from $-$10.8 d to 150.7 d post $V$-band maximum. SN~2021foa shows prominent He {\sc i} lines comparable in strength to the H$\alpha$ line around peak, placing SN~2021foa between the SN~IIn and SN~Ibn populations. The spectral comparison shows that it resembles the SN~IIn population at pre-maximum, becomes intermediate between SNe~IIn/Ibn and at post-maximum matches with SN~IIn 1996al. The photometric evolution shows a precursor at $-$50 d and a light curve shoulder around 17d. The peak luminosity and color evolution of SN 2021foa are consistent with most SNe~IIn and Ibn in our comparison sample. SN~2021foa shows the unique case of a SN~IIn where the narrow P-Cygni in H$\alpha$ becomes prominent at 7.2 days. The H$\alpha$ profile consists of a narrow (500 -- 1200 km s$^{-1}$) component, intermediate width (3000 -- 8000 km s$^{-1}$) and broad component in absorption. Temporal evolution of the H$\alpha$ profile favours a disk-like CSM geometry. Hydrodynamical modelling of the lightcurve well reproduces a two-component CSM structure with different densities ($\rho$ $\propto$ r$^{-2}$ -- $\rho$ $\propto$ r$^{-5}$), mass-loss rates (10$^{-3}$ -- 10$^{-1}$ M$_{\odot}$ yr$^{-1}$) assuming a wind velocity of 1000 km s$^{-1}$ and having a CSM mass of 0.18 M$_{\odot}$. The overall evolution indicates that SN~2021foa most likely originated from a LBV star transitioning to a WR star with the mass-loss rate increasing in the period from 5 to 0.5 years before the explosion or it could be due to a binary interaction. 
\end{abstract}

\begin{keywords}
spectroscopy, photometry, supernovae (SNe)
\end{keywords}



\section{Introduction}
\label{intro}

Massive stars that eventually undergo core collapse when surrounded by some dense circumstellar material (CSM) are known as Type IIn/Ibn supernovae (SNe) \citep{schlegel1990,Filippenko1997,Fraser2020}. This is signified in spectra by a bright, blue continuum and narrow emission lines at early times. SNe of Type IIn display H-emission lines with multi-component profiles showing narrow-width lines (NW), Intermediate-width (IW) lines and Broad width lines (BW). Narrow ($\sim$ 100 $-$ 500 km s$^{-1}$) components arise mostly in the photo-ionized, slow-moving CSM. Intermediate width emission lines ($\sim$ 1000 km s$^{-1}$) arise from either electron scattering of photons in narrower lines or emission from gas shocked by supernova (SN) ejecta. Some events also show very broad emission or absorption features ($\sim$ 10,000 km s$^{-1}$) arising from fast ejecta, typically associated with material ejected in the core-collapse explosion. An interesting extension of the Type IIn phenomenon was illustrated by SN 2006jc \citep{Foley2007,Pastorello2008}, which exhibited narrow He {\sc i} lines instead of NW/IW Hydrogen lines as spectral signatures of strong CSM interaction. Spectra of SN 2006jc had intermediate-width lines similar to those of SNe~IIn (2000 -- 3000 km s$^{-1}$), but seen mainly in He {\sc i} emission lines – there was only a trace amount of Hydrogen in the spectra. SN~2006jc is therefore referred to as a ‘SN~Ibn’ event, instead of a SN~IIn. It has important implications for understanding the broader class of SNe~IIn and Ibn with CSM interaction because it is also one of the few Type Ibn SN that was observed to have a non-terminal LBV-like outburst just 2 years prior to explosion \citep{Itagaki2006,Pastorello2007}. Other than the traditional definition, the existence of transitional events that change type between Type IIn and Type Ibn over time (e.g \citealp{Pastorello2015,reguitti2022}) suggests a continuum in the CSM properties of these events and consequently in the mass-loss history of their progenitor stars. We now also have a couple of new candidates of an interesting class of events named SNe~Icn, which show narrow narrow emission lines of C and O, but are devoid of H and He \citep{Gal-Yam2022,Perley2022,Fraser_21csp_2021,Pellegrino2022,Davis2023,Nagao2023}. \cite{Pursiainen2023emq} shows that the appearance of prominent narrow He emission lines in the spectra of SN~2023emq (Icn), around maximum light are typical of a SN~Ibn. The family of interacting SNe thus occupy a unique space, but probably linked by a continuum of outer envelopes.

The eruptive mass-loss process expected in these transients has been associated with multiple mechanisms.
The energy deposited in the envelope by waves driven by advanced nuclear burning phases \citep{Quataert2012,Fuller2017,FullerandRo2018}, the pulsational pair-instability or late-time instabilities \citep{Woosley2007,Woosley2017,Renzo2020}, an inflation of the progenitors radius that triggers violent binary interactions like collisions or mergers before the core-collapse event \citep{Soker2013,SmithArnett2014,McleySoker2014}, or just the expansion of the envelope pre-collapse in massive stars \citep{Liron2014}.  
In mass-loss caused by binary interactions, one expects highly asymmetric distributions of CSM (disk-like or bipolar), relevant to the asymmetric line profiles seen in interacting SNe along with high degrees of polarisation \citep{2017hcc_2020Nathan}. 
This brief period of enhanced mass loss likely influences the photometric evolution of the supernova (SN) (having single/double/multiple peaks), including duration and luminosity, along with the spectroscopic appearance of emission/absorption line profiles and their evolution (for example see \citealp{interaction_kurfurst_2019,Suzuki2019}). \\

The traditional LBV stars as progenitors of Type IIn SNe are generally bright, blue and varying \citep{2020Weis}. \cite{2011hw_smith2012,Pastorello2015,Pastorello2008} introduced the first class of these events SNe~2005la and 2011hw, which as per the interpretation has progenitor star exploded as core-collapse while transitioning from the LBV to WR star phase. Theoretical models show that the event rate of appropriate binary mergers may match the rate of SNe with immediate LBV progenitors; and that the progenitor birthrate is $\sim$ 1 $\%$ of the CCSNe rate \citep{Stephan2014}. Observationally, the LBV/SN~IIn connection inferred from properties of the CSM is reinforced by the detection of luminous LBV-like progenitors of three SNe~IIn \citep{2005gl_GalYam2009,Kochanek2011,2011MNRAS.415..773S}. Standard evolution models
instead suggest that massive stars are supposed to undergo only a very brief (10$^{4}$ – 10$^{5}$ yr) transitional LBV phase and then spend 0.5-–1 Myr in the core-He burning Wolf-Rayet (WR) phase before exploding as a stripped envelope SN~Ib/Ic \citep{Heger2003}. This discrepancy between observational and theoretical numbers most likely exists due to insufficient mass-loss rate estimates, not considering binary evolution scenario and also not taking into account the criticality of the LBV phase \citep{2011hw_smith2012}. This accounts for the fact that stellar evolution models are missing essential aspects of the end stages of massive stars.

While SNe~IIn explosions make up 8–9 percent of all core-collapse SNe in the Lick Observatory Supernova Search sample \citep{Li2011,Perley2020}, the Type Ibn events like SN~2006jc represent a substantially smaller fraction. The fraction of SN~2006jc like event in this case constituted only 1 percent of the core-collapse sample which agrees with an independent estimate of the fraction of SN Ibn events by \cite{Pastorello2008_2006jclike}. \cite{Perley2020} updated this fraction for the Zwicky Transient Factory and found that SNe~IIn consitute 14.2 percent of H-rich CCSNe while SNe~Ibn constitute 9.2 percent of H-poor CCSNe. Given their rare occurrence, additional examples are valuable to demonstrate the diversity of the subclass. Among this whole sample, there are very few members of this peculiar SN class like SNe~2005la and 2011hw which had prominent narrow H and He lines \citep{Pastorello2008,Kool2021,Farias2024}.
This motivates us to study another rare case of transitioning Type IIn/Ibn SNe, which belongs to the same category as SNe~2005la and 2011hw. 

SN~2021foa were already investigated by \citealt{reguitti2022} (hereafter R22).
Here we present further detailed photometric and spectroscopic observations of SN 2021foa, which exhibits both H and He emission features and shows similarities with both Type IIn and Type Ibn SNe at distinct phases of its evolution.
SN 2021foa show similarities in photometric evolution with both Type IIn and Type Ibn SNe \citep{Hosseinzadeh2017,Ransome2021}, but their diverse spectroscopic behaviour needs to be explored further to understand the division.

This work presents an extended analysis on SN~2021foa after R22. In their analysis, R22 showed that SN~2021foa belongs to sub-class of SN~IIn which are labelled as SN~IId \citep{1996al_benetti_2016} and shows prominent narrow H$\alpha$ early on with ejecta signatures later on. SN~2021foa, however, showed early prominent signatures of He {\sc i} 5876 \AA~ than other SNe~IIn like 2009ip, 2016jbu. R22 quoted that SN 2021foa may be part of a bridge connecting H-rich SN 2009ip-like and Type Ibn SNe, indicating the possible existence of a continuum in properties, mass-loss history and progenitor types between these two types of peculiar transients. In our paper, we did a more robust modelling of the lightcurve, spectra and derived the physical parameters associated with the explosion. R22 indicated that SNe~IId are probably connected to those objects by having similar a progenitor with LBVs transitioning to WR phase, but with a different mass-loss history or observed with a different orientation. We indeed, notice that our estimated mass-loss rates changes at different phases in the evolution of the SN, and spectral modelling shows an asymmetric CSM structure giving rise to H$\alpha$ and He {\sc i} at different strengths. Thus, our results are in concordance with what has been predicted by R22 and also an more elaborate description of it.

\section{Observations and Data Reduction}
\label{obsanddatareduction}
\subsection{Optical $\&$ Near-Infrared Observations}
We observed SN~2021foa in \textit{UBgVriocRIJHK} bands from day $-$34.3 to $\sim$150 d post \textit{V}-band maximum (see section \ref{sec:explosion_epoch}). The {\it oc}-band ATLAS data was reduced and calibrated using the techniques mentioned in \citep{Tonry2018}. Imaging observations in \textit{BVRIJHK} were carried out using the 1.5m Kanata telescope (KT; \citealp{2008SPIE.7014E..4LK}) of Hiroshima University; Japan. 
Several bias, dark, and twilight flat frames were obtained during the observing runs along with science frames. For the initial pre-processing, several steps, such as bias-subtraction, flat-fielding correction, and cosmic ray removal, were applied to raw images of the SN. We used the standard tasks available in the data reduction software IRAF\footnote{IRAF stands for Image Reduction and Analysis Facility distributed by the National Optical Astronomy Observatory, operated by the Association of Universities for Research in Astronomy (AURA) under a cooperative agreement with the National Science Foundation.} for carrying out the pre-processing. Multiple frames were taken on some nights and co-added in respective bands after the geometric alignment of the images to increase the signal-to-noise ratio.  

Given the proximity of SN 2021foa to its host galaxy, host galaxy contamination was removed by performing image subtraction using IRAF. For the templates, we used a set of deep images obtained on 2022 when the SN went beyond the detection limit of the telescope. 
For the optical photometry of KT data, local comparison star magnitudes were calibrated using the photometric standard stars \citep{landolt1992} observed on the same nights.
The zero point and the color terms were derived from these comparison stars to calibrate the instrumental magnitudes. We also observed SN~2021foa with the Las Cumbres Observatory (LCO) network of telescopes as part of the Global Supernova Project.
The pre-processing of LCO data was conducted using the BANZAI pipeline \citep{Banzai}. The photometry was performed using the \texttt{lcogtsnpipe}\footnote{\url{https://github.com/LCOGT/lcogtsnpipe/}} pipeline \citep{valenti_lcogtsnpipe}. The template subtraction was performed using PyZOGY library \citep{PyZOGY, ZOGY} implemented within the \texttt{lcogtsnpipe} pipeline. The {\it UBVgri} instrumental magnitudes were obtained from the difference images.
The {\it gri} apparent magnitudes of the local comparison stars were taken from the Sloan Digital Sky Survey (SDSS) catalog, and the {\it UBV} magnitudes of the local comparison stars were calibrated against standard Landolt fields observed on the same nights as the SN field. Then, instrumental magnitudes of the SN were converted to apparent magnitudes by using the zero point and color terms derived from these comparison stars.
Table~\ref{tab:lco_japan_photometry} reports the complete photometric lightcurve evolution of SN~2021foa taken from the LCO and the Japan Telescopes.

The near-infrared (NIR) data of SN~2021foa were obtained with the HONIR instrument of KT \citep{akitaya2014}. The sky-background subtraction was done using a template sky image obtained by dithering individual frames at different positions. We performed PSF photometry and calibrated the SN magnitudes using comparison stars in the 2MASS catalog \citep{1998AJ....116.3040G}. The final NIR magnitudes in the SN field are shown in Table~\ref{tab:nir_photometry}.

Low-resolution (R $\sim 400-700$) optical spectroscopic observations were carried out using the FLOYDS spectrographs mounted on the LCO 2m telescopes. The 1D wavelength and flux calibrated spectra were extracted using the \texttt{floydsspec}\footnote{\url{https://github.com/LCOGT/floyds_pipeline}} pipeline \citep{Valenti_floyds}. Spectroscopic observations were also carried out using the KOOLS-IFU \citep{2019PASJ...71..102M} on the Seimei Telescope. Our spectral coverage spans from $-$10.8 d to +69.5\,d. The spectra with KOOLS-IFU were taken through optical fibers and the VPH-blue grism. The data reduction was performed using the Hydra package in IRAF \citep{1994ASPC...55..130B} and a reduction software developed for KOOLS-IFU data\footnote{\url{http://www.o.kwasan.kyoto-u.ac.jp/inst/p-kools}} (Proposal numbers: 21A-N-CT02, 21A-O-0004, 2021-04-05 21A-K-CT02, 2021-04-14 21A-N-CT02, 2021-05-03 21A-O-0008). For each frame, we performed sky subtraction using a sky spectrum created by combining fibers to which the contributions from the object are negligible. Arc lamps of Hg, Ne, and Xe were used for wavelength calibration. Finally, the spectra were corrected for the heliocentric redshift of the host galaxy (see for reference Section~\ref{spec}). The slit loss corrections were done by scaling the spectra with respect to the SN photometry. 
The log of spectroscopic observations is reported in Table~\ref{tab:2021foa_spec_obs}.

Along with that, we also obtained high-resolution spectroscopic data with High Dispersion Spectrograph (HDS) mounted on Subaru Telescope on 2021 April 22 (UT). This observation was done as a part of the Subaru Proposal No S21A-014 (PI: Keiichi Maeda). The Echelle setup was chosen to cover the wavelength range of 5700–7100 \AA\ with a spectral resolution of $\sim$50,000 in the Red Cross Disperser mode. We followed standard procedures to reduce the data. The wavelength calibration was performed using Th–Ar lamps. A heliocentric velocity correction was applied to each spectrum. The sky subtraction was performed using data at an off-target position in the target frames. The spectra were not flux calibrated by a standard star but were scaled to photometric fluxes at similar epochs to account for any flux losses. 

\subsection{Radio Observations} 
The observing campaign of SN~2021foa was carried out using the Giant Meterwave Radio Telescope (GMRT), Pune, India in Bands 4 and 5 (PI: Poonam Chandra). There was no detection of the source on observations dated 11 January 2022 and 31 March 2022. Our rms obtained in Band 5 (1.265 GHz) and Band 4 (0.745 GHz) are 33 and 215 $\mu$Jy and the 3-sigma limits corresponding to the non-detections are 100 $\mu$Jy and 645 $\mu$Jy in the two bands. There was a nearby radio-bright galaxy with extended emission which was contaminating the SN location due to which our Band 4 rms are high. We also checked the Very-Large Array Telescope archive and did not find any detection of this source. The obtained radio luminosities at these phases for Band 5 and Band 4 are 1.455 $\times$ 10$^{26}$ erg sec$^{-1}$ and 9.385 $\times$ 10$^{26}$ erg sec$^{-1}$. Figure~18 of \cite{Gangopadhyay2023} shows the radio luminosity of a group of core-collapse SNe which includes SNe~IIn. We see that the minimum radio luminosity driving SNe~IIn are in between 10$^{27}$ - 10$^{29}$ erg sec$^{-1}$. We do expect a radio emission at a phase of 584 d (1st observation) if it were a SNe~IIn. Previous cases of SNe 2006jd, 2010jl \citep{Chandra2012,Chandra2015} have shown radio lightcurves peaking between 500 d - 700 d post explosion. Radio emission in SNe IIn is expected to be synchrotron emission, initially absorbed mainly by free–free absorption, while X-ray, optical emission is likely to have a thermal, radioactive origin. So, even if we expect the radio lightcurves to peak for SN~2021foa, given the fast decaying light curve of SN~2021foa, it's expected that the radio power will decrease, and thus, we get only radio upper limits.

\subsection{Estimation of explosion epoch} \label{sec:explosion_epoch}
\cite{2021TNSTR.767....1S} from the ASASSN team report the discovery of SN~2021foa (RA = 13:17:12.290; DEC = $-$17:15:24.19) on 2021-03-15 10:48:00 
(MJD = 59288.45) at a discovery AB mag of 15.9 using the $g$ filter. A non-detection of the source was reported on 2021-03-05 09:50:24 (MJD = 59278.41) at a limiting magnitude of 17.9 mag ($g$ band). \cite{classificationspec2021} report the classification of SN~2021foa from a spectrum taken with ALFOSC mounted on the Nordic Optical Telescope using gr4, which matches with an SN~IIn. 

To estimate the explosion epoch, we fitted a parabola function on the rising part of the $g$-band light curve. The early light curve shape is well-reproduced by a parabola. We performed the fit using 20000 iterations of Markov Chain Monte-Carlo (MCMC) simulations. Using this method, we find the explosion epoch to be MJD = $59284.8 \pm 0.2$, 3.6 days prior to the first detection. This estimate is consistent with the non-detection of the source, and we adopt it as the explosion epoch.

However, since it is often difficult to estimate the explosion epoch for the comparison SNe in the literature, we adopt the $V$-band maximum (MJD 59301.8) as the reference epoch. This value is in agreement with the estimate from \citetalias{reguitti2022}.

\subsection{Distance \& extinction} Adopting H$_{0}$ = 73 km s$^{-1}$ Mpc$^{-1}$, $\Omega_{m}$ = 0.27 and $\Omega_{\Lambda}$ = 0.73, we obtain a distance of 34.89 $\pm$ 2.44 Mpc ($\mu$ = 32.71 $\pm$ 0.15) corrected for Virgo, Shapley and GA (corresponding to a redshift z=0.00839 \footnote{\url{https://ned.ipac.caltech.edu/byname?objname=IC0863&hconst=73&omegam=0.27&omegav=0.73&wmap=1&corr_z=2}}) for SN~2021foa. This value is the same as that adopted by \citetalias{reguitti2022}. The Milky Way extinction along the line of sight of SN 2021foa is $A_V = 0.224$~mag \citep{milkyway_reddening}. We see a conspicuous dip at 5892.5 \AA~ from \ion{Na}{1}D in the spectra of SN~2021foa taken on 2021-03-20, 2021-03-23 and 2021-03-25. For estimating the extinction due to the host galaxy, we estimate equivalent widths of the \ion{Na}{1}D line iteratively three times in the combined spectra of these three dates to increase the signal-to-noise ratio. Using the formulation by \cite{2012MNRAS.426.1465P}, we estimate host galaxy $A_V$ = 0.40 $\pm$ 0.19~mag. We multiply this reddening value by 0.86 to be consistent with the recalibration of Milky Way extinction by \cite{milkyway_reddening}.
Thus, we adopt a total extinction of  $A_V$ = 0.57 $\pm$ 0.16 mag. We use these values of distance and extinction throughout the paper, which is also consistent with the values quoted by \citetalias{reguitti2022}.

\begin{figure*}
	\begin{center}
		\includegraphics[width=\linewidth]{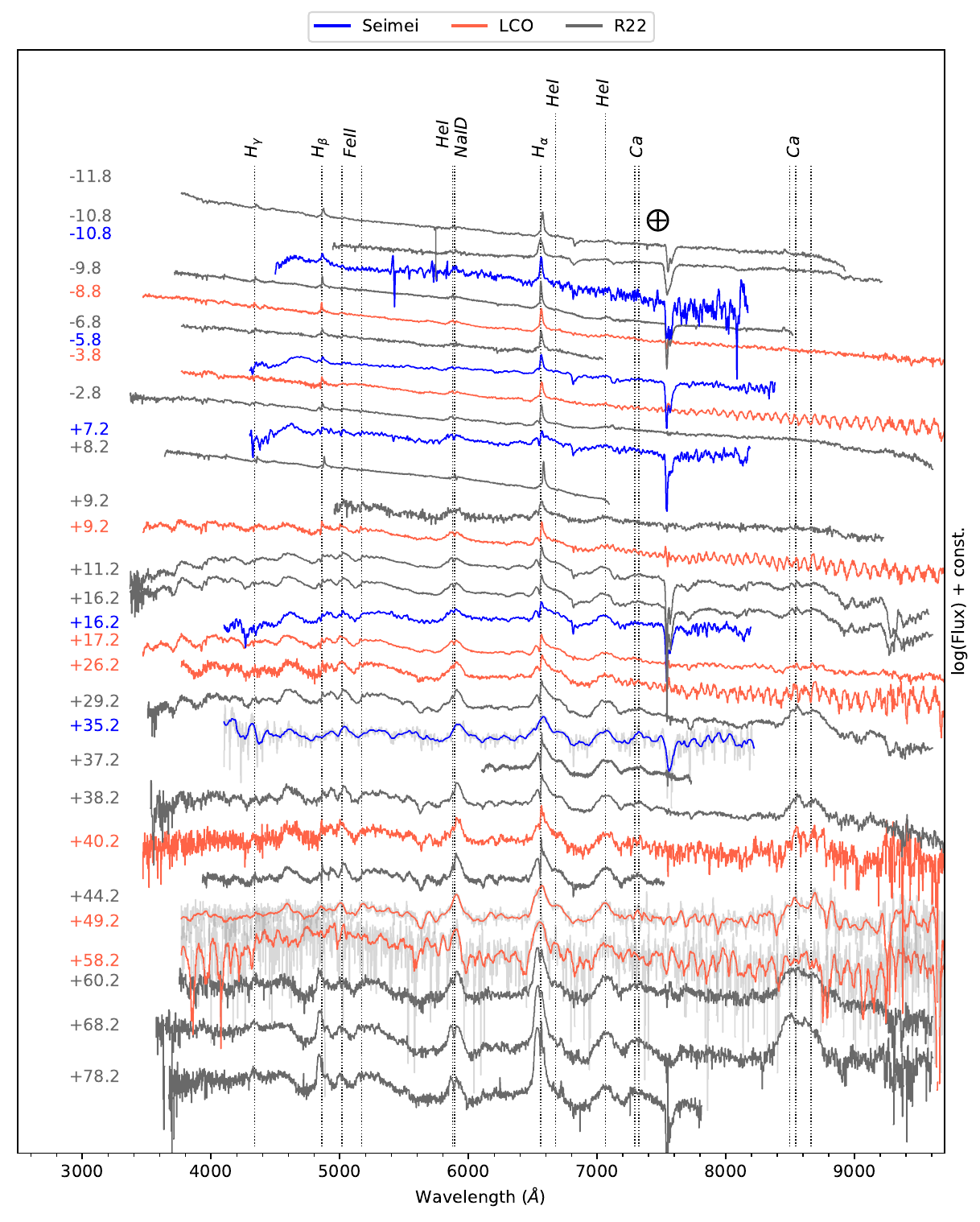}
	\end{center}
	\caption{{\it Figure shows the complete spectral evolution of SN 2021foa from $-$12 to 78 d post maximum. The spectra colored blue and red are our data, while the grey-colored spectra are plotted from \citetalias{reguitti2022}. The early spectra show the characteristic evolution of a SN~IIn, which later on shows prominent narrow emission lines of He {\sc i}.}}
	\label{fig:spectra_2021foa}
\end{figure*} 

\section{Spectroscopic Evolution}
\label{spec}

We conducted the spectroscopic follow-up of SN~2021foa from $-$10.8 d to 58.6 d post $V$-band maximum. The complete spectral evolution of SN~2021foa, which includes our data and those published in \citetalias{reguitti2022}, are shown in Figure~\ref{fig:spectra_2021foa}. The early time spectral sequence shows prominent lines of H$\alpha$ 6563 \AA~ along with H$\beta$ 4861 \AA~ and H$\gamma$ 4340 \AA. From $-$10.8 d to $-3.2$ d, the He 5876~\AA~ line also starts developing but is not very prominent. The H$\beta$ profile initially shows narrow emission, but develops a narrow P-Cygni feature at around $-$5.8 d in our spectra. The spectra from \citetalias{reguitti2022} show an even earlier appearance of this narrow P-Cygni feature, which can be attributed to the higher resolution of their spectrum. 
The H$\alpha$ also develops this narrow P-Cygni feature in the $-$3.2 d spectra. The early spectral sequence till $-3.2$ d does not show prominent lines of Fe {\sc ii} 4924, 5018, and 5169 \AA, which are characteristic of a typical SN ejecta.

Post $-3.2$ d, the He {\sc i} 5876~\AA~ line becomes prominent. The other He {\sc i} 6678, 7065 \AA~ lines start developing at this phase. The narrow P-Cygni on top of H$\alpha$ is clearly seen with the blue wing extending up to $-$1900 km s$^{-1}$.  
From 7.2 d to 17.2 d, the H$\alpha$ and H$\beta$ show very complex profiles. This is also the phase where we see a second shoulder appearing in the light curve of SN~2021foa (c.f.r Section~\ref{phot}). 

From 17.2 d, we see that the spectrum transforms significantly. This marks the onset of the phase where we see that the flux of He {\sc i} 5876 \AA~ is comparable to H$\alpha$. The higher excitation H-lines like H$\beta$ and H$\gamma$ no longer show narrow emission lines. However, narrow emission and the corresponding narrow P-Cygni feature are still significant in the H$\alpha$ line. We also see a red wing developing in the H$\alpha$ profile, most likely due to He {\sc i} 6678 \AA. The blue part of the spectrum in this phase is also mostly dominated by the He lines, along with the Fe group of elements \citep{Pastorello2007,Anupama2009,Dessart2022}. 

From 26.8 d to 58.6 d, we see that both H$\alpha$ and He {\sc i} grow in strength. The narrow component of H$\beta$ seen at 44.2 d is due to poor resolution and is an artifact.
This also marks the phase where we see ejecta signatures in the spectral evolution. The [Ca {\sc ii}] 7291, 7324 \AA~ lines emerge at 7300 \AA, which could also be blended with He {\sc i} 7281 \AA. This phase also marks the appearance of the broad Ca II NIR triplet. At late times ($>60$ d), the intermediate to broad-width H-lines develop a slight blueshift in the observed profiles.

Overall, the spectral behavior shows a striking similarity with an interacting SNe~IIn early on, which later on is overtaken by a SN~Ibn like behaviour with He lines. The very late spectra shows ejecta signatures of Ca emerging along with H$\alpha$ and He {\sc i}.

\subsection{Line luminosities and line ratios:}
\label{linelum}
\begin{figure}
	\begin{center}
		\includegraphics[width=\columnwidth]{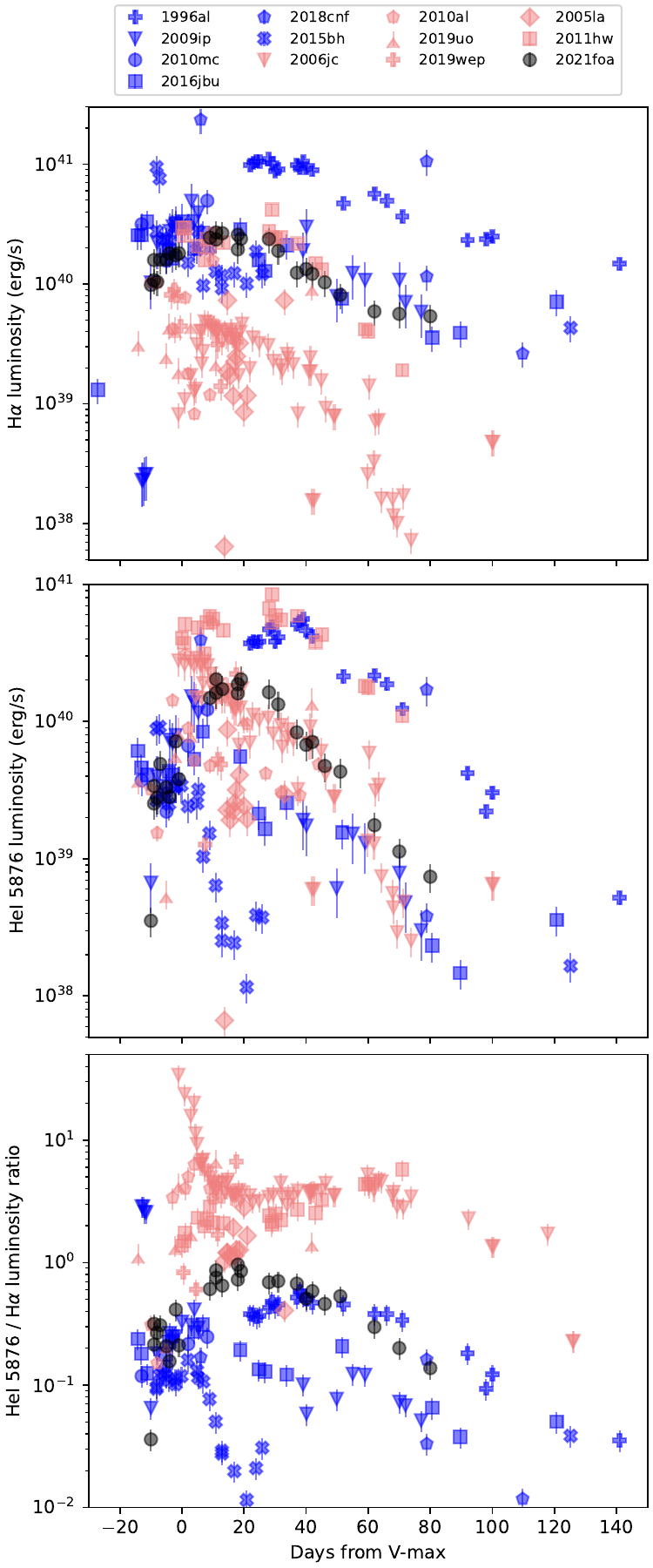}
	\end{center}
	\caption{{\it The top panel shows the H$\alpha$ luminosity of a sample of SNe~IIn (red) and SNe~Ibn (blue); the middle panel shows the luminosities of He {\sc i} 5876 \AA~ for the similar group of SNe~IIn (red) and SNe~Ibn (blue); the third panel shows the line luminosity ratios of H$\alpha$ and He {\sc i} and clearly indicates that initially it matches with SNe~IIn, lies intermediate between SNe~IIn and SNe~Ibn and at mid spectral epochs (from pre-maximum to 20 days after maximum) and become similar to SNe~IIn with ejecta signatures (like SN 1996al). The errors in the luminosities are calculated by considering 10\% error in flux values along with the error in the distances being propagated in quadrature. }}
	\label{fig:lumratio_2021foa}
\end{figure} 
 
We observed well-developed He {\sc i} features in the spectra, and the line-luminosity of the He {\sc i} 5876 \AA~ line becomes comparable to the H$\alpha$ line at about 17.2 d. To study the evolution of the He {\sc i} 5876 \AA ~line in comparison to the H$\alpha$ line, we estimate the line luminosities of H$\alpha$ and He {\sc i} over the evolution of the SN. 

To compare with other well-studied SNe, we selected a group of SNe~IIn having diversity in the luminosity distribution and some having precursor detections, similar to SN~2021foa. We also include a set of classical, bright SNe~Ibn, along with some that have some residual H-envelope. The sample includes- SNe~IIn: 1996al \citep{1996al_benetti_2016}, 2009ip \citep{pastorello2013}, 2010mc \citep{ofek2014}, 2015bh \citep{Nancy2016}, 2016jbu \citep{Brennan2022}, 2018cnf \citep{Pastorello2019} and SNe~Ibn: 2005la \citep{Pastorello2008}, 2006jc \citep{Pastorello2007}, 2010al \citep{Pastorello2015}, 2011hw \citep{Pastorello2015}, 2019uo \citep{Gangopadhyay2020}, 2019wep \citep{Gangopadhyay2022}.

\begin{figure}
	\begin{center}
		\includegraphics[width=\columnwidth]{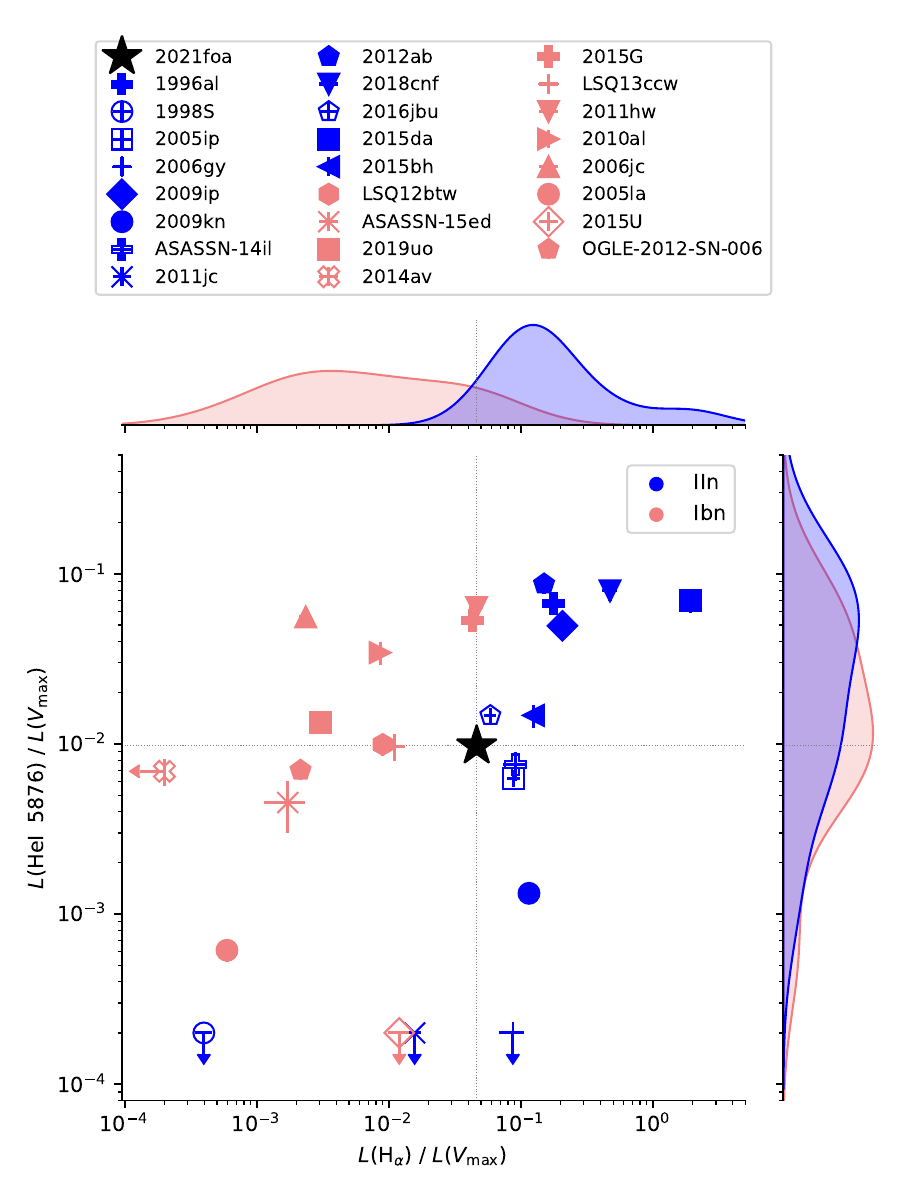}
	\end{center}
	\caption{{\it The Figure shows the luminosity ratio of H$\alpha$ to the optical luminosity (H$\alpha$/V$_{\rm max}$) versus luminosity ratio of He {\sc i} to the optical luminosity (He {\sc i}/V$_{\rm max}$) for a group of SNe~IIn and SNe~Ibn. The distributions imply that it is the strength of H$\alpha$ that separates the SN~IIn from the SN~Ibn class. All the measurements for the luminosities are measured around maximum light. We put arrows for those SNe~IIn/Ibn in our sample where H/He lines are not detected properly and we put only upper limits in the evolution. The upper limit points are not considered in the marginal histograms.}}
	\label{fig:peak_line_lum_ratio}
\end{figure} 

To study whether SN~2021foa belongs to the SNe~IIn or the SNe~Ibn regime, we compare the evolution of H$\alpha$ and He {\sc i} line luminosities with the sample. The line luminosities are estimated by integrating the continuum-subtracted line regions of the de-reddened spectra. The top panel of the Figure~\ref{fig:lumratio_2021foa} shows the H$\alpha$ luminosities of SNe~IIn (blue) and SNe~Ibn (pink) sample and SN~2021foa marked in black. The H$\alpha$ luminosities of SNe~IIn and SNe~Ibn are well separated in luminosity scales. SN~2021foa shows similarity with SN~2011hw in the H$\alpha$ space, which is an SN~Ibn with a significant amount of residual Hydrogen \citep{Pastorello2015}. On the contrary, in the He {\sc i} luminosity scale (middle panel of Figure \ref{fig:lumratio_2021foa}), SNe~IIn and SNe~Ibn do not show a clear distinction. The He {\sc i} 5876 \AA~ luminosity of SN~2021foa matches with SN~2006jc \citep{Pastorello2007} over the evolution. The lower panel of Figure~\ref{fig:lumratio_2021foa} shows the luminosity ratios of He {\sc i} 5876 \AA~ to H$\alpha$. The distinction between these two classes of objects is the most prominent in this plot, however, we want to mention that there might be some residual H$\alpha$ luminosity from the host galaxy in case of SNe~Ibn (possibly for SNe~2006jc, 2010al, 2014av and 2019uo from our sample; being located in the spiral arms of the galaxy.). 

The line-luminosity ratio of SN~2021foa, from $-$10.8 d to about 17.2 d, rises from 0.3 to 0.7, placing it in the intermediate region between SNe~IIn and SNe~Ibn. From 40 d, SN~2021foa shows similarities with SN~1996al, which is a SN showing signs of ejecta signatures and interaction signatures simultaneously for 15 years \citep{1996al_benetti_2016}. This also marks the onset of the phase where we see ejecta signatures arising for both SN~1996al and our object SN~2021foa.
To summarize, SN~2021foa shows a complete demarcation and lies intermediate between the SN~IIn and SN~Ibn population from $-$20 d to 20 d, which is taken over by its match with SN~1996al at late phases mostly dominated by the appearing ejecta signatures. Also, there is a probability of H$\alpha$ to He {\sc i} reaching a straight line for SNe~IIn but would be affected by the sample size.

Figure~\ref{fig:peak_line_lum_ratio} aims to show how much the H$\alpha$ and He {\sc i} luminosity contributes to the total optical luminosity ($V$-band; L$_{v}$) at peak for a group of SNe~IIn and SNe~Ibn. To avoid possible biases in the distribution, we added more diverse SNe~IIn and SNe~Ibn sample for this comparison plot to highlight the position of SN~2021foa in the phase space. The additional SNe~IIn used for this comparison plot are : SNe~1998S \citep{Fassia2001}, 2005ip \citep{2005ip_2006jd_Stritzinger2012}, 2006gy \citep{2007ApJ...671L..17S,Agnoletto2009}, 2006tf \citep{2008Smith}, 2009kn \citep{2012MNRAS.424..855K}, 2011ht \citep{2013MNRAS.431.2599M}, PTF11oxu/2011jc \citep{Nyholm2020}, ASASSN-14il \citep{14il}, 2015da \citep{2015dalate_smith2024} and ASASSN-15ua \citep{2024Dickinson}. The additional SNe~Ibn used for the comparison are : SNe~OGLE-2012-SN-006 \citep{Pastorello-OGLE}, LSQ12btw \citep{Pastorello_2015_LSQ12btw_LSQ13ccw}, LSQ13ccw \citep{Pastorello_2015_LSQ12btw_LSQ13ccw}, ASASSN-15ed \citep{Pastorello-15ed}, SNe~2014av \citep{Pastorello2016}, 2015U \citep{Shivvers2016} and 2015G \citep{Hosseinzadeh2017}. We added SNe~IIn/Ibn of different CSM configurations, long-lived, short-lived, with and without precursor to diversify this plot. Also, we chose only those SNe for which {\it V}-band observations around maximum are available with a good signal-to-noise ratio spectrum of min 10. We plot the ratio of H$\alpha$ to the peak $V$-band optical luminosity against the ratio of He {\sc i} to the peak $V$-band optical luminosity for SN~2021foa and the comparison sample. The surrounding axes shows the marginalized distribution with respect to the axes variable. However, these SNe are in general have asymmetric CSM geometries \citep{smith2017_interacting,Fraser2020} and the peak $V$-band line luminosities may be slightly affected by our viewing angle. We see that the SNe~IIn and SNe~Ibn in our sample are well separated in this phase space, with SN~2021foa (black star) lying in between the two sub-classes around peak luminosity. SN~2021foa shares remarkable similarities with SN~2011hw in this space as well. \cite{Pastorello2015} have shown that SN~2011hw is also a SN~Ibn with significant residual H$\alpha$. The distributions of SNe~IIn and SNe~Ibn also help in deciphering the fact that it is the H$\alpha$ contributing to the optical luminosity that demarcates the two classes. He {\sc i} distribution is blended for the SNe~IIn and SNe~Ibn population. However, we want to remark that a statistically decent sample would help in further verification of this. Nonetheless, SN~2021foa clearly lies at the junction between the two populations at this phase.  

\subsection{Spectral Comparison}
\label{speccomp}
In this section, we compare and classify SN~2021foa with a group of SNe~IIn and SNe~Ibn in the pre-maximum, about 20 d post maximum and around 40 d post-maximum to see the changing trend in the evolution of the SN.

\begin{figure*}
	\begin{center}
		\includegraphics[scale=0.45]{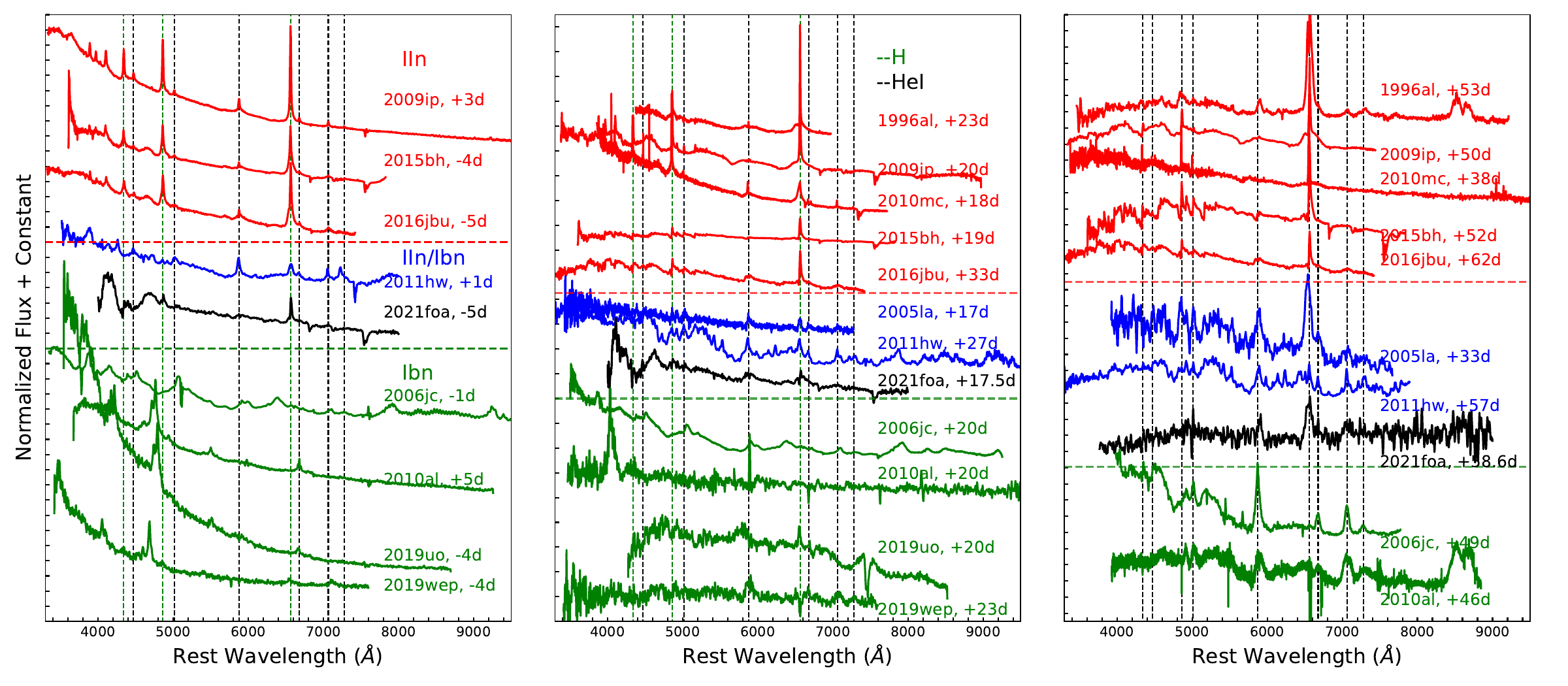}
	\end{center}
	\caption{{\it The left panel shows early spectral comparison of SN~2021foa with a group of SNe~IIn and SNe~Ibn. The spectra shows similarity with a traditional SNe~IIn with prominent H$\alpha$ lines. The middle panel shows comparison spectral sequence at 20 d. The spectrum at this phase shows more similarity with SNe~Ibn in terms of He {\sc i} features, however, H$\alpha$ 6563 \AA~ is still significant than all SNe~Ibn of our sample. This phase shows that SN~2021foa lies in a beautiful boundary between SNe~IIn and SNe~Ibn. The third panel of Figure shows the 50 d comparison of the spectra of SN~2021foa. This phase marks the onset of ejecta signatures in SN~2021foa similar to SN~1996al.}}
	\label{fig:spectra_comb}
\end{figure*}

The first panel of Figure~\ref{fig:spectra_comb} shows the pre-maximum spectral comparison of SN~2021foa with a group of SNe~IIn and SNe~Ibn. The pre-maximum spectral profile of SN~2021foa looks very similar to all the SNe~IIn in our comparison sample. However, the H-lines are more prominent in the SNe~IIn compared to SN~2021foa. In contrast, the SNe~Ibn show little to no hydrogen in their spectra.
In SN~2011hw, an SN~Ibn with significant residual hydrogen, both H and He lines are visible at this phase, while for SN~2021foa He {\sc i} lines are not seen. 
Most SNe~Ibn in our comparison sample at this phase show flash features, which is absent in our observed profile. Overall, at this phase, SN~2021foa behaves more like a SN~IIn. 

Middle panel of Figure~\ref{fig:spectra_comb} shows the spectral comparison at 20 d after the $V$-maximum. This phase marks the remarkable transition of SN~2021foa, showing prominent lines of H$\alpha$ and He {\sc i} simultaneously. We also see narrow P-Cygni of H$\alpha$ in SN~2021foa at this phase, similar to SNe~1996al and 2016jbu. The strength of H$\alpha$ remains lower in strength for SN~2021foa than other SNe~IIn; however, He {\sc i} shows similar strength with most SNe~Ibn in the sample. This separates SN~2021foa with the SNe~IIn and SNe~Ibn population, again justifying our case of SN~2021foa having strong H$\alpha$ and He {\sc i} emission simultaneously at mid epochs and at similar strengths.

We also show the late-time spectral comparison of SN~2021foa in the third panel of Figure~\ref{fig:spectra_comb}.  
The spectral evolution of SN~2021foa at this phase matches very well with SNe~2005la and 2011hw, which are SNe~Ibn with residual Hydrogen. SNe~Ibn at this phase shows very strong He {\sc i} lines, unlike SN~2021foa. Similarly, the He {\sc i} lines are more prominent than traditional long-lasting SNe~IIn. The H$\alpha$ and He {\sc i} line profiles of SN~2021foa also show similarities with SN~1996al at this phase, in addition to the similarities in the line-luminosity ratio, noted earlier.

Overall, we conclude that the early time spectral evolution is similar to that of traditional SNe~IIn followed by a phase where the spectral evolution is intermediate between those of SNe~IIn and SNe~Ibn. During late times, the SN shows spectral evolution similar to SNe~IIn that show ejecta signatures at late phases or with SNe~Ibn having a residual Hydrogen envelope.

\section{Photometric Evolution}
\label{phot}
\begin{figure}
	\begin{center}
		\includegraphics[width=\columnwidth]{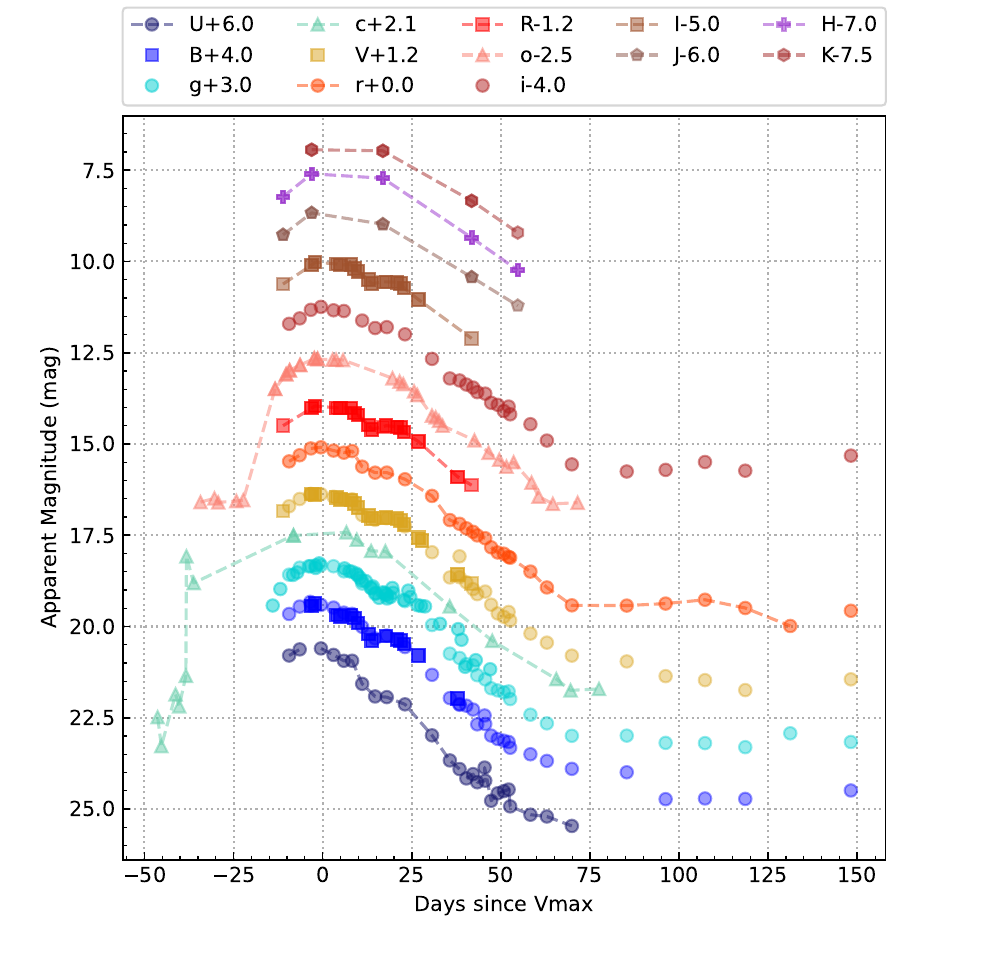}
	\end{center}
	\caption{{\it UBgVRIJHKco} light curve evolution of SN~2021foa. The light curve shows a multi-staged behaviour with early rise reaching a peak, a shoulder, decay, and late-time flattening.}
	\label{fig:lc_2021foa}
\end{figure}

We present the complete photometric evolution of SN~2021foa from $-$50 d to about 150 d post maximum, which extends the current published dataset. Our light curve spans about 30 d more than the time evolution presented in \citetalias{reguitti2022}, after which the object went behind the sun. The rise and peak of the light curve are well sampled in the $V$-band, and the adopted value to $V$-band maximum (MJD 59301.8) is the same as \citetalias{reguitti2022}. 
SN~2021foa showed prominent signatures of precursor from $-$50 d to about $-$23 d in ATLAS $c$ and $o$-bands. 

The precursor for SN~2021foa lasted for a shorter duration than SNe~2009ip, 2016jbu and 2018cnf where the precursor event was observed about $-$90 d to $-$200 d before maximum. This precursor activity can be mainly attributed to the mass-loss eruptions from the progenitor star that occurred months-years before the explosion (For example, SN~2009ip; \citep{smith2014} or SN~2006jc; \citealp{Pastorello2007}). This has been interpreted as LBV stars undergoing eruptions, but, since our CSM is a combination of Hydrogen and Helium, this could be attributed to a LBV star transitioning to a WR phase \citep{Pastorello2008}. For SN~2021foa, the non-detection of the precursor before 50 d could be attributed to ATLAS upper limits reaching 20.4 mag (3$\sigma$ detection), so, there is a chance the precursor activity might have lasted longer than the timescale of detections.

After the precursor, the lightcurve rose to peak in most of the optical bands in 6 d - 9 d which is consistent with SNe~Ibn \citep{Hosseinzadeh2017} and fast rising sample of SNe~IIn \citep{Nyholm2020}. From the peak to about 14 d, it did not have much evolution and changed by only 0.3 - 0.5 mag in the optical wavebands. At $\sim$ 14 d, we see a shoulder in the light curve of SN~2021foa. This also marks the phase where we see the He {\sc i} features developing and strengthening of a narrow P-Cygni on top of H$\alpha$ (c.f.r Section~\ref{spec}). 
The bump or the shoulder in the lightcurve of SN~2021foa is weaker in the redder bands than in the bluer bands. After this phase, the lightcurve drops sharply at a decline rate of about 3 mag in 50 days. 

Post 75 d, the lightcurve shows a flattening lasting from 75 d to 150 d post maximum. The flattening in the light curve of SN~2021foa was also noticed by \citetalias{reguitti2022}. The late time flattening in the redder bands have been attributed to the formation of dust \citep{Anupama2009}, but, we do not have any NIR observations to verify this scenario \citep{smith2009, 1996al_benetti_2016, 2015dalate_smith2024}. The late time flattening could also be due to interaction with a uniform density CSM as we also see in our lightcurve modelling section (c.f.r Section~\ref{hydro}). The double peak or hump seen in the lightcurve of SN~2021foa can also be reproduced overall by the grid of models by \cite{Suzuki2019} which are based on a CSM of mass $\sim$ 10 M$_{\odot}$ assuming a disk-like geometry of the CSM.

\begin{figure}
	\begin{center}
		\includegraphics[width=\columnwidth]{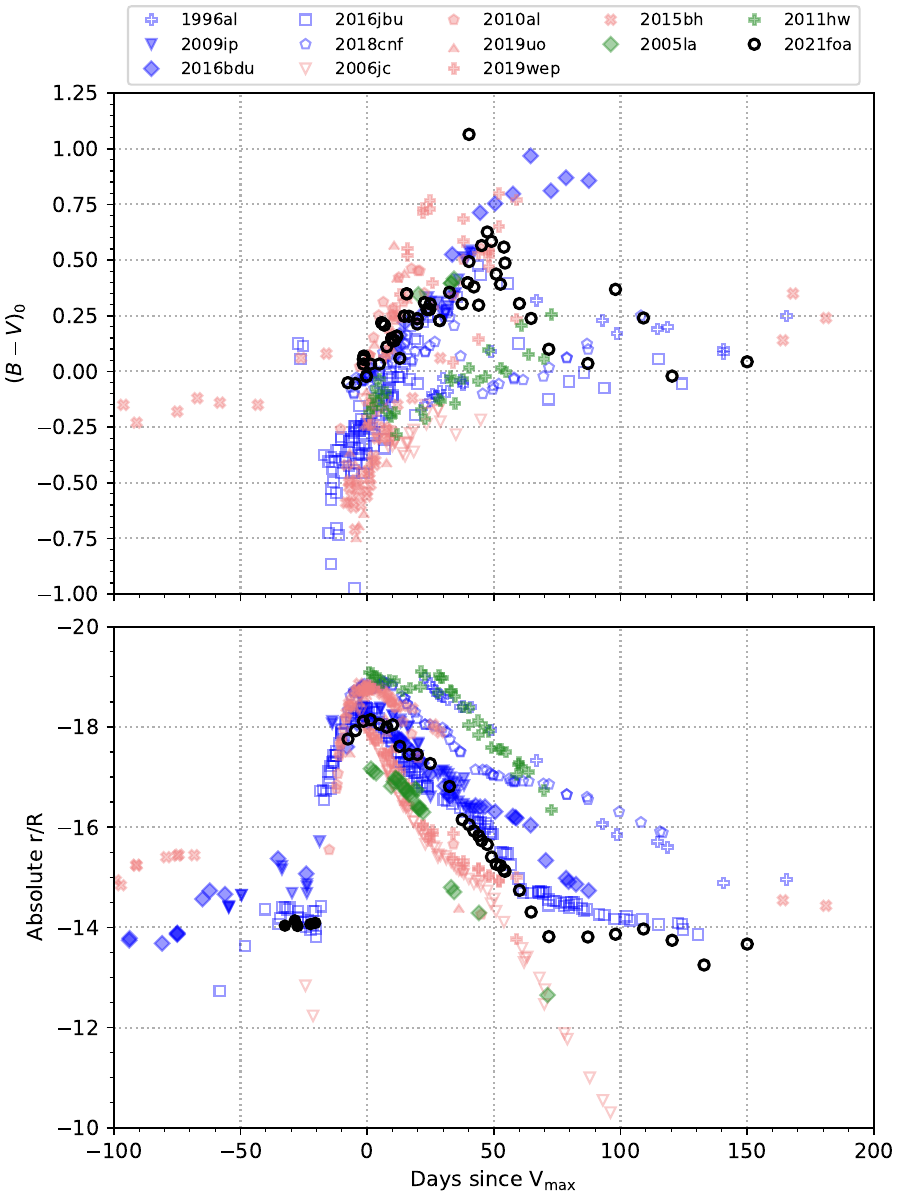}
	\end{center}
	\caption{{\it Figure shows the absolute magnitude and color curve comparison of SN~2021foa with a sample of SNe~IIn and SNe~Ibn. The two category of the SNe in the sample are marked by open and closed symbols.} SN~2021foa lies fairly intermediate between SNe~IIn and SNe~Ibn in our observation sample.}
	\label{fig:lcfit_2021foa}
\end{figure}

Figure~\ref{fig:lcfit_2021foa} shows the {\it (B-V)$_{0}$} color evolution and the absolute magnitude lightcurve of SN~2021foa with other members of the SNe~IIn and SNe~Ibn subclass. 
We have used near-simulatenous observations ($ <0.2$ day) in {\it B} and {\it V} filter without any interpolations, to generate the color curve. The plotted epoch is the mean of observation epochs.
The {\it (B-V)$_{0}$} color curve of SN~2021foa increases by $\sim$ 0.5 mag in colors from $-$10 d to about 40 d post maximum. In the phase space of color evolution, we see two sectors of events. One set of SNe (1996al, 2006jc, 2016jbu, 2018cnf; open symbols; Category 1) in our plot, increases from red to blue from $-$10 d to about maximum and then becomes flatter in the color curve evolution while for the latter, we see a constant rising of red to blue colors upto 30-40 d post maximum (SNe; closed symbols; Category 2) and then drops in the color evolution. For SNe 2019uo, 2019wep we cannot conclusively say anything due to limited data sets. Our SN~2021foa follows Category 2 in the evolution. The SNe of Category 1 are the SNe~IIn which had a long term precursor activity. Also, for these events the flattening in color evolution is seen after the second peak in the SN light curve \citep{Brennan2022}. This is in contrary to SN~2021foa which shows a short term precursor and becomes red upto 50 d post maximum. Post 50 d, the colorcurve again becomes blue and continues upto 150 d. This marks also the phase where we see a change in the mass-loss rates of the evolution (see subsection~\ref{mass-lossrate}). 

We compare the absolute magnitude ($r/R$-band) lightcurve of SN~2021foa with a group of SNe~IIn and SNe~Ibn. For the cases where $r$-band in not available, we use Johnson Cousin $R/r$-band. The absolute magnitude lightcurve of SN~2021foa behaves similarly with other events having precursor activities. 
The precursor lightcurve (Event A) had an absolute magnitude $\sim$ $-$14 mag (\citetalias{reguitti2022}) similar to SNe~2009ip, 2016jbu and 2021qqp \citep{Hiramatsu2024}. The second peak in the lightcurve (Event B) lies fairly intermediate (M$_{V}$ = $-$17.8) among the SNe~IIn and SNe~Ibn (c.f.r Figure 1 of \citetalias{reguitti2022}). \cite{Kiewe2012}, through their studies have found have found that most SNe~IIn with precursor events typically rises to second peak around $\sim$ 17 days. The event B typically rises to a maximum with absolute magnitude $\sim$ $-$18 mag $\pm$ 0.5 \citep{Kiewe2012,Nyholm2020} followed by a bumpy decline. Our SN~2021foa also reaches a peak mag at around $\sim$ 20 days, however, SN~2021foa has a low luminosity compared to the sample of \cite{Kiewe2012, Nyholm2020}. Overall, the light curve of SN~2021foa is similar in luminosity to both SNe~IIn and SNe~Ibn population with L $\sim$ 10$^{42}$ - 10$^{43}$ erg sec$^{-1}$, but the lightcurve resembles more those of SNe~IIn given the heterogeneity. The SNe~Ibn in our comparison sample have instead
more short-lived and less bumpy lightcurves, in accordance with the sample presented by \cite{Hosseinzadeh2017}. Also, interestingly, some events with precursor activity like SNe~2009ip and 2018cnf show a light curve shoulder similar to that of SN~2021foa, 20 d post $V$-band maximum. This is most likely associated with the change in the mass-loss rate happening years before explosion. This is also affected by the opacity effects influencing the lightcurve behaviour. 

\begin{figure*}
	\begin{center}
		\includegraphics[scale=0.5]{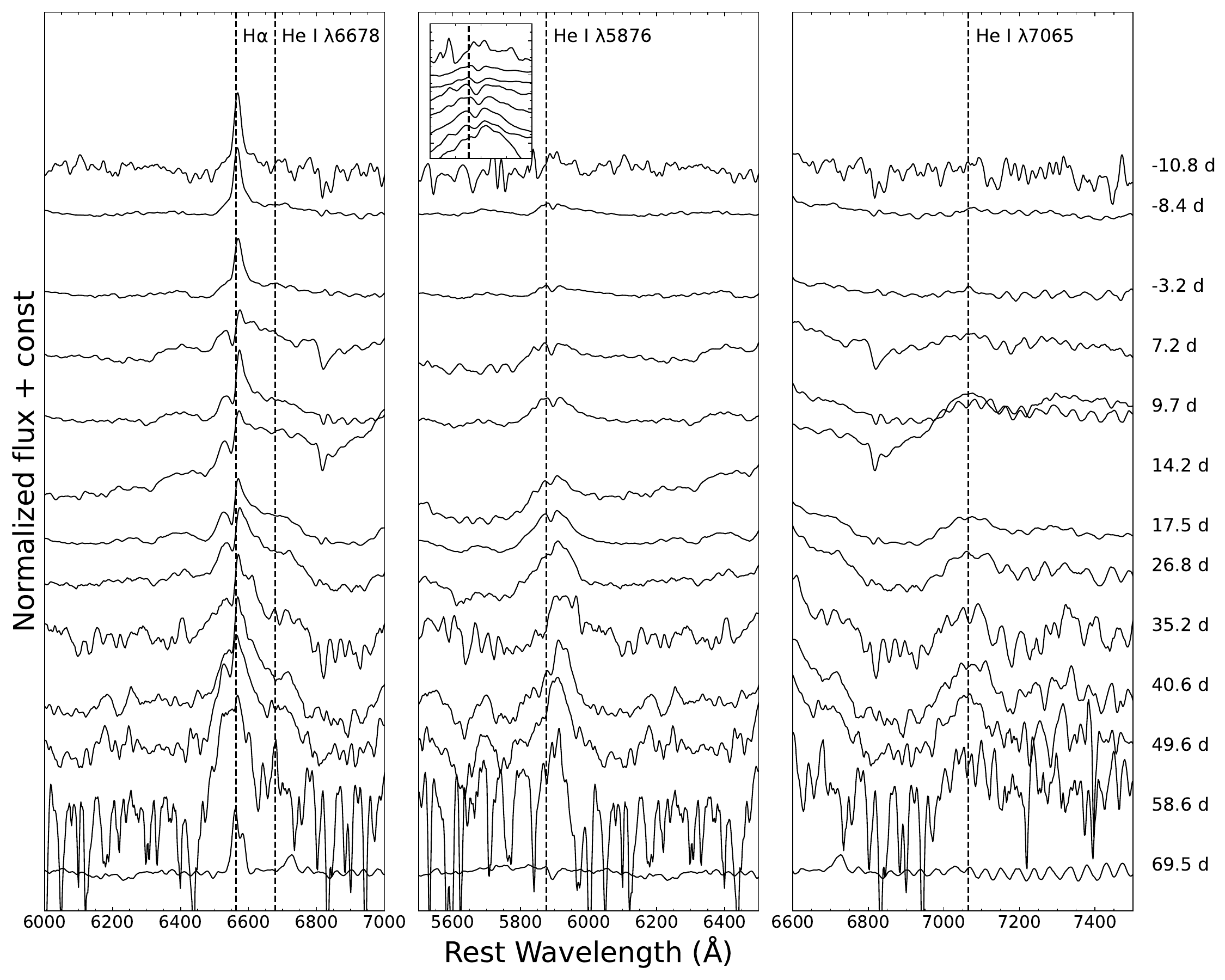}
	\end{center}
	\caption{{\it Figure shows the narrow P-Cygni H$\alpha$ profile, He {\sc i} 5876, 6678, 7065 \AA~ profiles for SN~2021foa. All the spectra have been continuum normalised. The inset in the middle panel highlights the narrow component of He {\sc i} 5876 \AA~ line in the spectral evolution.}}
	\label{fig:zoomedplot}
\end{figure*} 

\section{Spectral and Lightcurve modelling}
\label{spec-lcmodels}

Figure~\ref{fig:zoomedplot} shows the zoomed-in spectral evolution of the line profiles of H$\alpha$, and He {\sc i} 5876, 6678, 7065 \AA. H$\alpha$ shows a very complex profile throughout the evolution. Initially, H$\alpha$ shows a narrow line on the top of a broad component. Around $-$3.2 d, we see a narrow P-Cygni component appearing on top of a broad H$\alpha$ profile. Thereafter, the H$\alpha$ profile is complex and highly asymmetric. Post 14.2 d, the red part of H$\alpha$ starts developing, possibly due to contamination from He {\sc i} 6678 \AA. In the inset plot in second panel, we see a narrow He {\sc i} 5876 \AA\ component followed by the dip which is most likely due to Na I D. The He {\sc i} 5876 \AA~ line shows evolution, with the FWHM varying between 2500 -- 4000 km s$^{-1}$ which is the narrow to intermediate width component in accordance with \cite{Pastorello2008}. The He {\sc i} 5876 \AA\ grows in strength and by 7.2 d its luminosity becomes comparable to that of H$\alpha$. The He {\sc i} 7065 \AA\ line develops later, at 14.2 d, and grows in strength thereafter.  
Interestingly, the FWHM of the He {\sc i} 5876 \AA~ line is similar to that of the H$\alpha$ line throughout the evolution, which again may indicate a mixed composition of the CSM. The implications of these line profiles with regards to the geometry of the SN is discussed in Sect.~\ref{halphadecomp}. 

\begin{figure}
	\begin{center}
		\includegraphics[width=\columnwidth]{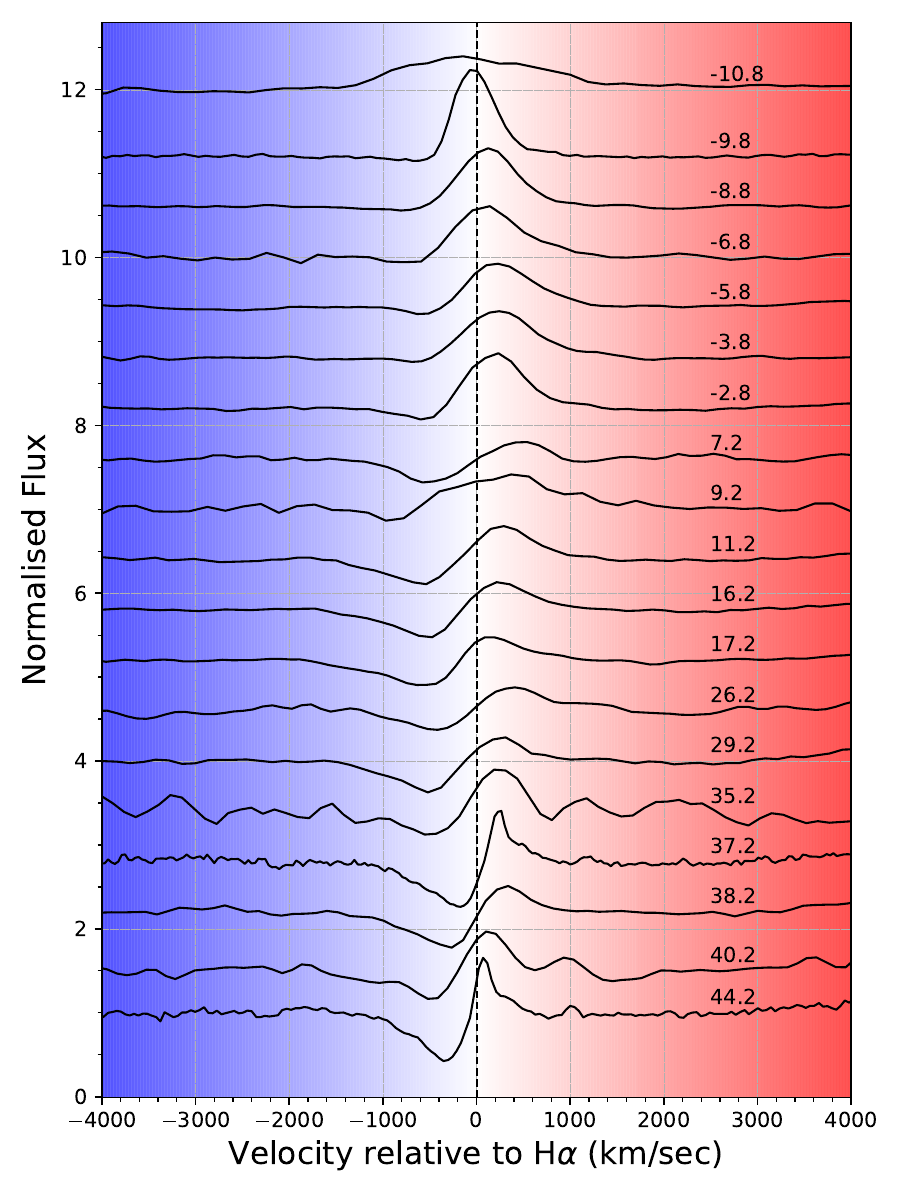}
	\end{center}
	\caption{{\it Figure shows the zoomed narrow emission profile seen in the evolution of SN~2021foa. The narrow profile is obtained by fitting and subtracting the broad component from the H$\alpha$ spectral profile. The narrow features are not clearly resolved in some spectra due to limited resolution. }}
	\label{fig:zoomedpcygninarrow}
\end{figure} 

\begin{figure*}
	\begin{center}
		\includegraphics[scale=0.5]{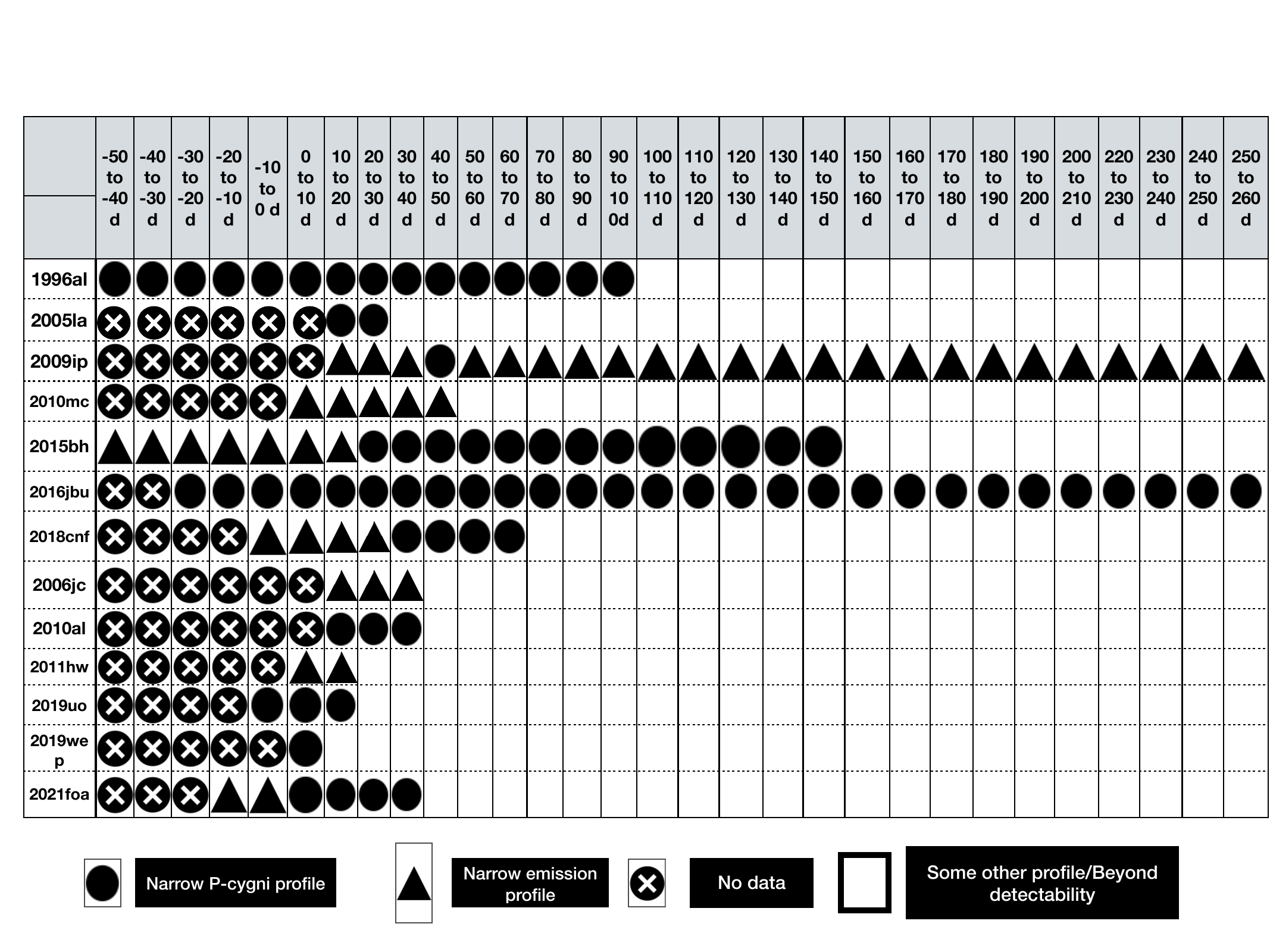}
	\end{center}
	\caption{{\it Figure shows a distribution of the appearance of narrow emission or narrow P-Cygni profile of H$\alpha$ and He {\sc i} in the spectral profiles of a set of SNe IIn and SNe Ibn along with comparison of SN 2021foa. The phase has been chosen with respect to the epoch of maximum for all the SNe in the comparison. For all the SNe IIn in the chart, the explosion occurred before -10 days, however, for SNe Ibn, the explosion epochs are sometimes at a later date as quoted in \citep{Hosseinzadeh2017}}}
	\label{fig:narrowlines}
\end{figure*} 

To discern the origin of the narrow P-Cygni profile of H$\alpha$, we selected the broad emission and absorption components as a continuum and normalize the spectra with respect to it.  
Figure~\ref{fig:zoomedpcygninarrow} shows that in the continuum normalized spectra, from $-$10.8 d to 49.5 d, we still see a narrow emission of FWHM $\sim$ 500 km s$^{-1}$. We want to remark that N II lines as well as the narrow H$\alpha$ and H$\beta$ lines appears to be redshifted by nearly 100 km sec$^{-1}$ in all the the high-resolution spectra, including our spectra from SUBARU. Since it is noticed in multiple spectra from different telescopes, it is likely not an issue with the wavelength calibration and is most likely associated with the incorrect estimation of the redshift of the SN/host-galaxy. However, we kept the redshift same with \citep{reguitti2024,Farias2024} to be consistent.
No absorption features are seen in the spectra before $-$8.3 d. The narrow P-Cygni absorption starts to appear at $-$5.8 d with its blue edge reaching up to velocities of $\sim$ $-$1000 km s$^{-1}$. The narrow P-Cygni becomes really prominent at 7.2 d. The FWHM of the P-Cygni profile typically varied between 500 km s$^{-1}$ to 800 km s$^{-1}$.

To compare the origin of the appearance of narrow P-cygni absorption/emission profiles in interacting SNe, we compared the time evolution of the appearance of P-Cygni profiles for our group of SNe~IIn and SNe~Ibn (see Figure~\ref{fig:narrowlines}). 

For most of the events in our comparison sample (SNe~1996al, 2005la, 2010al, 2016jbu, 2019uo, 2019wep), the narrow P-Cygni lines is present from the beginning of the observed evolution of the SN.
This arises due to the presence of pre-shock CSM along the line of sight \citep{smith2017_interacting}. 
However, some objects (SNe~2006jc, 2010mc, 2010al, 2011hw) show only narrow emission profiles. In this case, we are not seeing the ionized pre-shock CSM along the line of sight. 
In addition to that, SN 2015bh shows a delayed onset of a P-Cygni profile similar to what is seen for SN~2021foa \citep{Nancy2016,Thone2017}, and SN~2009ip shows narrow emission early on, followed by narrow P-Cygni at intermediate times and then narrow emission lines again at a later stage of the evolution. \cite{smith2009} suggested that in SN~2009ip, narrow P-Cygni initially arose when the star was in the LBV phase. Just after the explosion, it showed narrow emission lines due to interaction with a dense CSM and thereafter from interacting with another shell moving at $-$2000 km s$^{-1}$. \cite{smith2009} also associated the origin of the narrow P-Cygni as due to outbursts a few decades prior to a ``hyper eruption" or the final core-collapse. For the case of SN~2021foa, we also see that the velocity of the blue edge extends up to $-$1000 km s$^{-1}$ and we see narrow P-Cygni features developing, which indicate the presence of a shell/disk of CSM along the line of sight. A detailed interpretation of this associated with the geometry of the CSM is also described in Sect.~\ref{physcenario}.

\begin{table*}
    \centering
    \caption{The table lists the FWHM of the components of H$\alpha$ obtained by the spectral fitting. The H$\alpha$ profile is decomposed into a Narrow width, Intermediate width and Broad Width component at different stages of its evolution. \bf{The uncertainties in the table represent the fitting uncertainties. Uncertainties due to the resolution element ($\sim 500$ km s$^{-1}$) should be added in quadrature to find the true uncertainties.}}
    \label{tab:halpha_spectral_decomposition}
    \begin{tabular}{| c | c c c | c c c | c c c |}
\hline \hline
Phase  & \multicolumn{2}{c|}{Lorentzian} & \multicolumn{2}{c|}{Narrow Emission} & \multicolumn{2}{c|}{Narrow P-Cygni} & \multicolumn{2}{c|}{Broad Absorption} \\ 
epoch & center & fwhm & center & fwhm & center & fwhm & center & fwhm \\ 
\hline
(d) & (km s$^{-1}$) & (km s$^{-1}$) & (km s$^{-1}$) & (km s$^{-1}$) & (km s$^{-1}$) & (km s$^{-1}$) & (km s$^{-1}$) & (km s$^{-1}$)\\ 
\hline
-10.8 & 196$\pm$9 & 3638$\pm$703 & 196$\pm$9 & 767$\pm$35 & -- & -- & -4535$\pm$175 & 3532$\pm$ -- \\
-8.8 & 598$\pm$47 & 3631$\pm$116 & 128$\pm$3 & 740$\pm$10 & -- & -- & -3678$\pm$315 & 3532$\pm$ -- \\
-5.8 & 290$\pm$16 & 3612$\pm$764 & 290$\pm$16 & 777$\pm$63 & -- & -- & -4150$\pm$318 & 3532$\pm$ -- \\
-3.8 & 269$\pm$6 & 4477$\pm$326 & 269$\pm$6 & 841$\pm$21 & -- & -- & -3876$\pm$89 & 3532$\pm$ -- \\
7.2 & 450$\pm$49 & 7843$\pm$857 & 450$\pm$49 & 653$\pm$83 & -506$\pm$32 & 653$\pm$83 & -3688$\pm$178 & 3304$\pm$407 \\
9.2 & 247$\pm$25 & 5475$\pm$231 & 247$\pm$25 & 714$\pm$77 & -328$\pm$302 & 1152$\pm$349 & -3171$\pm$100 & 3285$\pm$240 \\
16.2 & 303$\pm$57 & 7265$\pm$437 & 303$\pm$57 & 701$\pm$122 & -644$\pm$47 & 850$\pm$116 & -3432$\pm$114 & 3092$\pm$264 \\
17.2 & 110$\pm$29 & 6529$\pm$162 & 110$\pm$29 & 738$\pm$92 & -436$\pm$232 & 1151$\pm$240 & -3420$\pm$53 & 3301$\pm$113 \\
26.2 & 300$\pm$58 & 7363$\pm$259 & 300$\pm$58 & 804$\pm$203 & -228$\pm$577 & 1177$\pm$343 & -3251$\pm$97 & 4136$\pm$175 \\
35.2 & 426$\pm$181 & 6648$\pm$260 & 225$\pm$28 & 467$\pm$64 & -523$\pm$58 & 706$\pm$328 & -2100$\pm$ -- & 3000$\pm$ -- \\
40.2 & 89$\pm$18 & 5200$\pm$137 & 89$\pm$18 & 511$\pm$94 & -381$\pm$261 & 942$\pm$356 & -2085$\pm$217 & 3120$\pm$348 \\
49.2 & -27$\pm$27 & 4601$\pm$91 & -27$\pm$27 & 652$\pm$112 & -778$\pm$29 & 566$\pm$71 & -- & -- \\
58.2 & -377$\pm$73 & 3986$\pm$175 & -- & -- & -- & -- & -- & -- \\
\hline
\end{tabular}

\end{table*}

\subsection{H$\alpha$ decomposition}
\label{halphadecomp}
To decipher the origin of the complex H$\alpha$ structure, we tried to deconvolve the line profiles of this SN. 

\begin{figure*}
    \centering
    \includegraphics[width=0.40\linewidth]{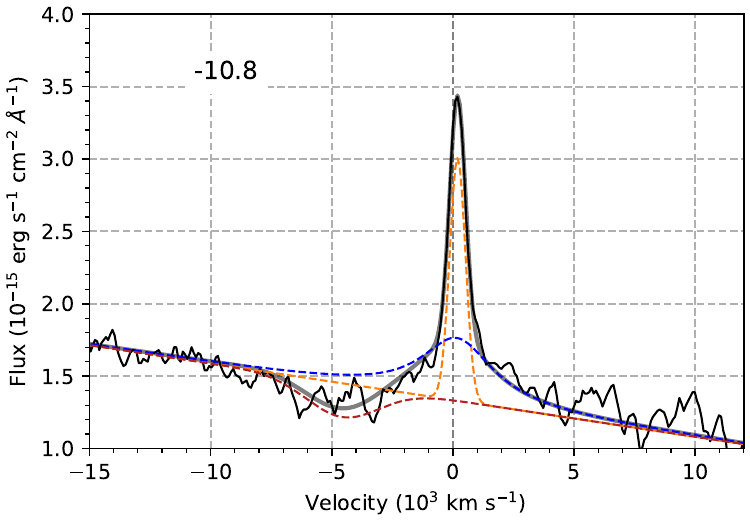}%
    \includegraphics[width=0.40\linewidth]{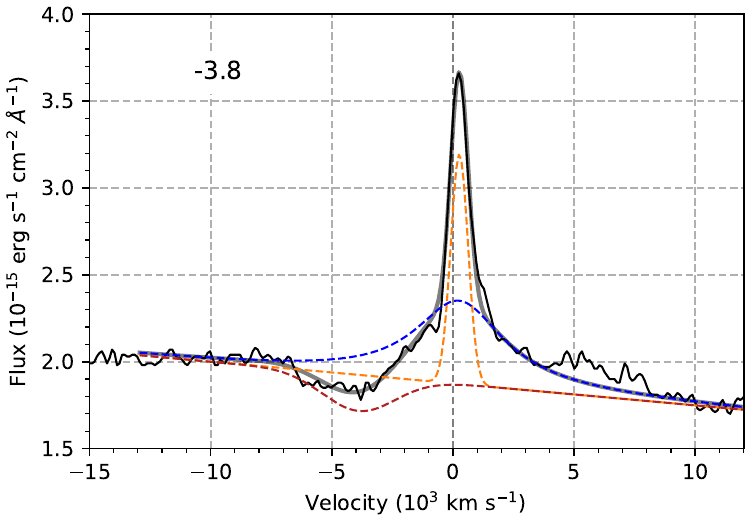}

    \includegraphics[width=0.40\linewidth]{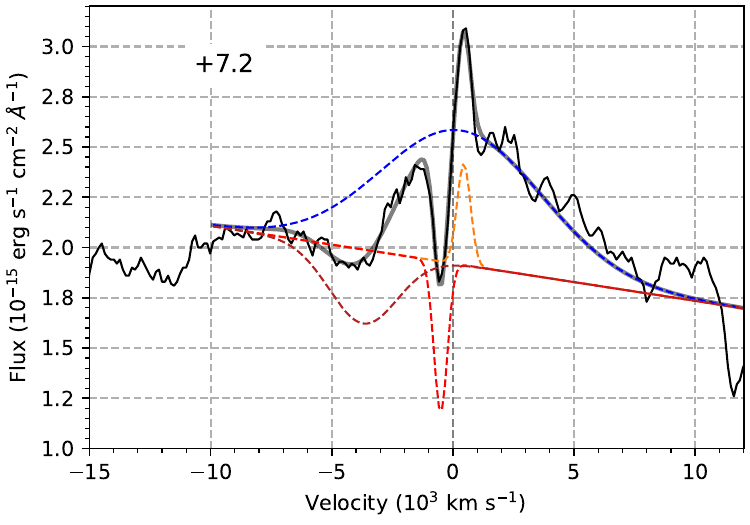}%
    \includegraphics[width=0.40\linewidth]{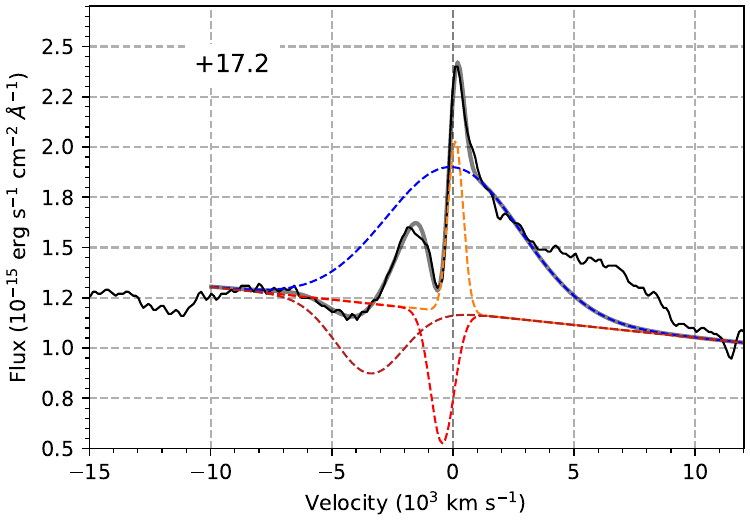}
    
    \includegraphics[width=0.40\linewidth]{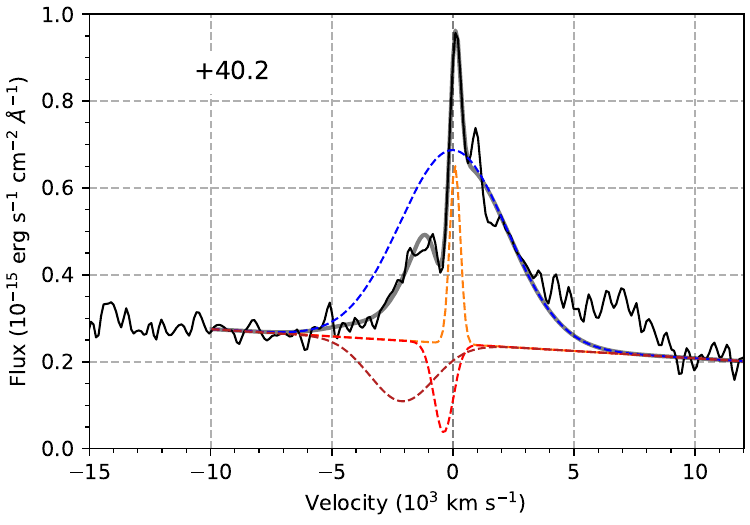}%
    \includegraphics[width=0.40\linewidth]{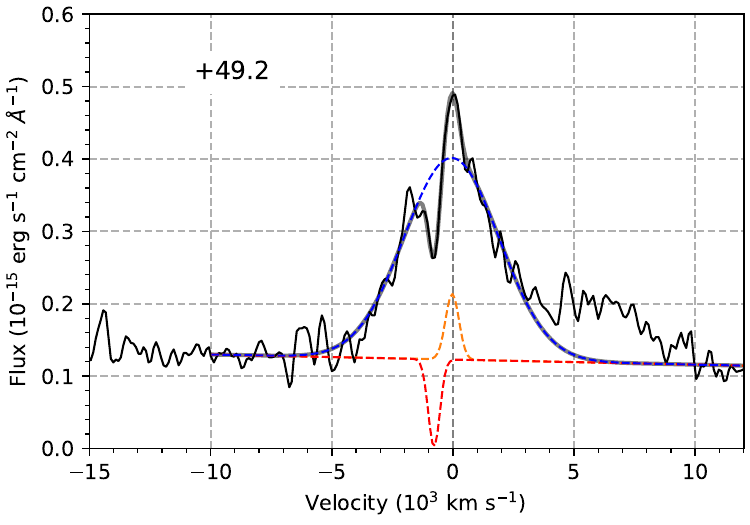}
    \caption{{\it Figure shows the H$\alpha$ evolution of SN~2021foa being fitted with multiple Lorentzian and Gaussian components. The host galaxy contribution is subtracted, and the continuum is selected by masking the line regions. SN~2021foa shows a very complex geometry of the spectral profiles. We see the narrow, intermediate, and broad components, respectively, in the spectral profile of H$\alpha$ at different stages of the evolution.}}
    \label{fig:halpha}
\end{figure*}

\begin{figure}
	\begin{center}
		\includegraphics[width=\columnwidth]{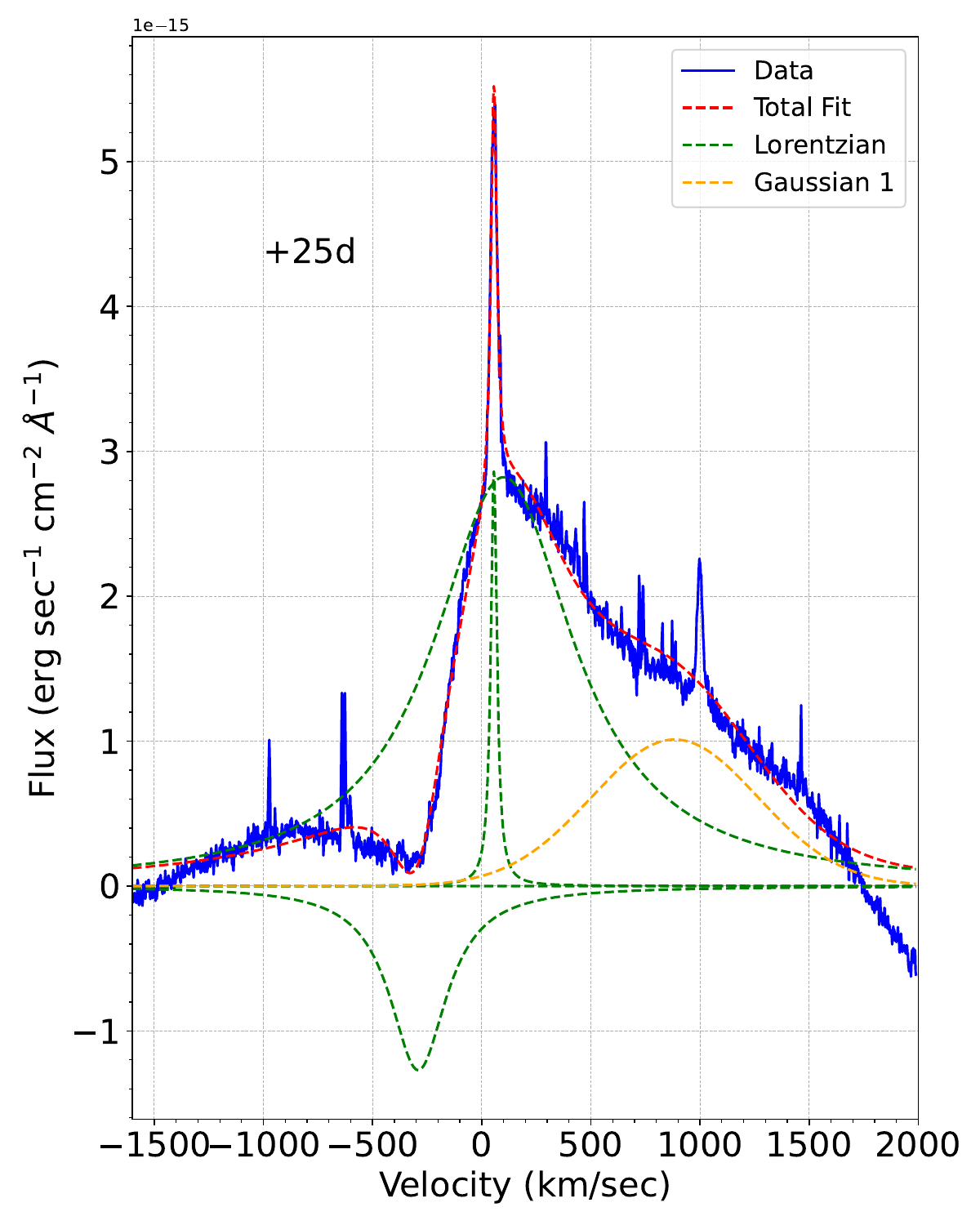}
	\end{center}
	\caption{{\it Figure shows the deconvolved high-resolution Subaru HDS spectrum of SN~2021foa. The narrow component is well resolved for SN~2021foa with FWHMs of 319 and 782 km sec$^{-1}$ in absorption and emission.}}
	\label{fig:subaru}
\end{figure}

From the typical explosion circumstances of interacting SNe \citep{smith2017_interacting}, we expect a narrow component from the unshocked CSM, an intermediate width component either from the e-scattering of the narrow line photons or from the cold, dense shell (CDS), a broad width component from the uninterrupted ejecta which sometimes may show an associated absorption component as well. Therefore, we try to deconstruct the H$\alpha$ profile in terms of these components.
The H$\alpha$ profile was fitted with combinations of different line profiles in order to reproduce the overall line profile seen in SN~2021foa at different stages of its evolution. We used i) A narrow Gaussian component that is slightly redshifted from the center ii) A Lorentzian/Gaussian intermediate width component that is redshifted from the center, iii) A broad Gaussian component in absorption. At early times, the overall H$\alpha$ profile is better represented by a Lorentzian intermediate width component as the emission lines are dominated by electron scattering. On top of that, we have a narrow emission component of H$\alpha$ mostly arising from pre-shocked CSM. 
During the middle phases, the narrow emission is replaced by a narrow P-Cygni profile, and at late times ($>$30 d), a Gaussian intermediate width component better reproduces the overall line profile as it evolves into a more complex multi-component structure. The choice of the continuum is very critical for performing these fits. The continuum is selected to be far from the line region by at least 50~\AA. The continuum is varied between 50 $\pm$ few \AA~ to check the consistency of the fits. The spectral evolution and the corresponding fits at representative epochs are shown in \autoref{fig:halpha}. The parameters of the components obtained from the fitting at all epochs are given in \autoref{tab:halpha_spectral_decomposition}. The errors listed represent only fitting errors, and other uncertainties like resolution matching, subtraction of host-spectra, and imperfections in wavelength calibration have not been included. 

During $-$10.8 d to $-$3.2 d, the blue edge of the line profile extends up to $-$4500 km s$^{-1}$ with an estimated FWHM of 3532 km sec$^{-1}$. We want to remark that the fitting FWHM from $-$10.8 to $-$3.2 generates a limiting FWHM of 3500 km sec$^{-1}$ which is our prior in fitting. An increase in prior shifts the absorption center to the redder wavelength, which is unphysical for a P-Cygni profile. Thus, we want to remark that at pre-maximum times, there is a shallow component in absorption mostly due to the freely expanding SN ejecta. Diverse geometry ranging from a disk-like CSM in SN~2012ab \citep{Gangopadhyay2020} or a clumpy CSM like in SN~2005ip \citep{smith2009} can facilitate direct line of sight to the freely expanding ejecta. In addition to that, we have an intermediate width Lorentzian profile varying in FWHM between 3600 - 4500 km sec$^{-1}$ whose wings are mostly dominated by electron scattering and arising due to the ejecta interacting with dense CSM \citep{smith2014}. We also see a narrow emission varying in FWHM between 700 - 800 km sec$^{-1}$. This indicates that the photosphere lies in the unshocked CSM at this phase. The UV (and bolometric) light curves peak after this phase, which ionizes the unshocked CSM \citep{smith2014}. An emission component was fitted in the early spectrum of SN~2021foa at this phase as the P-Cygni was not resolved.

From 7.2 d to 40.2 d, we notice a significant increase in the Lorentzian emission FWHM of H$\alpha$, indicating an enhanced interaction between the SN ejecta and the CSM. The FWHM of the emission component typically varied between 5500 km sec$^{-1}$ - 7800 km sec$^{-1}$. The absorption component at this phase decreases with a reduction in the systematic blueshift. The 26.2 d spectrum marks the onset of the optically thin regime where we no longer see absorption in the H$\alpha$ profile. Since we have prominent He {\sc i} emission in the line profiles of SN~2021foa, post 17.2 d, we do not fit the right bump of the profile of H$\alpha$ which arises possibly due to He {\sc i} 6678 \AA. The beginning of 7.2 d also marks the prominent strength of narrow P-Cygni features appearing in the spectral evolution of SN~2021foa. 

After day 40.2, we see only the intermediate width Lorentzian component in the line profiles of SN~2021foa with FWHM varying between 3900 km sec$^{-1}$ - 5200 km sec$^{-1}$. However, after 40.2 d, a blueshift can be noticed in the intermediate width component of H$\alpha$, which has now centered between $\sim$ $-$27 to $-$377~km~s$^{-1}$. The late-time blueshift can be explained by dust formation in the post-shock CSM or ejecta (similar to SNe 2005ip, 2010jl, and 2015da; \citealp{smith2009, 2016MNRAS.456.2622J, 2015dalate_smith2024}). The narrow P-Cygni has also now reduced in FWHM between 500 km sec$^{-1}$ - 600 km sec$^{-1}$.

Figure~\ref{fig:subaru} shows the deconvolved high resolution H$\alpha$ region spectrum of SN~2021foa. The narrow component in the spectra of SN~2021foa can be well reproduced by two Lorentzians in absorption and emission with FWHMs of 319 $\pm$ 20 km sec$^{-1}$ and 782 $\pm$ 15 km sec$^{-1}$. Along with that, we see an additional IW H$\alpha$ Lorentzian component of FWHM 900 km sec$^{-1}$. We see a very narrow component of FWHM 33 km sec$^{-1}$ in the H$\alpha$ profile which is most likely from the host galaxy contribution. We, thus, see that the narrow component seen in our high resolution spectrum is in concordance with our model fittings validating the narrow emission to narrow P-cygni transition in SN~2021foa. 

The detailed physical interpretation corresponding to the line geometries is explained in Section~\ref{disc-summary}. 

\subsection{Radius and Temperature Evolution}
\label{rad-temp}

\begin{figure}
	\begin{center}
		\includegraphics[width=\columnwidth]{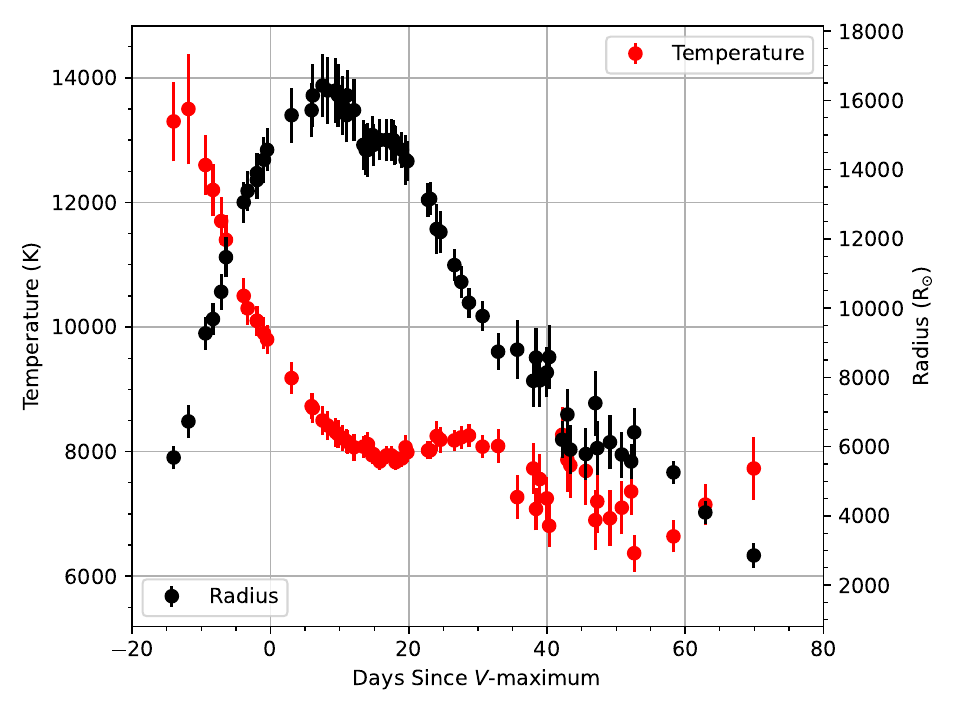}
	\end{center}
	\caption{{\it The complete radius and temperature evolution of SN~2021foa.}}
	\label{fig:rad-temp}
\end{figure} 

As the SN ejecta is expanding, the shock breakout from the surface of the progenitor is 
followed by a rapid cooling due to the rapid expansion driven by the shock \citep{falk1977}. This would lead to a rapidly increasing photospheric radius and a decrease in the temperature of the SN ejecta within a couple of hours of the explosion. However, this will be affected for extended stars with larger radius. For the case of interacting SNe, this is not always the case as the ejecta is masked by CSM \citep{Irani2023}. Figure~\ref{fig:rad-temp} shows the radius and temperature evolution of SN~2021foa. The temperature and radius are obtained from the blackbody fits including the UV to IR data. For estimating the values, {\it Stefan-Boltzman} law was used to estimate the corresponding parameters. SN~2021foa first shows a decrease in temperature evolution from 14000~K to about 8000~K. We then see a slight rise in the temperature evolution of SN~2021foa from 7300~K to 8300~K between 17 d and 23 d post maximum, during the shoulder in the lightcurve, and then the temperature evolution becomes flatter. This rise in the temperature indicates an injection of energy in the cooling ejecta, perhaps due to interaction with additional CSM or interaction with regions of enhanced CSM density. This is in turn affected by opacity effects in the CSM ejecta interacting zone. 
 
The black body radius of SN~2021foa increases from 5500 R$_{\odot}$ to 15500 R$_{\odot}$ at $\sim$ 10 d past maximum light. From 7 d to $\sim$ 20 d past max, the radius stays at 16000 R$_{\odot}$. The radius evolution then shows a small shoulder similar to the temperature evolution, fluctuating between the values of 16000 R$_{\odot}$ to 15000 R$_{\odot}$ and then decreases to a value of 2800 R$_{\odot}$. The late time radius evolution is flat, as is the temperature evolution and the luminosity as well, but we do not show that in the plot because blackbody approximation fails there.

We thus see that the temperature increases at the point in time when we expect an interaction to occur, a few days after maximum light and when the H$\alpha$ and He {\sc i} lines start appearing at similar strength in the spectral evolution. A similar flattening behaviour is also noticed in the radius evolution of SN~2021foa at this phase. The radius and temperature evolution are well in synergy with the light-curve evolution.

\subsection{Hydrodynamical Modelling}
\label{hydro}
To construct the bolometric light curve of SN~2021foa between UV to IR bands, we used the \texttt{SuperBol} code \citep{2018RNAAS...2..230N}. The missing UV and NIR data was supplemented by extrapolating the Specrtral Energy Distributions (SEDs) using the blackbody approximation and direct integration method as described in \cite{2017PASP..129d4202L}. A linear extrapolation was performed in the UV regime at late times.
We conducted light-curve modeling of SN~2021foa using the one-dimensional multi-frequency radiation hydrodynamics code \texttt{STELLA} \citep{1998ApJ...496..454B,2000ApJ...532.1132B,2006A&A...453..229B}. Because \texttt{STELLA} treats radiation hydrodynamics in multi-frequencies, \texttt{STELLA} can construct pseudo-bolometric light curves that can be directly compared with the observed ones. 

\begin{figure}
	\begin{center}
         \includegraphics[width=\columnwidth]{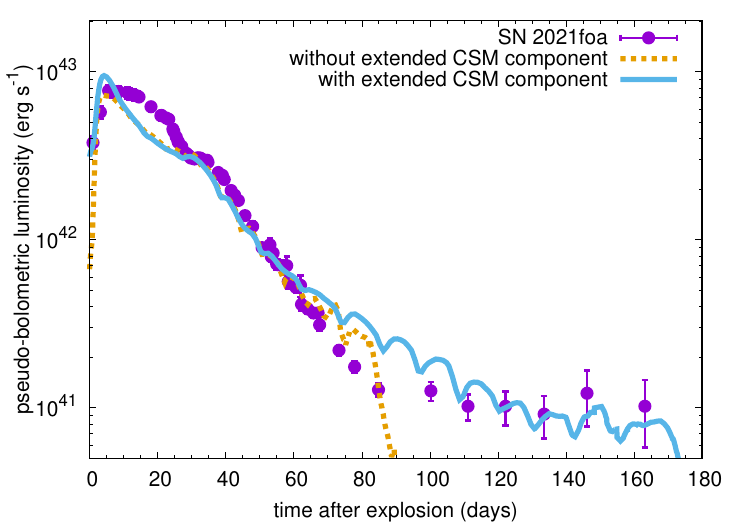}
         \includegraphics[width=\columnwidth]{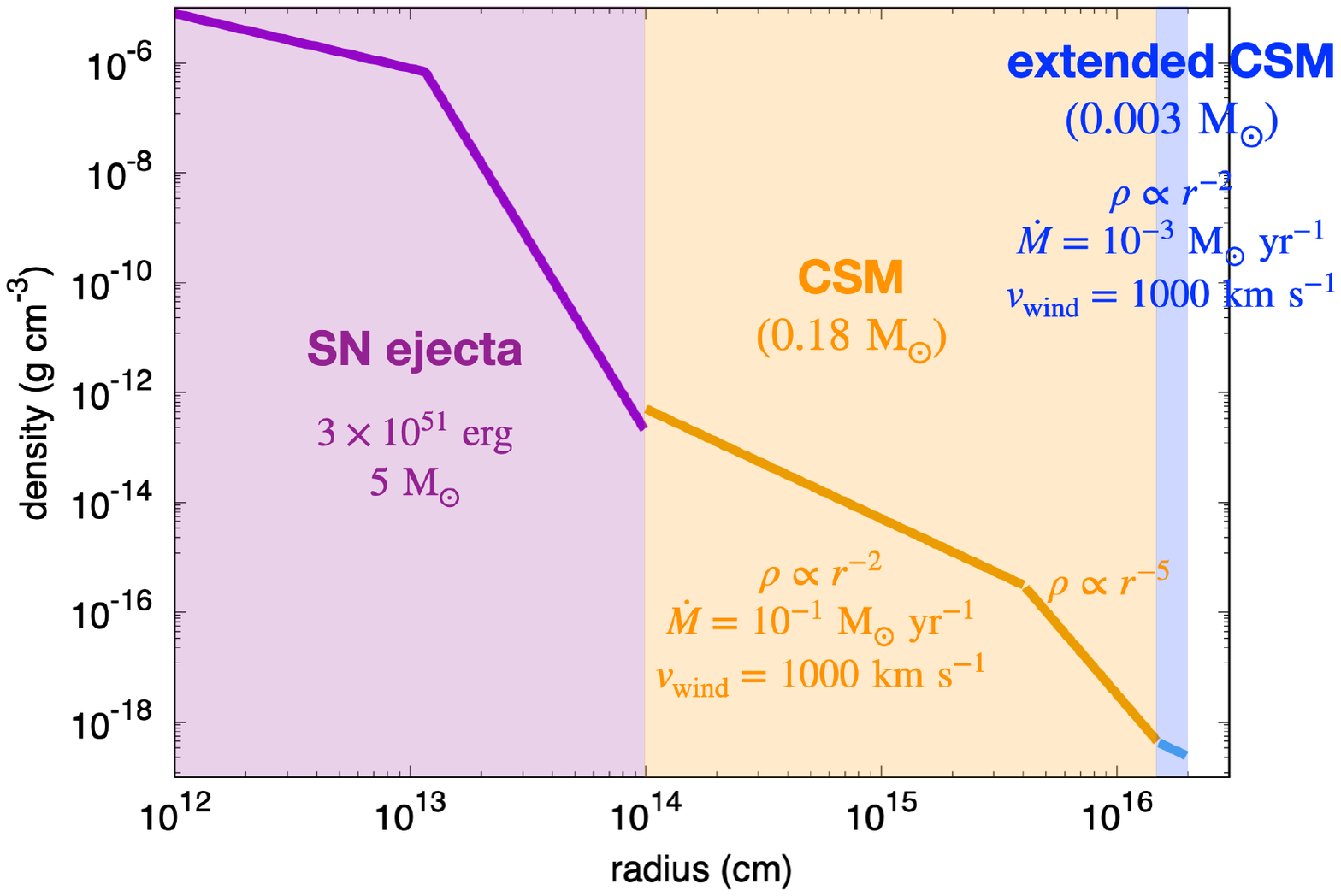}
	\end{center}
	\caption{{\it Bolometric light-curve models (top) and the initial density structure of the models (bottom).}}
	\label{fig:hydro}
\end{figure}

Figure~\ref{fig:hydro} presents our light-curve models and the initial density structure that can reproduce the overall light-curve properties of SN~2021foa assuming a spherically symmetric configuration. We approximate the SN ejecta by using the double power-law density structure ($\rho_\mathrm{ejecta}\propto r^{-1}$ inside and $\rho_\mathrm{ejecta}\propto r^{-7}$ outside, e.g., \citealt{1999ApJ...510..379M}). The SN ejecta are assumed to expand homologously. The SN ejecta start to interact with CSM at $10^{14}~\mathrm{cm}$. This radius is chosen arbitrarily but it is small enough not to affect the overall light-curve properties. The SN ejecta have an explosion energy of $3\times 10^{51}~\mathrm{erg}$ and a mass of $5~\mathrm{M_\odot}$. We want to remark here that there is a degeneracy in the ejecta mass and energy, and thus the particular set taken here ($5~\mathrm{M_\odot}$) is an assumption.

We found that the CSM with two power-law density components can reproduce the overall light-curve properties of SN~2021foa. The wind velocity is assumed to be $1000~\mathrm{km~s^{-1}}$ adopted from the blue edge of the narrow absorption component of the CSM (see Figure~\ref{fig:narrowlines}). The inner CSM component has $\rho_\mathrm{CSM}\propto r^{-2}$. The CSM with $10^{-1}~\mathrm{M_\odot~yr^{-1}}$ can account for the early-time luminosity around the light-curve peak. After around 30~days, the luminosity decline becomes faster than expected from interaction with a CSM with $\rho_\mathrm{CSM}\propto r^{-2}$. We found that the fast luminosity decline can be reproduced when the CSM density structure follows $\rho_\mathrm{CSM}\propto r^{-5}$ from $3\times 10^{15}~\mathrm{cm}$. This CSM component has mass of $0.18~\mathrm{M_\odot}$.

The pseudo-bolometric light-curve of SN~2021foa flattens from around 80~days. In order to reproduce the luminosity flattening, an extended CSM component that is flatter than $\rho_\mathrm{CSM}\propto r^{-5}$ is required. We found that if the extended CSM with $\rho_\mathrm{CSM}\propto r^{-2}$ is attached above $1.5\times 10^{15}~\mathrm{cm}$, we can reproduce the flat phase in the light curve when $10^{-3}~\mathrm{M_\odot~yr^{-1}}$ is assumed. Assuming a wind velocity of $v_\mathrm{wind}=1000~\mathrm{km~s^{-1}}$, this kind of the CSM structure can be achieved if the mass-loss rate of the progenitor gradually increase from $10^{-3}~\mathrm{M_\odot~yr^{-1}}$ to $10^{-1}~\mathrm{M_\odot~yr^{-1}}$ from about 5~years to 1~year before the explosion, and the mass-loss rate is kept at $10^{-1}~\mathrm{M_\odot~yr^{-1}}$ in the final year before the explosion.

Also, we want to mention that the model is stable around the peak time, however, the latter part of the lightcurve is not well-established due to numerical limitations of \texttt{STELLA}. Hence, a two zone CSM for SN~2021foa well-reproduces the lightcurve until 80 d. We want to remark that even if we obtain a disk-like CSM geometry from spectral modelling in Section~\ref{spec-lcmodels}, our simplistic approximations of spherical geometry well reproduces the order of magnitude of the obtained physical parameters in lightcurve modelling.

\section{Mass-loss rates}
\label{mass-lossrate}
The mass-loss rates are governed by the ejecta-CSM interaction in SNe~IIn/Ibn and can be estimated from spectral profiles as well \citep{Gangopadhyay2020}. Assuming that the luminosity of the ejecta-CSM interaction is fed by energy at the shock front, the progenitor mass-loss rate $\dot{M}$ can be calculated using the relation of \cite{chugai_danziger_1994}: 

\begin{equation}
\dot{M}=\frac{2L}{\epsilon}\frac{v_w}{v_{SN}^{3}}
\end{equation}

where $\epsilon$ ($<1$) is the efficiency of conversion of the shock's kinetic energy into optical radiation (an uncertain quantity), $v_w$ is the velocity of the pre-explosion stellar wind, $v_{\mathrm{SN}}$ is the velocity of the post-shock shell, and $L$ is the bolometric luminosity of the SN. The above equation is derived assuming a spherical symmetry and assuming M$_{ej}$ $\gg$ M$_{csm}$. We see changes in the spectral line profile for SN~2021foa around maximum light, so we estimate the mass-loss rates at both $-$5.8 d and at 7.2 d post maximum.

Since we see narrow emission lines of both H$\alpha$ and He {\sc i} at different stages of the evolution, we assume a typical unshocked wind velocity as observed for LBV winds of $v_{w}$$\sim$100 km s$^{-1}$ and $v_{w}$$\sim$1000 km s$^{-1}$ for the WR stars. The shock velocity is inferred from the IW component. We want to remark, however, that the first phase might be affected by electron scattering, and there may be contamination by the ejecta signatures at these epochs. We take the shock velocities to be 3612 km sec$^{-1}$ and 7843 km sec$^{-1}$ at $-$5.8 d and 7.2 d. Using the bolometric luminosity at day $-$5.8 (L $ = 3.77 \times 10^{42}$ erg sec$^{-1}$), wind speeds of 100 (and 1000) km sec$^{-1}$ and assuming 50$\%$ conversion efficiency ($\epsilon=0.5$), the estimated mass-loss rate is found to be  0.05 (and 0.5) M$_{\odot}$ yr$^{-1}$. The estimated mass-loss rate at 7.2 d when the narrow P-Cygni line profile becomes prominent in the spectra are estimated similarly for the wind speed of 100 (and 1000) km sec$^{-1}$ and assuming L=5.48 $\times$ 10$^{42}$ erg sec$^{-1}$ gives 0.007 (and 0.07) M$_{\odot}$ yr$^{-1}$. 

The estimated value of mass-loss rate are consistent with the typical LBV winds \citep{Smith_mass_loss_2014} and are consistent with most SNe~IIn, which are of the order of 0.1 M$_{\odot}$ yr$^{-1}$ as observed in some giant eruptions of LBVs \citep{Chugai_1994W, Kiewe2012}. These values are higher than those of normal-luminosity SNe~IIn like SN~2005ip ($2-4\times10^{-4}$ M$_{\odot}$ yr$^{-1}$; \citealp{smith2009}). It is also much larger than the typical values of RSG and yellow hypergiants ($10^{-4}-10^{-3}$ M$_{\odot}$ yr$^{-1}$; \citealp{Smith_mass_loss_2014}), and quiescent winds of LBV (10$^{-5}-10^{-4}$ M$_{\odot}$ yr$^{-1}$, \citealp{Vink2018}). The obtained mass-loss rate in SN~2021foa indicates the probable progenitor to be an LBV star transitioning to a WR star that underwent an eruptive phase and transitioning mass-loss rates. The CSM may be the result of interaction with a binary companion as well, which would explain the asymmetry we see in the line profiles of SN~2021foa. 

In the region between the forward shock and the reverse shock, there exists a contact discontinuity. This boundary separates the shocked CSM and the shocked ejecta. At this interface, material cools, mixes via Rayleigh-Taylor instabilities, and accumulates. This region is referred to as the cold, dense shell (CDS). The velocity at this boundary is denoted as v$_{CDS}$.
When examining the narrow component of the H$\alpha$ line, its evolution did not show significant changes over the lifetime of the SN. However, v$_{CDS}$ showed noticeable variation at the boundary around 7.2 days in the H$\alpha$ evolution.

Considering these two epochs of time at $-$5.8 d and 7.2 d, estimating the time periods preceding the explosion when the progenitor showed some activity is calculated by t = t$_{\mathrm{obs}}$ x (v$_{\mathrm{w}}$/v$_{\mathrm{CDS}}$). Assuming v$_{\mathrm{w}}$ from the narrow component of H$\alpha$ at these two phases indicates that the progenitor had undergone a change in the eruptive activity at two stages corresponding to 0.5 -- 1 year before exploding as an SN. This is much lower than the traditional SNe~IIn and super luminous SNe~IIn like 2006gy, 2010jl and 2017hcc \citep{2007ApJ...671L..17S, 2016MNRAS.456.2622J,2017hcc_2020Nathan} which showed eruptive activity 6-12 years before the explosion. On the contrary, this is quite similar to SN~2019uo \citep{Gangopadhyay2020}, resulting in a shorter-lived light curve as SNe~Ibn. This has also been observed previously for the case of SN~2020oi which is a SN~Ic showing radio interaction signatures and showed progenitor activity/ expelled shells 1 yr before the explosion \cite{2020oi_Maeda2021}.

\section{Discussion}
\label{disc-summary}

We have presented the photometric and spectroscopic analysis of SN~2021foa, and hereafter, we discuss the major properties of the SN and summarise our results. 

In this paper, we present the unique case of SN~2021foa, where we see line luminosity ratios intermediate between SNe~IIn and SNe~Ibn. We also come to the conclusion that it is the H$\alpha$ line luminosity that better separates the two populations, while the two classes cannot be segregated based on He {\sc i} 5876~\AA ~line luminosities. SN~2021foa exhibits the classic evolution of line profile shapes that is common in strongly interacting SNe~IIn, which transition from symmetric Lorentzian profiles at early times (before and during peak), to irregular, broader, and asymmetric shapes at late times well after peak. The phenomena is understood as a shift from narrow CSM lines broadened by electron scattering to emission lines formed in the post-shock cold dense shell CDS \citep{smith2017_interacting,Dessart2015}. In addition, for SN~2021foa, we also see asymmetric He {\sc i} lines which broaden over time from 2000 km sec$^{-1}$ to 5000 km sec$^{-1}$ and show line luminosities comparable to H$\alpha$ around the lightcurve peak. 

The spectral evolution also shows narrow and intermediate-width H lines at pre-maximum times. Around $-$3.8 d, we see narrow P-Cygni appearing in H$\alpha$. The narrow P-Cygni in H$\alpha$ appears later than the narrow P-Cygni in H$\beta$. Furthermore, a recent study by \cite{CSM_structure_Ishii_24} explores the relations between line shapes and CSM structure using Monte Carlo radiative transfer codes. They find that a narrow line exhibits a P-Cygni profile only when an eruptive mass-loss event forms the CSM. The CSM structure from a steady mass loss will have a negative velocity gradient after the SN event due to radiative acceleration. Therefore, an H$\alpha$ photon emitted at the deeper CSM layers, traveling outwards, will never be able to undergo another H$\alpha$ transition. However, if there is an eruptive mass-loss that comes into play after a steady wind scenario, then a positive velocity gradient would give rise to narrow P-Cygni lines formed along the line of sight. We see from subsection~\ref{hydro} and Section~\ref{mass-lossrate}, the mass-loss rates typically varies from 10$^{-3}$  - 10$^{-1}$ M$_{\odot}$ yr$^{-1}$ during this phase. Also, the light curve modelling helps us to infer that the density-radius variation occurs from $\rho$ $\propto$ r$^{-2}$ to $\rho$ $\propto$ r$^{-5}$. This validates our scenario that probably the change in this mass-loss rates and possible eruption would have given rise to this P-Cygni arising along the line of sight. The changing mass-loss rates and the eruptive activity seen at 0.5 -- 5 years before the explosion typically also govern the dual peaked light curve and appearance of P-Cygni seen in the light curve and spectral evolution of SN~2021foa. \cite{reguitti2024} checked the pre-cursor activities of SN~2021foa and have detected it only upto $-$50 days pre-explosion. This implies that even though there is an activity in the progenitor driving the peak luminosities at 0 and 17 d post maximum, it was not significant enough to be detected as a precursor in the light curve a year before the explosion. \cite{Nancy2016,Thone2017} have explained the origin of double-peak due to the interaction with a shell at a later point in time, which can also be a possible case of SN~2021foa. But, we find that a disk-like geometry better explains our overall line profiles, and we discuss below in detail the physical scenario (see subsection~\ref{physcenario}) governing the CSM and the explosion geometry.

\subsection{Physical scenario and asymmetry}
\label{physcenario}
\begin{figure}
	\begin{center}
		\includegraphics[trim={80 50 160 30},clip,width=\columnwidth]{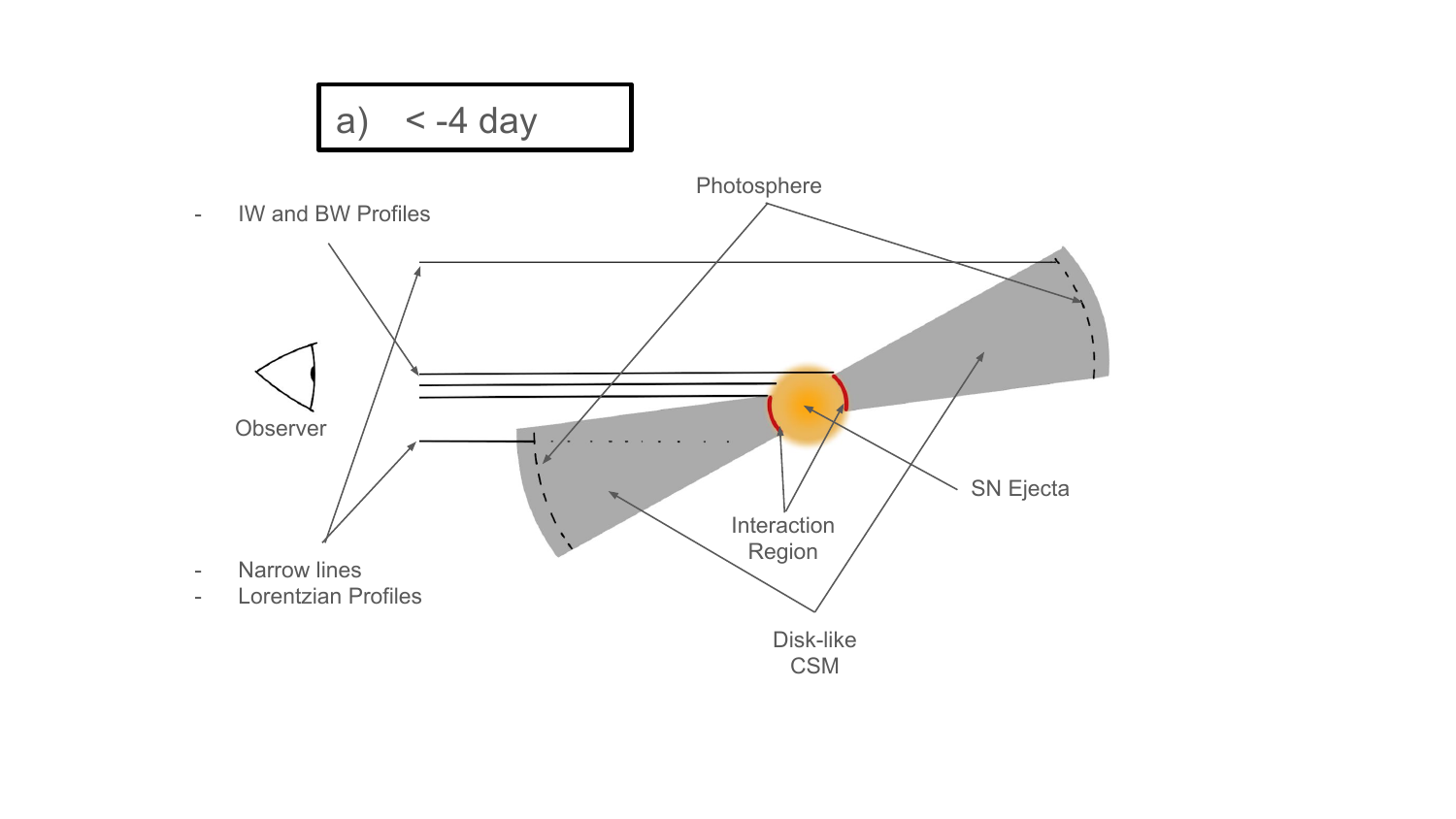}
        \includegraphics[trim={80 50 160 30},clip,width=\columnwidth]{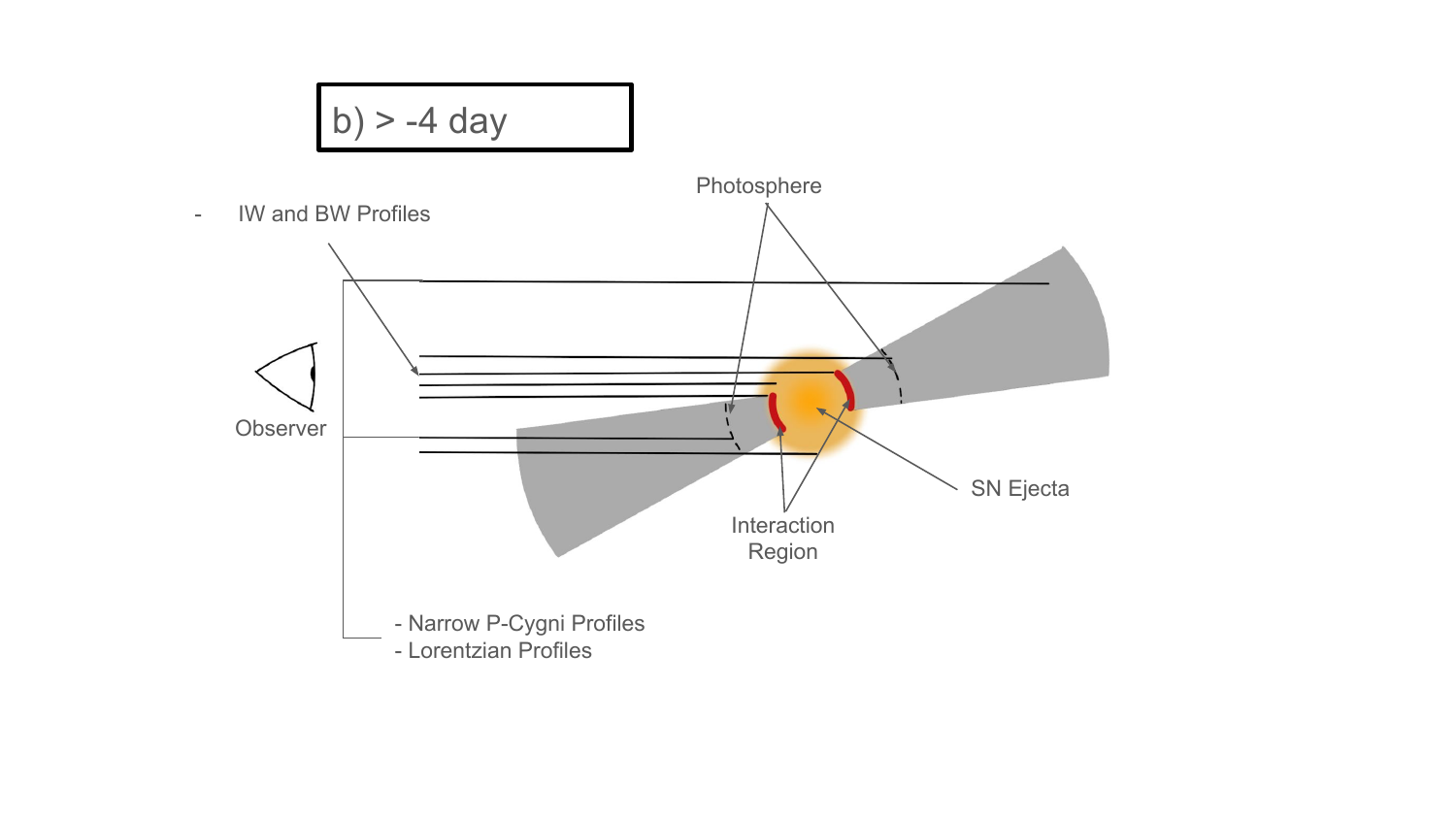}
	\end{center}
	\caption{{\it Cartoon diagram representing the evolution of SN~2021foa through various distinct phases. Representative line-of-sights are shown by solid lines facing towards the observer. A dense disk-like CSM is located at an angle towards the observer. Interaction with dense CSM gives rise to most of the luminosity while the absorption early on, comes from the SN ejecta along the line of line-of-sight. As time passes, the photosphere recedes, and we see some interaction directly along the line-of-sight giving NW P-Cygni along with IW due to the interaction of ejecta-disk from other regions.}}
	\label{fig:cartoon}
\end{figure}

Figure~\ref{fig:cartoon} describes a physical scenario for SN~2021foa, although not unique \citep{smith2017_interacting}, which explains all the observable of SN~2021foa through various stages of its evolution. In this scenario, the ejecta of the SN interacts with a disk-like CSM structure. The viewing angle of the observer is located at an angle from the horizontal plane of the disk. 

Figure~\ref{fig:cartoon}(a) describes a scenario where the photosphere lies in the pre-shock CSM surrounding the interaction region. The early profiles $\leq$ $-$4 d are characterised by a narrow line emission. These early profiles have broad wings that follow a symmetric Lorentzian shape, which is most likely due to incoherent electron scattering of narrow emission from pre-shock gas \citep{Chugai2001, smith2017_interacting}. At these particular epochs, pre-SN mass-loss speeds are mostly determined from the width of the line profiles while the line wings are caused by thermal broadening and not reflecting the expansion velocities. So, as shown in Panel (a) of our cartoon diagram, at $\leq$ $-$4 d the continuum photosphere is in the CSM ahead of the shock, which will hide the emission from the SN ejecta and the CDS. However, we mentioned in Section~\ref{halphadecomp} that we do see ejecta signatures as well, in the form of broad absorption at FWHM $\sim$ 3500 km sec$^{-1}$. This indicates the geometry of the CSM is asymmetric and the viewing angle is not inclined to a plane but rather at an angle. This is also the phase where we see Lorentzian profiles in the spectral evolution along with absorption components, which further indicates an asymmetric CSM configuration. So, up to $-$3.2 d, a disk-like CSM configuration with the observer placed at a viewing angle which also justifies our case where we see the freely expanding ejecta, narrow emission lines from pre-shocked material and intermediate width components due to interaction of the SN ejecta with the disk-like CSM.

Figure~\ref{fig:cartoon}(b) marks the onset of narrow P-Cygni features along with IW Lorentzian profiles as seen in the spectral profiles. As the photosphere recedes, the interaction zone of the photosphere and the disk comes along our line of sight, giving rise to a narrow P-Cygni profile. This phase also marks the onset of the phase where we see strong IW/BW interaction signatures with FWHM $\sim$ 3000 -- 5000 km s$^{-1}$, which most likely arises from the combination of both post-shock gas in the CDS and freely expanding ejecta. The absorption component in H$\alpha$ vanishes at $\sim$ 26 d. The transition occurs for a phase where the ejecta becomes optically thin and the essential properties of this transition are reproduced in the radiative transfer simulation of \cite{Dessart2015}. These simulations also indicate that the early time data is driven by electron scattering in the CSM and the late time data is driven by emission from post-shock gas at later times. An interesting aspect of the broad component that we see is given by the fact that even though we see it in the absorption profiles of spectral evolution, we cannot separate the contribution of CDS or ejecta in the emission profiles. Seeing strong emission lines for ejecta in SNe~IIn are not common as continuum optical depths often hide emission from underlying ejecta or the IW component due to interaction with a dense CSM dominates the line profiles. Previous examples of SN~2010jl \citep{2012Smith,Fransson2014} and SN~2006tf \citep{2008Smith} have only showed IW H$\alpha$ emission and weak He {\sc i} absorption at fast blueshifts. We also see the case of SN~2017hcc where we see freely expanding SN ejecta in emission which mostly arises due to viewing angle from polar regions \citep{2017hcc_2020Nathan}. Thus, similar explosion/CSM geometry have been proposed for SNe~2010jl, 2006tf, 2015da, and 2017hcc \citep{2012Smith,2008Smith,2017hcc_2020Nathan} but with different viewing angles. \cite{Suzuki2019} have generated the lightcurves of SNe~IIn using a grid of CSM masses, different viewing angles and assuming a disk-like geometry. We see that disk-like geometry fairly reproduces the double-peaked lightcurve (seen for SN~2021foa) with a CSM mass of 10 M$_{\odot}$ and a viewing angle between 30 -- 60 degrees. 

The H$\alpha$ profile of SN~2021foa at late phases shows a deficit in the flux in the red side of the wing and a systematic blueshift in the line centers from $-27$ km s$^{-1}$ to 377 km s$^{-1}$. This deficit is visible as an overall blueshift of the line after day 40 and this increases with time.

Blueshifts can arise due to various reasons in a SN~IIn/Ibn. Radiative acceleration, asymmetric CSM or lop-sided ejecta, and obscuration of the receding material by the continuum photosphere can all give rise to 
asymmetric CSM. Blueshifted line profiles that become more prominent with time can also arise from the formation or re-growth of dust grains in either the post-shock zone of the CDS or in the unshocked SN ejecta. Both ejecta and CSM components have a differing relative contribution to the total dust at different times in the evolution of SNe~IIn/Ibn. The optical to IR analysis of SN 2010jl at early and late times demonstrated dust formation in the post-shock CDS and continual grain growth in the SN ejecta \citep{2012Smith, 2016MNRAS.456.2622J,Chugai2018}.

To investigate the case of SN~2021foa, we see that the blueshift increases with time. In the case of a blueshift caused by the radiative acceleration of the CSM, the blueshift should decrease as the luminosity drops, and we expect no significant wavelength dependence for this case. Additionally, in the case of a blueshift caused by the radiative acceleration scenario, the original narrow line photon source should be blueshifted as well which is not true for SN~2021foa ~\citep{Dessart2015}. For obscuration by the continuum photosphere, the blueshift should be strongest in early times and decrease later on as continuum optical depth drops. For a lopsided CSM structure as well, blueshift should be present from early times which is not consistent with observations of SN~2021foa.  \cite{Farias2024} in their observations ruled out the scenario for the new dust formation in the CDS because for that case, the blueshift of the
emission line profiles and red-blue asymmetry exhibits a measurable wavelength dependence, with bluer emission lines
exhibiting bluer redshifts than the redder wavelengths. This is not observed both in our and their spectral evolution. Additionally, ejecta dust formation occurs typically $\sim$ 1 yr post explosion reaching temperatures of 1500-2000 K which is also not seen in our spectral evolution. So, we propose that the arising blueshift is most-likely arising due to dust or due to occultation by the photosphere. This dust is most-likely pre-existing at some distance and not newly formed. We do see a late-time flattening in the optical light curves of SN~2021foa. However, we should remark that other signatures of dust formation which involves increase in NIR flux is not seen in the case of SN~2021foa due to lack of observations and also because it went behind the sun. Nonetheless, dust formation is very common in SNe~IIn/Ibn 2006tf \citep{2008Smith}, 2006jc \citep{Smith2008_2006jc}, 2010jl \citep{Maeda2013_2010jl} and could be a probable case for SN~2021foa.

\cite{Thomas2022_2014C} investigated the case of SN~2014C, a SN~Ib, which showed narrow emission lines of Hydrogen about 127 d post-explosion. They proposed a torus-like geometry which is also consistent with \cite{Suzuki2019}. Our proposed disk-like scenario also supports a case similar to SN~2014C as well. While some asymmetries may be produced by single stars, they proposed a binary evolution \citep{Sun2020} that led to a common envelope phase that was responsible for the formation of the Hydrogen-rich CSM. The likely distribution of matter in a system that has undergone binary evolution with the ejection of a common envelope is that the Hydrogen-rich envelope material substantially will be confined to the equatorial plane with a He-rich star as the progenitor. SN~2021foa could also be in such a scenario where the progenitor would be a star that stripped both H and He in the CSM and blew a fast wind that interacted with the main-sequence secondary that facilitated the past expulsion of the progenitor’s outer H, and He layers in a common envelope interaction. The secondary blows a slower Hydrogen-rich wind that would be entrained by the fast Hydrogen + Helium wind of the primary, thus forming a torous/disk-like structure in the equatorial plane.

An interesting point and open question to the behaviour of the SN is if the intermediate/narrow lines are from the CSM, what is going on behind the SNe~IIn to SNe~Ibn transition is indeed hard to explain. If this would simply reflect the CSM composition, it requires that the fraction of He in the mass-loss wind/ejecta is decreasing toward the SN, which is in accordance with the standard picture. Basically, it could be just ionisation effect and composition effect. Further modelling is required to say more about this. Even though the claims of a single star exploding while transitioning from a LBV to WR phase is explained in past studies, the He envelope in the CSM could also be due to a star in binary composition giving rise to narrow emission lines. A detailed theoretical interpretation is essential to describe the plausible scenario giving rise to these kind of SNe.

\section{Summary}
\label{summary}
\begin{enumerate}
    \item SN~2021foa is a unique member in the transitional SN~IIn to SN~Ibn subclass with H$\alpha$ to He {\sc i} line ratios intermediate between those of SNe~IIn and SNe~Ibn around maximum light. 
    \item Early time spectral comparison shows that SN~2021foa is similar to SNe in the Type~IIn class while the mid and late-time spectral evolution indicates its similarity with SNe~Ibn. At $\sim$ 7 -- 14 d, we also see that the He {\sc i} line luminosity is of comparable strength to the H$\alpha$ luminosity, justifying the transitional nature of the SN. 
    \item SN~2021foa shows a dramatic lightcurve evolution with a precursor activity (M$_{\rm v}$ $\sim$ $-$14 mag) and reaching a secondary maximum at $-$17.8 mag, a shoulder at about $\sim$ 17 d and a late-time flattening. The SN lies in the middle of the luminosity distribution of SNe~IIn and SNe~Ibn. Even though the light curve shows a short-duration precursor, the colors are more similar to SNe without precursor activity.
    \item The H$\alpha$ evolution is complex having a NW (500 -- 1000 km s$^{-1}$), IW component in emission (2000 -- 4000 km s$^{-1}$) and a BW component in absorption at $\sim$ 3500 km s$^{-1}$. We see a narrow P-Cygni profile in the H$\alpha$ line arising after the line is seen in emission, which could be due to either precursor activity, viewing angle and geometrical effects of the CSM, or interaction with another shell of CSM. 
    \item We propose that the shoulder in the lightcurve arises due to the geometry of the CSM and the late-time flattening is most-likely arising from the dust or occulation \citep{Farias2024} as seen from the systematic blueshift in the H$\alpha$ profile at late phases. The dust also is pre-existing at some distance.
    \item Hydrodynamical lightcurve modelling using \texttt{STELLA} indicates that the lightcurve until 80 d can be reproduced by a two-component CSM with $\rho_\mathrm{CSM}\propto r^{-2}$ -- $\rho_\mathrm{CSM}\propto r^{-5}$ starting from $3\times 10^{15}~\mathrm{cm}$, with a CSM mass of 0.18 M$_{\odot}$ and mass-loss rate of 10$^{-1}$ M$_{\odot}$. If the extended CSM with $\rho_\mathrm{CSM}\propto r^{-2}$ is attached above $1.5\times 10^{15}~\mathrm{cm}$, then we can also reproduce the late time flattening of the lightcurve.   
    \item Combining spectral and lightcurve modelling, the mass-loss rates would have increased from $10^{-3}~\mathrm{M_\odot~yr^{-1}}$ to $10^{-1}~\mathrm{M_\odot~yr^{-1}}$ from 5~years to 1~year before the explosion, and also varied between 0.05 -- 0.5 $~\mathrm{M_\odot~yr^{-1}}$ with mass expelled at both 0.5~year to 1~year before the explosion assuming a wind velocity of 1000 km sec$^{-1}$. This changing mass-loss rate is most probably an indicator of the precursor activity and also explains the shoulder appearing in the light curve of SN~2021foa. 
    \item We see that a disk-like geometry like SN~2009ip best reproduces our observed profiles but the composition of CSM is most likely mixed composition with both H$\alpha$ and He {\sc i}.
    \item The composition of the CSM, the line ratios, spectral and temporal evolution, mass-loss rates all points towards a scenario where  SN~2021foa most-likely arose from the explosion of an LBV star which was transitioning to its WR phase. However, we cannot completely rule out the possibility of a binary scenario as proposed for the case of SN~2014C \citep{Thomas2022_2014C}.
\end{enumerate}


\section*{Acknowledgements}
We thank the anonymous referee for their valuable comments/suggestions which have significantly helped in improving the manuscript.
This work makes used of data from the Las Cumbres Observatory global telescope network. The LCO group is supported by NSF grants AST-1911151 and AST-1911225. C. Pellegrino acknowledges support from ADAP program grant No. 80NSSC24K0180 and from NSF grant AST-2206657. 
ND and KM acknowledge the support from BRICS grant DST/ICD/BRICS/Call-5/CoNMuTraMO/2023 (G) funded by the Department of Science and Technology (DST), India.
MS acknowledges the financial support provided under the National Post Doctoral Fellowship (N-PDF; File Number: PDF/2023/002244) by the Science \& Engineering Research Board (SERB), Anusandhan National Research Foundation (ANRF), Government of India.
K. Maeda acknowledges the JSPS Kakenhi grant (JP24H01810, 24KK0070). 
S. Schulze is partially supported by LBNL Subcontract 7707915.
We thank the staff of the GMRT that made these observations possible. GMRT is run by the National Centre for Radio Astrophysics of the Tata Institute of Fundamental Research. The National Radio Astronomy Observatory is a facility of the National Science Foundation operated under cooperative agreement by Associated Universities, Inc.

\section*{Data Availability}
The data presented in this paper will be provided upon request. All the spectra will be made publicly available in WiseRep and Zenodo.



\bibliographystyle{mnras}
\bibliography{example} 

\begin{thebibliography}{}
\makeatletter
\relax
\def\mn@urlcharsother{\let\do\@makeother \do\$\do\&\do\#\do\^\do\_\do\%\do\~}
\def\mn@doi{\begingroup\mn@urlcharsother \@ifnextchar [ {\mn@doi@}
  {\mn@doi@[]}}
\def\mn@doi@[#1]#2{\def\@tempa{#1}\ifx\@tempa\@empty \href
  {http://dx.doi.org/#2} {doi:#2}\else \href {http://dx.doi.org/#2} {#1}\fi
  \endgroup}
\def\mn@eprint#1#2{\mn@eprint@#1:#2::\@nil}
\def\mn@eprint@arXiv#1{\href {http://arxiv.org/abs/#1} {{\tt arXiv:#1}}}
\def\mn@eprint@dblp#1{\href {http://dblp.uni-trier.de/rec/bibtex/#1.xml}
  {dblp:#1}}
\def\mn@eprint@#1:#2:#3:#4\@nil{\def\@tempa {#1}\def\@tempb {#2}\def\@tempc
  {#3}\ifx \@tempc \@empty \let \@tempc \@tempb \let \@tempb \@tempa \fi \ifx
  \@tempb \@empty \def\@tempb {arXiv}\fi \@ifundefined
  {mn@eprint@\@tempb}{\@tempb:\@tempc}{\expandafter \expandafter \csname
  mn@eprint@\@tempb\endcsname \expandafter{\@tempc}}}

\bibitem[\protect\citeauthoryear{{Agnoletto} et~al.,}{{Agnoletto}
  et~al.}{2009}]{Agnoletto2009}
{Agnoletto} I.,  et~al., 2009, \mn@doi [\apj] {10.1088/0004-637X/691/2/1348},
  \href {https://ui.adsabs.harvard.edu/abs/2009ApJ...691.1348A} {691, 1348}

\bibitem[\protect\citeauthoryear{{Akitaya} et~al.,}{{Akitaya}
  et~al.}{2014}]{akitaya2014}
{Akitaya} H.,  et~al., 2014, in {Ramsay} S.~K.,  {McLean} I.~S.,   {Takami} H.,
   eds,  Society of Photo-Optical Instrumentation Engineers (SPIE) Conference
  Series Vol. 9147, Ground-based and Airborne Instrumentation for Astronomy V.
  p. 91474O, \mn@doi{10.1117/12.2054577}

\bibitem[\protect\citeauthoryear{{Angus}}{{Angus}}{2021}]{classificationspec2021}
{Angus} C.,  2021, Transient Name Server Classification Report, \href
  {https://ui.adsabs.harvard.edu/abs/2021TNSCR1133....1A} {2021-1133, 1}

\bibitem[\protect\citeauthoryear{{Anupama}, {Sahu}, {Gurugubelli}, {Prabhu},
  {Tominaga}, {Tanaka}  \& {Nomoto}}{{Anupama} et~al.}{2009}]{Anupama2009}
{Anupama} G.~C.,  {Sahu} D.~K.,  {Gurugubelli} U.~K.,  {Prabhu} T.~P.,
  {Tominaga} N.,  {Tanaka} M.,   {Nomoto} K.,  2009, \mn@doi [\mnras]
  {10.1111/j.1365-2966.2008.14129.x}, \href
  {https://ui.adsabs.harvard.edu/abs/2009MNRAS.392..894A} {392, 894}

\bibitem[\protect\citeauthoryear{{Barden}}{{Barden}}{1994}]{1994ASPC...55..130B}
{Barden} S.~C.,  1994, in {Pyper} D.~M.,  {Angione} R.~J.,  eds,  Astronomical
  Society of the Pacific Conference Series Vol. 55, Optical Astronomy from the
  Earth and Moon. pp 130--138

\bibitem[\protect\citeauthoryear{{Benetti} et~al.,}{{Benetti}
  et~al.}{2016}]{1996al_benetti_2016}
{Benetti} S.,  et~al., 2016, \mn@doi [\mnras] {10.1093/mnras/stv2811}, \href
  {https://ui.adsabs.harvard.edu/abs/2016MNRAS.456.3296B} {456, 3296}

\bibitem[\protect\citeauthoryear{{Blinnikov}, {Eastman}, {Bartunov},
  {Popolitov}  \& {Woosley}}{{Blinnikov} et~al.}{1998}]{1998ApJ...496..454B}
{Blinnikov} S.~I.,  {Eastman} R.,  {Bartunov} O.~S.,  {Popolitov} V.~A.,
  {Woosley} S.~E.,  1998, \mn@doi [\apj] {10.1086/305375}, \href
  {https://ui.adsabs.harvard.edu/abs/1998ApJ...496..454B} {496, 454}

\bibitem[\protect\citeauthoryear{{Blinnikov}, {Lundqvist}, {Bartunov}, {Nomoto}
   \& {Iwamoto}}{{Blinnikov} et~al.}{2000}]{2000ApJ...532.1132B}
{Blinnikov} S.,  {Lundqvist} P.,  {Bartunov} O.,  {Nomoto} K.,   {Iwamoto} K.,
  2000, \mn@doi [\apj] {10.1086/308588}, \href
  {https://ui.adsabs.harvard.edu/abs/2000ApJ...532.1132B} {532, 1132}

\bibitem[\protect\citeauthoryear{{Blinnikov}, {R{\"o}pke}, {Sorokina},
  {Gieseler}, {Reinecke}, {Travaglio}, {Hillebrandt}  \&
  {Stritzinger}}{{Blinnikov} et~al.}{2006}]{2006A&A...453..229B}
{Blinnikov} S.~I.,  {R{\"o}pke} F.~K.,  {Sorokina} E.~I.,  {Gieseler} M.,
  {Reinecke} M.,  {Travaglio} C.,  {Hillebrandt} W.,   {Stritzinger} M.,  2006,
  \mn@doi [\aap] {10.1051/0004-6361:20054594}, \href
  {https://ui.adsabs.harvard.edu/abs/2006A&A...453..229B} {453, 229}

\bibitem[\protect\citeauthoryear{{Brennan} et~al.,}{{Brennan}
  et~al.}{2022}]{Brennan2022}
{Brennan} S.~J.,  et~al., 2022, \mn@doi [\mnras] {10.1093/mnras/stac1243},
  \href {https://ui.adsabs.harvard.edu/abs/2022MNRAS.513.5642B} {513, 5642}

\bibitem[\protect\citeauthoryear{{Chandra}, {Chevalier}, {Chugai}, {Fransson},
  {Irwin}, {Soderberg}, {Chakraborti}  \& {Immler}}{{Chandra}
  et~al.}{2012}]{Chandra2012}
{Chandra} P.,  {Chevalier} R.~A.,  {Chugai} N.,  {Fransson} C.,  {Irwin} C.~M.,
   {Soderberg} A.~M.,  {Chakraborti} S.,   {Immler} S.,  2012, \mn@doi [\apj]
  {10.1088/0004-637X/755/2/110}, \href
  {https://ui.adsabs.harvard.edu/abs/2012ApJ...755..110C} {755, 110}

\bibitem[\protect\citeauthoryear{{Chandra}, {Chevalier}, {Chugai}, {Fransson}
  \& {Soderberg}}{{Chandra} et~al.}{2015}]{Chandra2015}
{Chandra} P.,  {Chevalier} R.~A.,  {Chugai} N.,  {Fransson} C.,   {Soderberg}
  A.~M.,  2015, \mn@doi [\apj] {10.1088/0004-637X/810/1/32}, \href
  {https://ui.adsabs.harvard.edu/abs/2015ApJ...810...32C} {810, 32}

\bibitem[\protect\citeauthoryear{{Chugai}}{{Chugai}}{2001}]{Chugai2001}
{Chugai} N.~N.,  2001, \mn@doi [\mnras] {10.1111/j.1365-2966.2001.04717.x},
  \href {https://ui.adsabs.harvard.edu/abs/2001MNRAS.326.1448C} {326, 1448}

\bibitem[\protect\citeauthoryear{{Chugai}}{{Chugai}}{2018}]{Chugai2018}
{Chugai} N.~N.,  2018, \mn@doi [\mnras] {10.1093/mnras/sty2386}, \href
  {https://ui.adsabs.harvard.edu/abs/2018MNRAS.481.3643C} {481, 3643}

\bibitem[\protect\citeauthoryear{{Chugai} \& {Danziger}}{{Chugai} \&
  {Danziger}}{1994}]{chugai_danziger_1994}
{Chugai} N.~N.,  {Danziger} I.~J.,  1994, \mn@doi [\mnras]
  {10.1093/mnras/268.1.173}, \href
  {https://ui.adsabs.harvard.edu/abs/1994MNRAS.268..173C} {268, 173}

\bibitem[\protect\citeauthoryear{{Chugai} et~al.,}{{Chugai}
  et~al.}{2004}]{Chugai_1994W}
{Chugai} N.~N.,  et~al., 2004, \mn@doi [\mnras]
  {10.1111/j.1365-2966.2004.08011.x}, \href
  {https://ui.adsabs.harvard.edu/abs/2004MNRAS.352.1213C} {352, 1213}

\bibitem[\protect\citeauthoryear{{Davis} et~al.,}{{Davis}
  et~al.}{2023}]{Davis2023}
{Davis} K.~W.,  et~al., 2023, \mn@doi [\mnras] {10.1093/mnras/stad1433}, \href
  {https://ui.adsabs.harvard.edu/abs/2023MNRAS.523.2530D} {523, 2530}

\bibitem[\protect\citeauthoryear{{Dessart}, {Audit}  \& {Hillier}}{{Dessart}
  et~al.}{2015}]{Dessart2015}
{Dessart} L.,  {Audit} E.,   {Hillier} D.~J.,  2015, \mn@doi [\mnras]
  {10.1093/mnras/stv609}, \href
  {https://ui.adsabs.harvard.edu/abs/2015MNRAS.449.4304D} {449, 4304}

\bibitem[\protect\citeauthoryear{{Dessart}, {Hillier}  \&
  {Kuncarayakti}}{{Dessart} et~al.}{2022}]{Dessart2022}
{Dessart} L.,  {Hillier} D.~J.,   {Kuncarayakti} H.,  2022, \mn@doi [\aap]
  {10.1051/0004-6361/202142436}, \href
  {https://ui.adsabs.harvard.edu/abs/2022A&A...658A.130D} {658, A130}

\bibitem[\protect\citeauthoryear{{Dickinson}, {Smith}, {Andrews}, {Milne},
  {Kilpatrick}  \& {Milisavljevic}}{{Dickinson} et~al.}{2024}]{2024Dickinson}
{Dickinson} D.,  {Smith} N.,  {Andrews} J.~E.,  {Milne} P.,  {Kilpatrick}
  C.~D.,   {Milisavljevic} D.,  2024, \mn@doi [\mnras]
  {10.1093/mnras/stad3631}, \href
  {https://ui.adsabs.harvard.edu/abs/2024MNRAS.527.7767D} {527, 7767}

\bibitem[\protect\citeauthoryear{{Dukiya} et~al.,}{{Dukiya}
  et~al.}{2024}]{14il}
{Dukiya} N.,  et~al., 2024, \mn@doi [arXiv e-prints]
  {10.48550/arXiv.2404.04235}, \href
  {https://ui.adsabs.harvard.edu/abs/2024arXiv240404235D} {p. arXiv:2404.04235}

\bibitem[\protect\citeauthoryear{{Elias-Rosa} et~al.,}{{Elias-Rosa}
  et~al.}{2016}]{Nancy2016}
{Elias-Rosa} N.,  et~al., 2016, \mn@doi [\mnras] {10.1093/mnras/stw2253}, \href
  {https://ui.adsabs.harvard.edu/abs/2016MNRAS.463.3894E} {463, 3894}

\bibitem[\protect\citeauthoryear{{Falk}, {Lattimer}  \& {Margolis}}{{Falk}
  et~al.}{1977}]{falk1977}
{Falk} S.~W.,  {Lattimer} J.~M.,   {Margolis} S.~H.,  1977, \mn@doi [\nat]
  {10.1038/270700a0}, \href
  {https://ui.adsabs.harvard.edu/abs/1977Natur.270..700F} {270, 700}

\bibitem[\protect\citeauthoryear{{Farias}}{{Farias}}{2024}]{Farias2024}
{Farias} D. e.~a.,  2024, \mn@doi [\apj] {2409.01359}, \href
  {https://arxiv.org/pdf/2409.01359} {000, }

\bibitem[\protect\citeauthoryear{{Fassia} et~al.,}{{Fassia}
  et~al.}{2001}]{Fassia2001}
{Fassia} A.,  et~al., 2001, \mn@doi [\mnras]
  {10.1046/j.1365-8711.2001.04282.x}, \href
  {https://ui.adsabs.harvard.edu/abs/2001MNRAS.325..907F} {325, 907}

\bibitem[\protect\citeauthoryear{{Filippenko}}{{Filippenko}}{1997}]{Filippenko1997}
{Filippenko} A.~V.,  1997, in {Ruiz-Lapuente} P.,  {Canal} R.,   {Isern} J.,
  eds,  NATO Advanced Study Institute (ASI) Series C Vol. 486, Thermonuclear
  Supernovae. p.~1, \mn@doi{10.1007/978-94-011-5710-0_1}

\bibitem[\protect\citeauthoryear{{Foley}, {Smith}, {Ganeshalingam}, {Li},
  {Chornock}  \& {Filippenko}}{{Foley} et~al.}{2007}]{Foley2007}
{Foley} R.~J.,  {Smith} N.,  {Ganeshalingam} M.,  {Li} W.,  {Chornock} R.,
  {Filippenko} A.~V.,  2007, \mn@doi [\apjl] {10.1086/513145}, \href
  {https://ui.adsabs.harvard.edu/abs/2007ApJ...657L.105F} {657, L105}

\bibitem[\protect\citeauthoryear{{Fransson} et~al.,}{{Fransson}
  et~al.}{2014}]{Fransson2014}
{Fransson} C.,  et~al., 2014, \mn@doi [\apj] {10.1088/0004-637X/797/2/118},
  \href {https://ui.adsabs.harvard.edu/abs/2014ApJ...797..118F} {797, 118}

\bibitem[\protect\citeauthoryear{{Fraser}}{{Fraser}}{2020}]{Fraser2020}
{Fraser} M.,  2020, \mn@doi [Royal Society Open Science] {10.1098/rsos.200467},
  \href {https://ui.adsabs.harvard.edu/abs/2020RSOS....700467F} {7, 200467}

\bibitem[\protect\citeauthoryear{{Fraser} et~al.,}{{Fraser}
  et~al.}{2021}]{Fraser_21csp_2021}
{Fraser} M.,  et~al., 2021, \mn@doi [arXiv e-prints]
  {10.48550/arXiv.2108.07278}, \href
  {https://ui.adsabs.harvard.edu/abs/2021arXiv210807278F} {p. arXiv:2108.07278}

\bibitem[\protect\citeauthoryear{{Fuller}}{{Fuller}}{2017}]{Fuller2017}
{Fuller} J.,  2017, \mn@doi [\mnras] {10.1093/mnras/stx1314}, \href
  {https://ui.adsabs.harvard.edu/abs/2017MNRAS.470.1642F} {470, 1642}

\bibitem[\protect\citeauthoryear{{Fuller} \& {Ro}}{{Fuller} \&
  {Ro}}{2018}]{FullerandRo2018}
{Fuller} J.,  {Ro} S.,  2018, \mn@doi [\mnras] {10.1093/mnras/sty369}, \href
  {https://ui.adsabs.harvard.edu/abs/2018MNRAS.476.1853F} {476, 1853}

\bibitem[\protect\citeauthoryear{{Gal-Yam} \& {Leonard}}{{Gal-Yam} \&
  {Leonard}}{2009}]{2005gl_GalYam2009}
{Gal-Yam} A.,  {Leonard} D.~C.,  2009, \mn@doi [\nat] {10.1038/nature07934},
  \href {https://ui.adsabs.harvard.edu/abs/2009Natur.458..865G} {458, 865}

\bibitem[\protect\citeauthoryear{{Gal-Yam} et~al.,}{{Gal-Yam}
  et~al.}{2022}]{Gal-Yam2022}
{Gal-Yam} A.,  et~al., 2022, \mn@doi [\nat] {10.1038/s41586-021-04155-1}, \href
  {https://ui.adsabs.harvard.edu/abs/2022Natur.601..201G} {601, 201}

\bibitem[\protect\citeauthoryear{{Gangopadhyay} et~al.,}{{Gangopadhyay}
  et~al.}{2020}]{Gangopadhyay2020}
{Gangopadhyay} A.,  et~al., 2020, \mn@doi [\apj] {10.3847/1538-4357/ab6328},
  \href {https://ui.adsabs.harvard.edu/abs/2020ApJ...889..170G} {889, 170}

\bibitem[\protect\citeauthoryear{{Gangopadhyay} et~al.,}{{Gangopadhyay}
  et~al.}{2022}]{Gangopadhyay2022}
{Gangopadhyay} A.,  et~al., 2022, \mn@doi [\apj] {10.3847/1538-4357/ac6187},
  \href {https://ui.adsabs.harvard.edu/abs/2022ApJ...930..127G} {930, 127}

\bibitem[\protect\citeauthoryear{{Gangopadhyay} et~al.,}{{Gangopadhyay}
  et~al.}{2023}]{Gangopadhyay2023}
{Gangopadhyay} A.,  et~al., 2023, \mn@doi [\apj] {10.3847/1538-4357/acfa94},
  \href {https://ui.adsabs.harvard.edu/abs/2023ApJ...957..100G} {957, 100}

\bibitem[\protect\citeauthoryear{{Guevel}, {Hosseinzadeh}, {Bostroem}  \&
  {Burke}}{{Guevel} et~al.}{2021}]{PyZOGY}
{Guevel} D.,  {Hosseinzadeh} G.,  {Bostroem} A.,   {Burke} C.~J.,  2021,
  {dguevel/PyZOGY: v0.0.2}, \mn@doi{10.5281/zenodo.4570234}

\bibitem[\protect\citeauthoryear{{Gunn} et~al.,}{{Gunn}
  et~al.}{1998}]{1998AJ....116.3040G}
{Gunn} J.~E.,  et~al., 1998, \mn@doi [\aj] {10.1086/300645}, \href
  {https://ui.adsabs.harvard.edu/abs/1998AJ....116.3040G} {116, 3040}

\bibitem[\protect\citeauthoryear{{Heger}, {Fryer}, {Woosley}, {Langer}  \&
  {Hartmann}}{{Heger} et~al.}{2003}]{Heger2003}
{Heger} A.,  {Fryer} C.~L.,  {Woosley} S.~E.,  {Langer} N.,   {Hartmann} D.~H.,
   2003, \mn@doi [\apj] {10.1086/375341}, \href
  {https://ui.adsabs.harvard.edu/abs/2003ApJ...591..288H} {591, 288}

\bibitem[\protect\citeauthoryear{{Hiramatsu} et~al.,}{{Hiramatsu}
  et~al.}{2024}]{Hiramatsu2024}
{Hiramatsu} D.,  et~al., 2024, \mn@doi [\apj] {10.3847/1538-4357/ad2854}, \href
  {https://ui.adsabs.harvard.edu/abs/2024ApJ...964..181H} {964, 181}

\bibitem[\protect\citeauthoryear{{Hosseinzadeh} et~al.,}{{Hosseinzadeh}
  et~al.}{2017}]{Hosseinzadeh2017}
{Hosseinzadeh} G.,  et~al., 2017, \mn@doi [\apj] {10.3847/1538-4357/836/2/158},
  \href {https://ui.adsabs.harvard.edu/abs/2017ApJ...836..158H} {836, 158}

\bibitem[\protect\citeauthoryear{{Irani} et~al.,}{{Irani}
  et~al.}{2023}]{Irani2023}
{Irani} I.,  et~al., 2023, \mn@doi [arXiv e-prints]
  {10.48550/arXiv.2310.16885}, \href
  {https://ui.adsabs.harvard.edu/abs/2023arXiv231016885I} {p. arXiv:2310.16885}

\bibitem[\protect\citeauthoryear{{Ishii}, {Takei}, {Tsuna}, {Shigeyama}  \&
  {Takahashi}}{{Ishii} et~al.}{2024}]{CSM_structure_Ishii_24}
{Ishii} A.~T.,  {Takei} Y.,  {Tsuna} D.,  {Shigeyama} T.,   {Takahashi} K.,
  2024, \mn@doi [\apj] {10.3847/1538-4357/ad072b}, \href
  {https://ui.adsabs.harvard.edu/abs/2024ApJ...961...47I} {961, 47}

\bibitem[\protect\citeauthoryear{{Itagaki} et~al.,}{{Itagaki}
  et~al.}{2006}]{Itagaki2006}
{Itagaki} K.,  et~al., 2006, \iaucirc, \href
  {https://ui.adsabs.harvard.edu/abs/2006IAUC.8762....1I} {8762, 1}

\bibitem[\protect\citeauthoryear{{Jencson}, {Prieto}, {Kochanek}, {Shappee},
  {Stanek}  \& {Pogge}}{{Jencson} et~al.}{2016}]{2016MNRAS.456.2622J}
{Jencson} J.~E.,  {Prieto} J.~L.,  {Kochanek} C.~S.,  {Shappee} B.~J.,
  {Stanek} K.~Z.,   {Pogge} R.~W.,  2016, \mn@doi [\mnras]
  {10.1093/mnras/stv2795}, \href
  {http://adsabs.harvard.edu/abs/2016MNRAS.456.2622J} {456, 2622}

\bibitem[\protect\citeauthoryear{{Justham}, {Podsiadlowski}  \&
  {Vink}}{{Justham} et~al.}{2014}]{Stephan2014}
{Justham} S.,  {Podsiadlowski} P.,   {Vink} J.~S.,  2014, \mn@doi [\apj]
  {10.1088/0004-637X/796/2/121}, \href
  {https://ui.adsabs.harvard.edu/abs/2014ApJ...796..121J} {796, 121}

\bibitem[\protect\citeauthoryear{{Kankare} et~al.,}{{Kankare}
  et~al.}{2012}]{2012MNRAS.424..855K}
{Kankare} E.,  et~al., 2012, \mn@doi [\mnras]
  {10.1111/j.1365-2966.2012.21224.x}, \href
  {http://adsabs.harvard.edu/abs/2012MNRAS.424..855K} {424, 855}

\bibitem[\protect\citeauthoryear{{Kawabata} et~al.,}{{Kawabata}
  et~al.}{2008}]{2008SPIE.7014E..4LK}
{Kawabata} K.~S.,  et~al., 2008, in {McLean} I.~S.,  {Casali} M.~M.,  eds,
  Society of Photo-Optical Instrumentation Engineers (SPIE) Conference Series
  Vol. 7014, Ground-based and Airborne Instrumentation for Astronomy II. p.
  70144L, \mn@doi{10.1117/12.788569}

\bibitem[\protect\citeauthoryear{{Kiewe} et~al.,}{{Kiewe}
  et~al.}{2012}]{Kiewe2012}
{Kiewe} M.,  et~al., 2012, \mn@doi [\apj] {10.1088/0004-637X/744/1/10}, \href
  {https://ui.adsabs.harvard.edu/abs/2012ApJ...744...10K} {744, 10}

\bibitem[\protect\citeauthoryear{{Kochanek}, {Szczygiel}  \&
  {Stanek}}{{Kochanek} et~al.}{2011}]{Kochanek2011}
{Kochanek} C.~S.,  {Szczygiel} D.~M.,   {Stanek} K.~Z.,  2011, \mn@doi [\apj]
  {10.1088/0004-637X/737/2/76}, \href
  {https://ui.adsabs.harvard.edu/abs/2011ApJ...737...76K} {737, 76}

\bibitem[\protect\citeauthoryear{{Kool} et~al.,}{{Kool}
  et~al.}{2021}]{Kool2021}
{Kool} E.~C.,  et~al., 2021, \mn@doi [\aap] {10.1051/0004-6361/202039137},
  \href {https://ui.adsabs.harvard.edu/abs/2021A&A...652A.136K} {652, A136}

\bibitem[\protect\citeauthoryear{{Kurf{\"u}rst} \&
  {Krti{\v{c}}ka}}{{Kurf{\"u}rst} \&
  {Krti{\v{c}}ka}}{2019}]{interaction_kurfurst_2019}
{Kurf{\"u}rst} P.,  {Krti{\v{c}}ka} J.,  2019, \mn@doi [\aap]
  {10.1051/0004-6361/201833429}, \href
  {https://ui.adsabs.harvard.edu/abs/2019A&A...625A..24K} {625, A24}

\bibitem[\protect\citeauthoryear{{Landolt}}{{Landolt}}{1992}]{landolt1992}
{Landolt} A.~U.,  1992, \mn@doi [\aj] {10.1086/116243}, \href
  {https://ui.adsabs.harvard.edu/abs/1992AJ....104..372L} {104, 372}

\bibitem[\protect\citeauthoryear{{Li} et~al.,}{{Li} et~al.}{2011}]{Li2011}
{Li} W.,  et~al., 2011, \mn@doi [\mnras] {10.1111/j.1365-2966.2011.18160.x},
  \href {https://ui.adsabs.harvard.edu/abs/2011MNRAS.412.1441L} {412, 1441}

\bibitem[\protect\citeauthoryear{{Lusk} \& {Baron}}{{Lusk} \&
  {Baron}}{2017}]{2017PASP..129d4202L}
{Lusk} J.~A.,  {Baron} E.,  2017, \mn@doi [\pasp] {10.1088/1538-3873/aa5e49},
  \href {https://ui.adsabs.harvard.edu/abs/2017PASP..129d4202L} {129, 044202}

\bibitem[\protect\citeauthoryear{{Maeda} et~al.,}{{Maeda}
  et~al.}{2013}]{Maeda2013_2010jl}
{Maeda} K.,  et~al., 2013, \mn@doi [\apj] {10.1088/0004-637X/776/1/5}, \href
  {https://ui.adsabs.harvard.edu/abs/2013ApJ...776....5M} {776, 5}

\bibitem[\protect\citeauthoryear{{Maeda} et~al.,}{{Maeda}
  et~al.}{2021}]{2020oi_Maeda2021}
{Maeda} K.,  et~al., 2021, \mn@doi [\apj] {10.3847/1538-4357/ac0dbc}, \href
  {https://ui.adsabs.harvard.edu/abs/2021ApJ...918...34M} {918, 34}

\bibitem[\protect\citeauthoryear{{Matsubayashi} et~al.,}{{Matsubayashi}
  et~al.}{2019}]{2019PASJ...71..102M}
{Matsubayashi} K.,  et~al., 2019, \mn@doi [\pasj] {10.1093/pasj/psz087}, \href
  {https://ui.adsabs.harvard.edu/abs/2019PASJ...71..102M} {71, 102}

\bibitem[\protect\citeauthoryear{{Matzner} \& {McKee}}{{Matzner} \&
  {McKee}}{1999}]{1999ApJ...510..379M}
{Matzner} C.~D.,  {McKee} C.~F.,  1999, \mn@doi [\apj] {10.1086/306571}, \href
  {https://ui.adsabs.harvard.edu/abs/1999ApJ...510..379M} {510, 379}

\bibitem[\protect\citeauthoryear{{Mauerhan} et~al.,}{{Mauerhan}
  et~al.}{2013}]{2013MNRAS.431.2599M}
{Mauerhan} J.~C.,  et~al., 2013, \mn@doi [\mnras] {10.1093/mnras/stt360}, \href
  {https://ui.adsabs.harvard.edu/abs/2013MNRAS.431.2599M} {431, 2599}

\bibitem[\protect\citeauthoryear{{McCully}, {Volgenau}, {Harbeck}, {Lister},
  {Saunders}, {Turner}, {Siiverd}  \& {Bowman}}{{McCully}
  et~al.}{2018}]{Banzai}
{McCully} C.,  {Volgenau} N.~H.,  {Harbeck} D.-R.,  {Lister} T.~A.,  {Saunders}
  E.~S.,  {Turner} M.~L.,  {Siiverd} R.~J.,   {Bowman} M.,  2018, in {Guzman}
  J.~C.,  {Ibsen} J.,  eds,  Society of Photo-Optical Instrumentation Engineers
  (SPIE) Conference Series Vol. 10707, Software and Cyberinfrastructure for
  Astronomy V. p. 107070K (\mn@eprint {arXiv} {1811.04163}),
  \mn@doi{10.1117/12.2314340}

\bibitem[\protect\citeauthoryear{{Mcley} \& {Soker}}{{Mcley} \&
  {Soker}}{2014a}]{McleySoker2014}
{Mcley} L.,  {Soker} N.,  2014a, \mn@doi [\mnras] {10.1093/mnras/stu1952},
  \href {https://ui.adsabs.harvard.edu/abs/2014MNRAS.445.2492M} {445, 2492}

\bibitem[\protect\citeauthoryear{{Mcley} \& {Soker}}{{Mcley} \&
  {Soker}}{2014b}]{Liron2014}
{Mcley} L.,  {Soker} N.,  2014b, \mn@doi [\mnras] {10.1093/mnras/stu1952},
  \href {https://ui.adsabs.harvard.edu/abs/2014MNRAS.445.2492M} {445, 2492}

\bibitem[\protect\citeauthoryear{{Nagao} et~al.,}{{Nagao}
  et~al.}{2023}]{Nagao2023}
{Nagao} T.,  et~al., 2023, \mn@doi [\aap] {10.1051/0004-6361/202346084}, \href
  {https://ui.adsabs.harvard.edu/abs/2023A&A...673A..27N} {673, A27}

\bibitem[\protect\citeauthoryear{{Nicholl}}{{Nicholl}}{2018}]{2018RNAAS...2..230N}
{Nicholl} M.,  2018, \mn@doi [Research Notes of the American Astronomical
  Society] {10.3847/2515-5172/aaf799}, \href
  {https://ui.adsabs.harvard.edu/abs/2018RNAAS...2..230N} {2, 230}

\bibitem[\protect\citeauthoryear{{Nyholm} et~al.,}{{Nyholm}
  et~al.}{2020}]{Nyholm2020}
{Nyholm} A.,  et~al., 2020, \mn@doi [\aap] {10.1051/0004-6361/201936097}, \href
  {https://ui.adsabs.harvard.edu/abs/2020A&A...637A..73N} {637, A73}

\bibitem[\protect\citeauthoryear{{Ofek} et~al.,}{{Ofek}
  et~al.}{2014}]{ofek2014}
{Ofek} E.~O.,  et~al., 2014, \mn@doi [\apj] {10.1088/0004-637X/789/2/104},
  \href {https://ui.adsabs.harvard.edu/abs/2014ApJ...789..104O} {789, 104}

\bibitem[\protect\citeauthoryear{{Pastorello} et~al.,}{{Pastorello}
  et~al.}{2007}]{Pastorello2007}
{Pastorello} A.,  et~al., 2007, \mn@doi [\nat] {10.1038/nature05825}, \href
  {https://ui.adsabs.harvard.edu/abs/2007Natur.447..829P} {447, 829}

\bibitem[\protect\citeauthoryear{{Pastorello} et~al.,}{{Pastorello}
  et~al.}{2008a}]{Pastorello2008_2006jclike}
{Pastorello} A.,  et~al., 2008a, \mn@doi [\mnras]
  {10.1111/j.1365-2966.2008.13602.x}, \href
  {https://ui.adsabs.harvard.edu/abs/2008MNRAS.389..113P} {389, 113}

\bibitem[\protect\citeauthoryear{{Pastorello} et~al.,}{{Pastorello}
  et~al.}{2008b}]{Pastorello2008}
{Pastorello} A.,  et~al., 2008b, \mn@doi [\mnras]
  {10.1111/j.1365-2966.2008.13603.x}, \href
  {https://ui.adsabs.harvard.edu/abs/2008MNRAS.389..131P} {389, 131}

\bibitem[\protect\citeauthoryear{{Pastorello} et~al.,}{{Pastorello}
  et~al.}{2013}]{pastorello2013}
{Pastorello} A.,  et~al., 2013, \mn@doi [\apj] {10.1088/0004-637X/767/1/1},
  \href {https://ui.adsabs.harvard.edu/abs/2013ApJ...767....1P} {767, 1}

\bibitem[\protect\citeauthoryear{{Pastorello} et~al.,}{{Pastorello}
  et~al.}{2015a}]{Pastorello2015}
{Pastorello} A.,  et~al., 2015a, \mn@doi [\mnras] {10.1093/mnras/stu2745},
  \href {https://ui.adsabs.harvard.edu/abs/2015MNRAS.449.1921P} {449, 1921}

\bibitem[\protect\citeauthoryear{{Pastorello} et~al.,}{{Pastorello}
  et~al.}{2015b}]{Pastorello-OGLE}
{Pastorello} A.,  et~al., 2015b, \mn@doi [\mnras] {10.1093/mnras/stu2621},
  \href {https://ui.adsabs.harvard.edu/abs/2015MNRAS.449.1941P} {449, 1941}

\bibitem[\protect\citeauthoryear{{Pastorello} et~al.,}{{Pastorello}
  et~al.}{2015c}]{Pastorello_2015_LSQ12btw_LSQ13ccw}
{Pastorello} A.,  et~al., 2015c, \mn@doi [\mnras] {10.1093/mnras/stv335}, \href
  {https://ui.adsabs.harvard.edu/abs/2015MNRAS.449.1954P} {449, 1954}

\bibitem[\protect\citeauthoryear{{Pastorello} et~al.,}{{Pastorello}
  et~al.}{2015d}]{Pastorello-15ed}
{Pastorello} A.,  et~al., 2015d, \mn@doi [\mnras] {10.1093/mnras/stv1812},
  \href {https://ui.adsabs.harvard.edu/abs/2015MNRAS.453.3649P} {453, 3649}

\bibitem[\protect\citeauthoryear{{Pastorello} et~al.,}{{Pastorello}
  et~al.}{2016}]{Pastorello2016}
{Pastorello} A.,  et~al., 2016, \mn@doi [\mnras] {10.1093/mnras/stv2634}, \href
  {https://ui.adsabs.harvard.edu/abs/2016MNRAS.456..853P} {456, 853}

\bibitem[\protect\citeauthoryear{{Pastorello} et~al.,}{{Pastorello}
  et~al.}{2019}]{Pastorello2019}
{Pastorello} A.,  et~al., 2019, \mn@doi [\aap] {10.1051/0004-6361/201935420},
  \href {https://ui.adsabs.harvard.edu/abs/2019A&A...628A..93P} {628, A93}

\bibitem[\protect\citeauthoryear{{Pellegrino} et~al.,}{{Pellegrino}
  et~al.}{2022}]{Pellegrino2022}
{Pellegrino} C.,  et~al., 2022, \mn@doi [\apj] {10.3847/1538-4357/ac8ff6},
  \href {https://ui.adsabs.harvard.edu/abs/2022ApJ...938...73P} {938, 73}

\bibitem[\protect\citeauthoryear{{Perley} et~al.,}{{Perley}
  et~al.}{2020}]{Perley2020}
{Perley} D.~A.,  et~al., 2020, \mn@doi [\apj] {10.3847/1538-4357/abbd98}, \href
  {https://ui.adsabs.harvard.edu/abs/2020ApJ...904...35P} {904, 35}

\bibitem[\protect\citeauthoryear{{Perley}, {Sollerman}, {Schulze}, {Yao},
  {Fremling}, {Gal-Yam}, {Ho}  \& {Yang}}{{Perley} et~al.}{2022}]{Perley2022}
{Perley} D.,  {Sollerman} J.,  {Schulze} S.,  {Yao} Y.,  {Fremling} U.,
  {Gal-Yam} A.,  {Ho} A.,   {Yang} Y.,  2022, in American Astronomical Society
  Meeting \#240. p. 232.08

\bibitem[\protect\citeauthoryear{{Poznanski}, {Prochaska}  \&
  {Bloom}}{{Poznanski} et~al.}{2012}]{2012MNRAS.426.1465P}
{Poznanski} D.,  {Prochaska} J.~X.,   {Bloom} J.~S.,  2012, \mn@doi [\mnras]
  {10.1111/j.1365-2966.2012.21796.x}, \href
  {https://ui.adsabs.harvard.edu/abs/2012MNRAS.426.1465P} {426, 1465}

\bibitem[\protect\citeauthoryear{{Pursiainen} et~al.,}{{Pursiainen}
  et~al.}{2023}]{Pursiainen2023emq}
{Pursiainen} M.,  et~al., 2023, \mn@doi [\apjl] {10.3847/2041-8213/ad103d},
  \href {https://ui.adsabs.harvard.edu/abs/2023ApJ...959L..10P} {959, L10}

\bibitem[\protect\citeauthoryear{{Quataert} \& {Shiode}}{{Quataert} \&
  {Shiode}}{2012}]{Quataert2012}
{Quataert} E.,  {Shiode} J.,  2012, \mn@doi [\mnras]
  {10.1111/j.1745-3933.2012.01264.x}, \href
  {https://ui.adsabs.harvard.edu/abs/2012MNRAS.423L..92Q} {423, L92}

\bibitem[\protect\citeauthoryear{{Ransome}, {Habergham-Mawson}, {Darnley},
  {James}, {Filippenko}  \& {Schlegel}}{{Ransome} et~al.}{2021}]{Ransome2021}
{Ransome} C.~L.,  {Habergham-Mawson} S.~M.,  {Darnley} M.~J.,  {James} P.~A.,
  {Filippenko} A.~V.,   {Schlegel} E.~M.,  2021, \mn@doi [\mnras]
  {10.1093/mnras/stab1938}, \href
  {https://ui.adsabs.harvard.edu/abs/2021MNRAS.506.4715R} {506, 4715}

\bibitem[\protect\citeauthoryear{{Reguitti} et~al.,}{{Reguitti}
  et~al.}{2022}]{reguitti2022}
{Reguitti} A.,  et~al., 2022, \mn@doi [\aap] {10.1051/0004-6361/202243340},
  \href {https://ui.adsabs.harvard.edu/abs/2022A&A...662L..10R} {662, L10}

\bibitem[\protect\citeauthoryear{{Reguitti}, {Pignata}, {Pastorello},
  {Dastidar}, {Reichart}, {Haislip}  \& {Kouprianov}}{{Reguitti}
  et~al.}{2024}]{reguitti2024}
{Reguitti} A.,  {Pignata} G.,  {Pastorello} A.,  {Dastidar} R.,  {Reichart}
  D.~E.,  {Haislip} J.~B.,   {Kouprianov} V.~V.,  2024, \mn@doi [arXiv
  e-prints] {10.48550/arXiv.2403.10398}, \href
  {https://ui.adsabs.harvard.edu/abs/2024arXiv240310398R} {p. arXiv:2403.10398}

\bibitem[\protect\citeauthoryear{{Renzo}, {Farmer}, {Justham}, {G{\"o}tberg},
  {de Mink}, {Zapartas}, {Marchant}  \& {Smith}}{{Renzo}
  et~al.}{2020}]{Renzo2020}
{Renzo} M.,  {Farmer} R.,  {Justham} S.,  {G{\"o}tberg} Y.,  {de Mink} S.~E.,
  {Zapartas} E.,  {Marchant} P.,   {Smith} N.,  2020, \mn@doi [\aap]
  {10.1051/0004-6361/202037710}, \href
  {https://ui.adsabs.harvard.edu/abs/2020A&A...640A..56R} {640, A56}

\bibitem[\protect\citeauthoryear{{Schlafly} \& {Finkbeiner}}{{Schlafly} \&
  {Finkbeiner}}{2011}]{milkyway_reddening}
{Schlafly} E.~F.,  {Finkbeiner} D.~P.,  2011, \mn@doi [\apj]
  {10.1088/0004-637X/737/2/103}, \href
  {https://ui.adsabs.harvard.edu/abs/2011ApJ...737..103S} {737, 103}

\bibitem[\protect\citeauthoryear{{Schlegel}}{{Schlegel}}{1990}]{schlegel1990}
{Schlegel} E.~M.,  1990, in Bulletin of the American Astronomical Society.
  p.~1214

\bibitem[\protect\citeauthoryear{{Shivvers} et~al.,}{{Shivvers}
  et~al.}{2016}]{Shivvers2016}
{Shivvers} I.,  et~al., 2016, \mn@doi [\mnras] {10.1093/mnras/stw1528}, \href
  {https://ui.adsabs.harvard.edu/abs/2016MNRAS.461.3057S} {461, 3057}

\bibitem[\protect\citeauthoryear{{Smith}}{{Smith}}{2014}]{Smith_mass_loss_2014}
{Smith} N.,  2014, \mn@doi [\araa] {10.1146/annurev-astro-081913-040025}, \href
  {https://ui.adsabs.harvard.edu/abs/2014ARA&A..52..487S} {52, 487}

\bibitem[\protect\citeauthoryear{{Smith}}{{Smith}}{2017}]{smith2017_interacting}
{Smith} N.,  2017, in {Alsabti} A.~W.,  {Murdin} P.,  eds, , Handbook of
  Supernovae.
p.~403, \mn@doi{10.1007/978-3-319-21846-5_38}

\bibitem[\protect\citeauthoryear{{Smith} \& {Andrews}}{{Smith} \&
  {Andrews}}{2020}]{2017hcc_2020Nathan}
{Smith} N.,  {Andrews} J.~E.,  2020, \mn@doi [\mnras] {10.1093/mnras/staa3047},
  \href {https://ui.adsabs.harvard.edu/abs/2020MNRAS.499.3544S} {499, 3544}

\bibitem[\protect\citeauthoryear{{Smith} \& {Arnett}}{{Smith} \&
  {Arnett}}{2014}]{SmithArnett2014}
{Smith} N.,  {Arnett} W.~D.,  2014, \mn@doi [\apj]
  {10.1088/0004-637X/785/2/82}, \href
  {https://ui.adsabs.harvard.edu/abs/2014ApJ...785...82S} {785, 82}

\bibitem[\protect\citeauthoryear{{Smith} \& {McCray}}{{Smith} \&
  {McCray}}{2007}]{2007ApJ...671L..17S}
{Smith} N.,  {McCray} R.,  2007, \mn@doi [\apjl] {10.1086/524681}, \href
  {http://adsabs.harvard.edu/abs/2007ApJ...671L..17S} {671, L17}

\bibitem[\protect\citeauthoryear{{Smith}, {Foley}  \& {Filippenko}}{{Smith}
  et~al.}{2008a}]{Smith2008_2006jc}
{Smith} N.,  {Foley} R.~J.,   {Filippenko} A.~V.,  2008a, \mn@doi [\apj]
  {10.1086/587860}, \href
  {https://ui.adsabs.harvard.edu/abs/2008ApJ...680..568S} {680, 568}

\bibitem[\protect\citeauthoryear{{Smith}, {Chornock}, {Li}, {Ganeshalingam},
  {Silverman}, {Foley}, {Filippenko}  \& {Barth}}{{Smith}
  et~al.}{2008b}]{2008Smith}
{Smith} N.,  {Chornock} R.,  {Li} W.,  {Ganeshalingam} M.,  {Silverman} J.~M.,
  {Foley} R.~J.,  {Filippenko} A.~V.,   {Barth} A.~J.,  2008b, \mn@doi [\apj]
  {10.1086/591021}, \href
  {https://ui.adsabs.harvard.edu/abs/2008ApJ...686..467S} {686, 467}

\bibitem[\protect\citeauthoryear{{Smith} et~al.,}{{Smith}
  et~al.}{2009}]{smith2009}
{Smith} N.,  et~al., 2009, \mn@doi [\apj] {10.1088/0004-637X/695/2/1334}, \href
  {https://ui.adsabs.harvard.edu/abs/2009ApJ...695.1334S} {695, 1334}

\bibitem[\protect\citeauthoryear{{Smith}, {Li}, {Silverman}, {Ganeshalingam}
  \& {Filippenko}}{{Smith} et~al.}{2011}]{2011MNRAS.415..773S}
{Smith} N.,  {Li} W.,  {Silverman} J.~M.,  {Ganeshalingam} M.,   {Filippenko}
  A.~V.,  2011, \mn@doi [\mnras] {10.1111/j.1365-2966.2011.18763.x}, \href
  {https://ui.adsabs.harvard.edu/abs/2011MNRAS.415..773S} {415, 773}

\bibitem[\protect\citeauthoryear{{Smith}, {Silverman}, {Filippenko}, {Cooper},
  {Matheson}, {Bian}, {Weiner}  \& {Comerford}}{{Smith}
  et~al.}{2012a}]{2012Smith}
{Smith} N.,  {Silverman} J.~M.,  {Filippenko} A.~V.,  {Cooper} M.~C.,
  {Matheson} T.,  {Bian} F.,  {Weiner} B.~J.,   {Comerford} J.~M.,  2012a,
  \mn@doi [\aj] {10.1088/0004-6256/143/1/17}, \href
  {https://ui.adsabs.harvard.edu/abs/2012AJ....143...17S} {143, 17}

\bibitem[\protect\citeauthoryear{{Smith}, {Mauerhan}, {Silverman},
  {Ganeshalingam}, {Filippenko}, {Cenko}, {Clubb}  \& {Kandrashoff}}{{Smith}
  et~al.}{2012b}]{2011hw_smith2012}
{Smith} N.,  {Mauerhan} J.~C.,  {Silverman} J.~M.,  {Ganeshalingam} M.,
  {Filippenko} A.~V.,  {Cenko} S.~B.,  {Clubb} K.~I.,   {Kandrashoff} M.~T.,
  2012b, \mn@doi [\mnras] {10.1111/j.1365-2966.2012.21849.x}, \href
  {https://ui.adsabs.harvard.edu/abs/2012MNRAS.426.1905S} {426, 1905}

\bibitem[\protect\citeauthoryear{{Smith}, {Mauerhan}  \& {Prieto}}{{Smith}
  et~al.}{2014}]{smith2014}
{Smith} N.,  {Mauerhan} J.~C.,   {Prieto} J.~L.,  2014, \mn@doi [\mnras]
  {10.1093/mnras/stt2269}, \href
  {https://ui.adsabs.harvard.edu/abs/2014MNRAS.438.1191S} {438, 1191}

\bibitem[\protect\citeauthoryear{{Smith}, {Andrews}, {Milne}, {Filippenko},
  {Brink}, {Kelly}, {Yuk}  \& {Jencson}}{{Smith}
  et~al.}{2024}]{2015dalate_smith2024}
{Smith} N.,  {Andrews} J.~E.,  {Milne} P.,  {Filippenko} A.~V.,  {Brink} T.~G.,
   {Kelly} P.~L.,  {Yuk} H.,   {Jencson} J.~E.,  2024, \mn@doi [\mnras]
  {10.1093/mnras/stae726}, \href
  {https://ui.adsabs.harvard.edu/abs/2024MNRAS.530..405S} {530, 405}

\bibitem[\protect\citeauthoryear{{Soker}}{{Soker}}{2013}]{Soker2013}
{Soker} N.,  2013, \mn@doi [arXiv e-prints] {10.48550/arXiv.1302.5037}, \href
  {https://ui.adsabs.harvard.edu/abs/2013arXiv1302.5037S} {p. arXiv:1302.5037}

\bibitem[\protect\citeauthoryear{{Stanek} \& {Kochanek}}{{Stanek} \&
  {Kochanek}}{2021}]{2021TNSTR.767....1S}
{Stanek} K.~Z.,  {Kochanek} C.~S.,  2021, Transient Name Server Discovery
  Report, \href {https://ui.adsabs.harvard.edu/abs/2021TNSTR.767....1S}
  {2021-767, 1}

\bibitem[\protect\citeauthoryear{{Stritzinger} et~al.,}{{Stritzinger}
  et~al.}{2012}]{2005ip_2006jd_Stritzinger2012}
{Stritzinger} M.,  et~al., 2012, \mn@doi [\apj] {10.1088/0004-637X/756/2/173},
  \href {https://ui.adsabs.harvard.edu/abs/2012ApJ...756..173S} {756, 173}

\bibitem[\protect\citeauthoryear{{Sun}, {Maund}  \& {Crowther}}{{Sun}
  et~al.}{2020}]{Sun2020}
{Sun} N.-C.,  {Maund} J.~R.,   {Crowther} P.~A.,  2020, \mn@doi [\mnras]
  {10.1093/mnras/staa2277}, \href
  {https://ui.adsabs.harvard.edu/abs/2020MNRAS.497.5118S} {497, 5118}

\bibitem[\protect\citeauthoryear{{Suzuki}, {Moriya}  \& {Takiwaki}}{{Suzuki}
  et~al.}{2019}]{Suzuki2019}
{Suzuki} A.,  {Moriya} T.~J.,   {Takiwaki} T.,  2019, \mn@doi [\apj]
  {10.3847/1538-4357/ab5a83}, \href
  {https://ui.adsabs.harvard.edu/abs/2019ApJ...887..249S} {887, 249}

\bibitem[\protect\citeauthoryear{{Thomas} et~al.,}{{Thomas}
  et~al.}{2022}]{Thomas2022_2014C}
{Thomas} B.~P.,  et~al., 2022, \mn@doi [\apj] {10.3847/1538-4357/ac5fa6}, \href
  {https://ui.adsabs.harvard.edu/abs/2022ApJ...930...57T} {930, 57}

\bibitem[\protect\citeauthoryear{{Th{\"o}ne} et~al.,}{{Th{\"o}ne}
  et~al.}{2017}]{Thone2017}
{Th{\"o}ne} C.~C.,  et~al., 2017, \mn@doi [\aap] {10.1051/0004-6361/201629968},
  \href {https://ui.adsabs.harvard.edu/abs/2017A&A...599A.129T} {599, A129}

\bibitem[\protect\citeauthoryear{{Tonry} et~al.,}{{Tonry}
  et~al.}{2018}]{Tonry2018}
{Tonry} J.~L.,  et~al., 2018, \mn@doi [\pasp] {10.1088/1538-3873/aabadf}, \href
  {https://ui.adsabs.harvard.edu/abs/2018PASP..130f4505T} {130, 064505}

\bibitem[\protect\citeauthoryear{{Valenti} et~al.,}{{Valenti}
  et~al.}{2014}]{Valenti_floyds}
{Valenti} S.,  et~al., 2014, \mn@doi [\mnras] {10.1093/mnrasl/slt171}, \href
  {https://ui.adsabs.harvard.edu/abs/2014MNRAS.438L.101V} {438, L101}

\bibitem[\protect\citeauthoryear{{Valenti} et~al.,}{{Valenti}
  et~al.}{2016}]{valenti_lcogtsnpipe}
{Valenti} S.,  et~al., 2016, \mn@doi [\mnras] {10.1093/mnras/stw870}, \href
  {https://ui.adsabs.harvard.edu/abs/2016MNRAS.459.3939V} {459, 3939}

\bibitem[\protect\citeauthoryear{{Vink}}{{Vink}}{2018}]{Vink2018}
{Vink} J.~S.,  2018, \mn@doi [\aap] {10.1051/0004-6361/201833352}, \href
  {https://ui.adsabs.harvard.edu/abs/2018A&A...619A..54V} {619, A54}

\bibitem[\protect\citeauthoryear{{Weis} \& {Bomans}}{{Weis} \&
  {Bomans}}{2020}]{2020Weis}
{Weis} K.,  {Bomans} D.~J.,  2020, \mn@doi [Galaxies]
  {10.3390/galaxies8010020}, \href
  {https://ui.adsabs.harvard.edu/abs/2020Galax...8...20W} {8, 20}

\bibitem[\protect\citeauthoryear{{Woosley}}{{Woosley}}{2017}]{Woosley2017}
{Woosley} S.~E.,  2017, \mn@doi [\apj] {10.3847/1538-4357/836/2/244}, \href
  {https://ui.adsabs.harvard.edu/abs/2017ApJ...836..244W} {836, 244}

\bibitem[\protect\citeauthoryear{{Woosley}, {Blinnikov}  \& {Heger}}{{Woosley}
  et~al.}{2007}]{Woosley2007}
{Woosley} S.~E.,  {Blinnikov} S.,   {Heger} A.,  2007, \mn@doi [\nat]
  {10.1038/nature06333}, \href
  {https://ui.adsabs.harvard.edu/abs/2007Natur.450..390W} {450, 390}

\bibitem[\protect\citeauthoryear{{Zackay}, {Ofek}  \& {Gal-Yam}}{{Zackay}
  et~al.}{2016}]{ZOGY}
{Zackay} B.,  {Ofek} E.~O.,   {Gal-Yam} A.,  2016, \mn@doi [\apj]
  {10.3847/0004-637X/830/1/27}, \href
  {https://ui.adsabs.harvard.edu/abs/2016ApJ...830...27Z} {830, 27}

\makeatother
\end{thebibliography}




\appendix

\section{Photometry}

\begin{table*}
    \centering
    \begin{center}
\begin{tabular}{ccccccccccc}
\hline
MJD (59287.36 +) & U & B & V & R & I & g & r & i & source$^{\dagger}$ \\ 
(day) & (mag) & (mag) & (mag) & (mag) & (mag) & (mag) & (mag) & (mag) &  \\ 
\hline
5.9 & - & - & $15.64 \pm 0.04$ & $15.70 (0.04)$ & $15.62 (0.04)$ & - & - & - & japan \\
7.6 & $14.80 (0.01)$ & $15.66 (0.04)$ & $15.51 \pm 0.02$ & - & - & $15.58 (0.03)$ & $15.48 (0.04)$ & $15.71 (0.01)$ & 1m0-08 \\
10.6 & $14.63 (0.01)$ & $15.45 (0.02)$ & $15.31 \pm 0.02$ & - & - & $15.39 (0.01)$ & $15.31 (0.01)$ & $15.56 (0.06)$ & 1m0-05 \\
13.5 & - & $15.33 (0.01)$ & $15.19 \pm 0.02$ & - & - & - & - & - & 1m0-05 \\
13.7 & - & $15.43 (0.01)$ & $15.17 \pm 0.01$ & - & - & $15.35 (0.05)$ & $15.12 (0.01)$ & $15.33 (0.02)$ & 1m0-11 \\
13.9 & - & $15.44 (0.04)$ & $15.17 \pm 0.02$ & $15.22 (0.02)$ & $15.10 (0.02)$ & - & - & - & japan \\
14.9 & - & $15.37 (0.04)$ & $15.19 \pm 0.02$ & $15.16 (0.02)$ & $15.02 (0.02)$ & - & - & - & japan \\
16.5 & $14.60 (0.05)$ & $15.41 (0.02)$ & $15.17 \pm 0.02$ & - & - & $15.32 (0.02)$ & $15.09 (0.01)$ & $15.24 (0.06)$ & 1m0-04 \\
20.0 & $14.78 (0.02)$ & $15.48 (0.03)$ & $15.25 \pm 0.02$ & - & - & $15.35 (0.03)$ & $15.18 (0.01)$ & $15.34 (0.02)$ & 1m0-13 \\
21.0 & - & $15.68 (0.03)$ & $15.26 \pm 0.05$ & $15.21 (0.05)$ & $15.07 (0.03)$ & - & - & - & japan \\
22.0 & - & $15.73 (0.05)$ & $15.32 \pm 0.04$ & $15.21 (0.02)$ & $15.10 (0.03)$ & - & - & - & japan \\
22.9 & $14.94 (0.04)$ & $15.62 (0.01)$ & $15.31 \pm 0.02$ & - & - & $15.49 (0.05)$ & $15.24 (0.03)$ & $15.36 (0.04)$ & 1m0-11 \\
24.9 & - & $15.68 (0.04)$ & $15.33 \pm 0.02$ & $15.22 (0.03)$ & $15.08 (0.02)$ & - & - & - & japan \\
25.3 & $14.94 (0.05)$ & $15.66 (0.02)$ & $15.32 \pm 0.02$ & - & - & $15.50 (0.04)$ & $15.19 (0.01)$ & - & 1m0-05 \\
25.9 & - & $15.77 (0.02)$ & $15.43 \pm 0.02$ & $15.35 (0.02)$ & $15.19 (0.02)$ & - & - & - & japan \\
26.9 & - & $15.90 (0.02)$ & $15.54 \pm 0.02$ & $15.41 (0.02)$ & $15.27 (0.02)$ & - & - & - & japan \\
28.1 & $15.57 (0.04)$ & $16.01 (0.01)$ & $15.75 \pm 0.02$ & - & - & $15.82 (0.01)$ & $15.62 (0.01)$ & $15.62 (0.02)$ & 1m0-12 \\
29.9 & - & $16.21 (0.04)$ & $15.76 \pm 0.02$ & $15.68 (0.02)$ & $15.50 (0.02)$ & - & - & - & japan \\
30.8 & - & $16.39 (0.02)$ & $15.84 \pm 0.02$ & $15.80 (0.03)$ & $15.61 (0.02)$ & - & - & - & japan \\
31.7 & $15.92 (0.03)$ & $16.33 (0.02)$ & $15.88 \pm 0.01$ & - & - & $16.11 (0.01)$ & $15.79 (0.03)$ & $15.82 (0.02)$ & 1m0-03 \\
33.9 & - & - & $15.83 \pm 0.07$ & $15.73 (0.04)$ & $15.56 (0.04)$ & - & - & - & japan \\
35.0 & $15.94 (0.03)$ & $16.25 (0.02)$ & $15.83 \pm 0.03$ & - & - & $16.15 (0.09)$ & $15.78 (0.04)$ & $15.80 (0.01)$ & 1m0-10 \\
34.8 & - & $16.26 (0.02)$ & $15.81 \pm 0.02$ & $15.70 (0.02)$ & $15.56 (0.02)$ & - & - & - & japan \\
37.8 & - & $16.35 (0.03)$ & $15.84 \pm 0.03$ & $15.73 (0.02)$ & $15.57 (0.02)$ & - & - & - & japan \\
38.9 & - & $16.38 (0.03)$ & $15.90 \pm 0.02$ & $15.75 (0.03)$ & $15.61 (0.03)$ & - & - & - & japan \\
40.1 & $16.13 (0.03)$ & $16.56 (0.05)$ & $16.06 \pm 0.02$ & - & - & $16.31 (0.02)$ & $15.97 (0.02)$ & $16.00 (0.04)$ & 1m0-12 \\
39.8 & - & $16.48 (0.04)$ & $16.00 \pm 0.02$ & $15.87 (0.02)$ & $15.73 (0.02)$ & - & - & - & japan \\
43.8 & - & $16.80 (0.03)$ & $16.37 \pm 0.04$ & $16.13 (0.03)$ & $16.04 (0.03)$ & - & - & - & japan \\
44.8 & - & - & $16.44 \pm 0.09$ & - & - & - & - & - & japan \\
47.7 & $16.98 (0.10)$ & $17.32 (0.01)$ & $16.77 \pm 0.01$ & - & - & $16.96 (0.02)$ & $16.42 (0.01)$ & $16.67 (0.09)$ & 1m0-11 \\
52.7 & $17.67 (0.01)$ & $17.96 (0.03)$ & $17.46 \pm 0.01$ & - & - & $17.74 (0.03)$ & $17.09 (0.03)$ & $17.21 (0.02)$ & 1m0-03 \\
54.9 & - & $17.97 (0.06)$ & $17.37 \pm 0.04$ & $17.10 (0.04)$ & - & - & - & - & japan \\
55.4 & $17.91 (0.02)$ & $18.14 (0.05)$ & $17.45 \pm 0.03$ & - & - & $17.86 (0.02)$ & $17.19 (0.03)$ & $17.26 (0.02)$ & 1m0-04 \\
55.4 & - & $18.13 (0.10)$ & $16.88 \pm 0.15$ & - & - & - & - & - & 1m0-05 \\
57.3 & $18.16 (0.05)$ & $18.17 (0.03)$ & $17.59 \pm 0.01$ & - & - & $18.03 (0.01)$ & $17.31 (0.02)$ & $17.37 (0.04)$ & 1m0-05 \\
59.2 & $18.04 (0.01)$ & $18.28 (0.02)$ & $17.78 \pm 0.02$ & - & - & - & - & - & 1m0-12 \\
58.8 & - & - & $17.62 \pm 0.04$ & $17.33 (0.04)$ & $17.11 (0.04)$ & - & - & - & japan \\
59.2 & - & - & - & - & - & $18.06 (0.05)$ & $17.41 (0.03)$ & $17.45 (0.01)$ & 1m0-12 \\
60.4 & $18.27 (0.01)$ & $18.68 (0.03)$ & $17.91 \pm 0.04$ & - & - & $18.34 (0.02)$ & $17.51 (0.03)$ & $17.58 (0.03)$ & 1m0-04 \\
62.6 & $18.22 (0.09)$ & $18.67 (0.03)$ & $17.84 \pm 0.41$ & - & - & $18.45 (0.03)$ & $17.58 (0.02)$ & $17.62 (0.07)$ & 1m0-03 \\
62.4 & $17.86 (0.23)$ & $18.44 (0.12)$ & - & - & - & - & - & - & 1m0-05 \\
64.3 & $18.77 (0.25)$ & $18.99 (0.06)$ & $18.20 \pm 0.09$ & - & - & $18.69 (0.05)$ & $17.83 (0.02)$ & $17.87 (0.05)$ & 1m0-04 \\
66.1 & $18.57 (0.19)$ & $19.08 (0.14)$ & $18.44 \pm 0.05$ & - & - & $18.75 (0.03)$ & $17.97 (0.02)$ & $17.93 (0.02)$ & 1m0-13 \\
67.8 & $18.51 (0.02)$ & $19.13 (0.04)$ & $18.54 \pm 0.07$ & - & - & $18.80 (0.05)$ & $18.01 (0.02)$ & $18.10 (0.03)$ & 1m0-11 \\
69.2 & $18.47 (0.17)$ & $19.16 (0.16)$ & $18.40 \pm 0.24$ & - & - & $18.78 (0.19)$ & $18.09 (0.06)$ & $17.98 (0.09)$ & 1m0-13 \\
69.6 & $18.93 (0.13)$ & $19.32 (0.08)$ & $18.63 \pm 0.03$ & - & - & $18.99 (0.01)$ & $18.12 (0.03)$ & $18.19 (0.01)$ & 1m0-11 \\
75.3 & $19.16 (0.06)$ & $19.50 (0.09)$ & $18.99 \pm 0.03$ & - & - & $19.42 (0.07)$ & $18.50 (0.10)$ & $18.46 (0.08)$ & 1m0-05 \\
79.9 & $19.21 (0.22)$ & $19.68 (0.13)$ & $19.24 \pm 0.20$ & - & - & $19.65 (0.17)$ & $18.93 (0.02)$ & $18.91 (0.02)$ & 1m0-10 \\
86.9 & $19.46 (0.17)$ & $19.90 (0.14)$ & $19.60 \pm 0.47$ & - & - & $20.00 (0.05)$ & $19.42 (0.10)$ & $19.56 (0.25)$ & 1m0-12 \\
102.3 & - & $19.99 (0.05)$ & $19.76 \pm 0.14$ & - & - & $19.99 (0.04)$ & $19.43 (0.16)$ & $19.75 (0.20)$ & 1m0-04 \\
113.2 & - & $20.73 (0.17)$ & $20.16 \pm 0.10$ & - & - & $20.19 (0.08)$ & $19.37 (0.03)$ & $19.71 (0.27)$ & 1m0-09 \\
124.3 & - & $20.71 (0.05)$ & $20.27 \pm 0.09$ & - & - & $20.20 (0.18)$ & $19.27 (0.08)$ & $19.49 (0.05)$ & 1m0-09 \\
135.6 & - & $20.72 (0.10)$ & $20.54 \pm 0.49$ & - & - & $20.31 (0.04)$ & $19.49 (0.11)$ & $19.73 (0.10)$ & 1m0-03 \\
148.2 & - & - & - & - & - & $19.93 (0.42)$ & $19.99 (0.02)$ & - & 1m0-04 \\
165.2 & - & $20.49 (0.50)$ & $20.25 \pm 0.33$ & - & - & - & - & - & 1m0-04 \\
165.2 & - & - & - & - & - & $20.16 (0.23)$ & $19.57 (0.34)$ & $19.32 (0.75)$ & 1m0-04 \\
\hline
\end{tabular}
$\dagger$ -- japan is used for the Kanata Telescope of Hiroshima University while 1m0-* are used for the telescopes of LCO. \\
\end{center}

    \caption{Log of photometric observations in {\it UBVRIgri} bands for SN~2021foa being observed from telescopes of India and Japan. All the {\it grico} reported magnitudes are in AB magnitude system and the {\it UBVRI} are in Vega magnitude system.}
    \label{tab:lco_japan_photometry}
\end{table*}

\begin{table*}
    \centering
    \begin{tabular}{ccccc}
\hline
MJD (59287.36 +) & K & H & J & source  \\ 
(day) & (mag) & (mag) & (mag) & tel  \\ 
\hline
3.32 & -- & 15.23(0.04) & 15.27(0.03)  & HONIR \\
18.42 & 14.44(0.04) &14.60(0.02) &14.67(0.02) & HONIR\\
31.31 & 14.47(0.04) &14.72(0.03) &14.98(0.02) & HONIR \\
56.23 & 15.84(0.07) & 16.35(0.05) &16.43(0.03) & HONIR \\
76.73 & 16.71(0.13) &17.24(0.09) &17.21(0.05)  & HONIR \\
\hline
\end{tabular}
    \caption{The table reports the log of {\it JHK} observations taken from 1.5m Kanata Telescope, Japan. All the magnitudes reported are in Vega magnitude system. }
    \label{tab:nir_photometry}
\end{table*}
\begin{table}
\caption {Log of spectroscopic observations of SN~2021foa. The phase is measured with respect to maximum (MJD$_{Vmax}=59301.8$).
\label{tab:2021foa_spec_obs}}
\begin{center}
\begin{tabular}{c c c c c c c c}
\hline \hline
Phase  &    Telescope  &     Instrument  &		Range & Disperser & Slit size \\
   (days)    &               &                 &             (\AA)   \\
\hline    
-10.8     &   3.8m Seimei    &    KOOLS-IFU    &   4000-8000 & Grism-Red/Blue  & 1$\arcsec$     \\
-8.3      &   2m FTN    &    FLOYDS    &    3400-9500  & Grism-Red/Blue  & 1$\arcsec$ \\
-5.8      &   3.8m Seimei & KOOLS-IFU  &   4000-8000   & VPH-Blue & --            \\
-3.2      &  2m FTN & FLOYDS      &  3400-9500  &  Grism-Red/Blue  & 1$\arcsec$ \\
7.2       &   3.8m Seimei & KOOLS-IFU    &   4000-8000 & VPH-Blue & --\\
9.7       &    2m FTN    &    FLOYDS   &   3400-9500  &  Grism-Red/Blue  & 1$\arcsec$  \\
14.2      &   3.8m Seimei & KOOLS-IFU    &   4000-8000  & VPH-Blue & --           \\
17.5      &  2m FTN & FLOYDS     &   3400-9500  &  Grism-Red/Blue  & 1$\arcsec$ \\
26.8      &   2m FTN & FLOYDS    &   3400-9500 &  Grism-Red/Blue  & 1$\arcsec$ \\
35.2      &    3.8m Seimei & KOOLS-IFU     &   4000-8000 & VPH-Blue & --       \\
40.6      &   2m FTN & FLOYDS    & 3400-9500  &  Grism-Red/Blue  & 1$\arcsec$  \\
49.5      &   2m FTN & FLOYDS      &   4000-8000 &  Grism-Red/Blue  & 1$\arcsec$            \\
58.6      &   2m FTN & FLOYDS    &   3400-9500  &  Grism-Red/Blue  & 1$\arcsec$ \\
69.5      &   2m FTN & FLOYDS    &   3400-9500 &  Grism-Red/Blue  & 1$\arcsec$ \\
\hline
\end{tabular}
\end{center}
\end{table}


\bsp	
\label{lastpage}
\end{document}